\documentclass[%
superscriptaddress,
twocolumn,
amsmath,amssymb,
aps,
prd,
floatfix,
]{revtex4}%{revtex4-2}

\usepackage{graphicx}
\usepackage{dcolumn}% Align table columns on decimal point
\usepackage{bm}% bold math
\usepackage{xcolor}
\usepackage[colorlinks=true, linkcolor=blue, citecolor=blue, urlcolor=blue]{hyperref}
\usepackage{cleveref}
\usepackage{diagbox}
\usepackage{multirow}
\usepackage{physics}
\usepackage{braket}

\DeclareRobustCommand{\fig}[1]{Fig.~\ref{fig:#1}}

\crefname{figure}{fig.}{figs.}
\Crefname{figure}{Fig.}{Figs.}
\crefname{table}{tab.}{tabs.}
\Crefname{table}{Tab.}{Tabs.}
\crefname{equation}{eq.}{eqns.}
\Crefname{equation}{Eq.}{Eqns.}
\crefname{listing}{lst.}{lsts.}
\Crefname{listing}{Lst.}{Lsts.}
\crefname{section}{sec.}{secs.}
\Crefname{section}{Sec.}{Secs.}
\crefname{appendix}{app.}{apps.}
\Crefname{appendix}{App.}{Apps.}

\newcommand{\pvec}{\vec{p}}
\newcommand{\qvec}{\vec{q}}
\newcommand{\xvec}{\vec{x}}
\newcommand{\yvec}{\vec{y}}
\newcommand{\zvec}{\vec{z}}
\newcommand{\ts}{t_{\rm sep}}
\newcommand{\ti}{t_{\rm ins}}
\newcommand{\mzero}{\overline{m}_0}

\newcommand{\qtilde}{\widetilde{q}}

\newcommand{\qcdcutoff}{\Lambda_{\rm QCD}}

\begin{document}

%\preprint{}

\title{Unpolarized proton PDF at NNLO from lattice QCD with physical quark masses}
%\thanks{}%

\author{Xiang Gao}
\email{gaox@anl.gov}
\affiliation{Physics Division, Argonne National Laboratory, Lemont, IL 60439, USA}

\author{Andrew D. Hanlon}
\email{ahanlon@bnl.gov}
\affiliation{Physics Department, Brookhaven National Laboratory, Bldg. 510A, Upton, New York 11973, USA}

\author{Jack Holligan}
\email{holligan@umd.edu}
\affiliation{Department of Physics, University of Maryland, College Park, MD 20742, USA}
\affiliation{Center for Frontier Nuclear Science, Stony Brook University, Stony Brook, NY 11794, USA}

\author{Nikhil Karthik}
\affiliation{American Physical Society, Hauppauge, New York 11788}
\affiliation{Department of Physics, Florida International University, Miami, Florida 33199}

\author{Swagato Mukherjee}
\affiliation{Physics Department, Brookhaven National Laboratory, Bldg. 510A, Upton, New York 11973, USA}

\author{Peter Petreczky}
\affiliation{Physics Department, Brookhaven National Laboratory, Bldg. 510A, Upton, New York 11973, USA}

\author{Sergey Syritsyn}
\affiliation{RIKEN-BNL Research Center, Brookhaven National Laboratory, Upton, New York 11973}
\affiliation{Department of Physics and Astronomy, Stony Brook University, Stony Brook, New York 11790}

\author{Yong Zhao}
\affiliation{Physics Division, Argonne National Laboratory, Lemont, IL 60439, USA}

%\collaboration{}

\date{\today}

\begin{abstract}
We present a lattice QCD calculation of the unpolarized isovector quark parton distribution function (PDF) of the proton utilizing a perturbative matching at next-to-next-to-leading-order (NNLO).
The calculations are carried out using a single ensemble of gauge configurations generated with $N_f = 2 + 1$ highly-improved staggered quarks with physical masses and a lattice spacing of $a = 0.076$ fm.
We use one iteration of hypercubic smearing on these gauge configurations, and the resulting smeared configurations are then used for all aspects of the subsequent calculation.
For the valence quarks, we use the Wilson-clover action with physical quark masses.
We consider several methods for extracting information on the PDF.
We first extract the lowest four Mellin moments using the leading-twist operator product expansion approximation.
Then, we determine the $x$ dependence of the PDF through a deep neural network within the pseudo-PDF approach and additionally through the framework of large-momentum effective theory utilizing a hybrid renormalization scheme.
This is the first application of the NNLO matching coefficients for the nucleon directly at the physical point.
\end{abstract}

%\keywords{}

\maketitle

%\tableofcontents

\section{Introduction}
\label{sec:intro}

Parton distribution functions (PDFs) are universal quantities that can be used to compute cross sections for various hard-scattering processes.
Their universality makes them particularly useful, in that they can be determined from some subset of processes and then used in predictions of other processes.
Additionally, their determination gives insights into the structure of hadrons.
The partonic internal-structure of the nucleon was established long ago from deep inelastic scattering experiments at SLAC.
This prompted significant experimental and theoretical efforts to reduce the uncertainty in these quantities.
Using the data provided from several experiments (e.g. the Tevatron, HERA, the LHC, etc.), the collinear structure for unpolarized and polarized nucleons has been determined to a few-percent accuracy from various recent global analyses~\cite{Alekhin:2017kpj,Hou:2019efy,Bailey:2020ooq,NNPDF:2021njg}.
The success of these programs has been significant,
yet there is still a need to further reduce the uncertainties on PDFs, and much remains unknown about the full internal structure of the nucleon.
For example, there is little experimental data for the distribution of a transversely polarized nucleon.
Further, even less is known about the generalized parton distributions (GPDs) and the transverse momentum dependent (TMD) PDFs.
These quantities will be better understood through data collected from the JLab 12 GeV upgrade~\cite{Dudek:2012vr} and the electron-ion collider~\cite{Accardi:2012qut,AbdulKhalek:2021gbh}.

Calculations of these quantities from first principles would be very useful.
There has been significant progress in the necessary theoretical developments for the determination of PDFs from lattice QCD.
This will allow for supplementing the existing experimental data in order to further reduce the global analysis uncertainties, as well as fill the gaps where little experimental data exists.
For recent reviews on the methods and progress on computing light-cone PDFs from lattice QCD see e.g. Refs.~\cite{Cichy:2018mum,Zhao:2018fyu,Radyushkin:2019mye,Ji:2020ect,Constantinou:2020pek,Constantinou:2020hdm,Cichy:2021lih}.

Although there are several methods that have been considered for the extraction of PDFs from the lattice,
two methods in particular have been extensively used in recent years with tremendous success.
These are the quasidistribution approach based on large-momentum effective theory (LaMET)~\cite{Ji:2013dva,Ji:2014gla}, and the pseudodistribution approach~\cite{Radyushkin:2017cyf,Orginos:2017kos} based on a short-distance factorization (SDF) that has also been applied in the current-current correlator approach~\cite{Braun:2007wv,Ma:2017pxb}.
The two methods become equivalent at infinite momentum, but differ in their systematics at finite momentum, and there has been much work regarding their differing advantages~\cite{Ji:2022ezo}.
SDF relies on the validity of the leading-twist operator product expansion (OPE), which can also be used to directly extract the first few Mellin moments.
While LaMET relies on large momentum to control the power corrections in the matching which become large for values of $x \sim 0$ and $x\sim 1$.

There have been several applications of these methods to the unpolarized isovector quark PDF of the nucleon, e.g. from the quasi-PDF approach~\cite{Alexandrou:2018pbm,Alexandrou:2019lfo,Chen:2018xof,LatticeParton:2018gjr,Fan:2020nzz,Alexandrou:2020qtt,Lin:2020fsj} and the pseudo-PDF approach~\cite{Orginos:2017kos,Joo:2019jct,Joo:2020spy,Bhat:2020ktg,Karpie:2021pap,Egerer:2021ymv,Bhat:2022zrw}.
In this paper, we extract this PDF using both methods, which allows us to better understand the different systematics between each method.
This is the first such calculation utilizing a matching kernel at next-to-next-to-leading order (NNLO) directly with physical quark masses.

The rest of the paper is organized as follows. First, in \Cref{sec:lattice} we describe the lattice setup.
Then, in \Cref{sec:analysis} we extract the bare matrix elements needed for the PDF determination.
Next, in \Cref{sec:leading_twist_OPE} we extract the first few Mellin moments via the leading-twist OPE approximation.
In \Cref{sec:dnn} we use a deep neural network (DNN) to overcome the inverse problem that arises within the pseudeo-PDF method.
Our final PDF extraction method is given in \Cref{sec:xspace} and uses a matching in $x$-space within the LaMET framework from hybrid renormalized matrix elements.
Finally, our conclusions are given in \Cref{sec:conc}.
\section{Lattice Details}
\label{sec:lattice}

The setup and calculations used here are very similar to our previous work on the pion valence PDF~\cite{Gao:2021dbh,Gao:2022iex}.
For convenience, we repeat the more salient details.

We use a single ensemble of $N_f = 2 + 1$ highly improved staggered quarks (HISQ)~\cite{Follana:2006rc} with physical masses, a size of $64^3 \times 64$, and a lattice spacing of $a = 0.076$ fm generated by the HotQCD collaboration~\cite{Bazavov:2019www}.
In the valence sector, we use the tree-level tadpole-improved Wilson-clover action with $c_{sw} = u_0^{-3/4} = 1.0372$ (where $u_0$ is the plaquette expectation value on the hypercubic (HYP) smeared gauge configurations), physical quark masses, and one step of HYP smearing~\cite{Hasenfratz:2001hp}.
This is the same ensemble used in our previous work~\cite{Gao:2022iex}.

Our calculations are carried out on GPUs with the Qlua software suite~\cite{qlua}, which utilizes the multigrid solver in QUDA~\cite{Clark:2009wm,Babich:2011np} for the quark propagators.
In order to reduce the computational cost, we employ all-mode averaging (AMA)~\cite{Shintani:2014vja},
which involves performing several low-precision (sloppy) solves in addition to a small number of high-precision (exact) solves to correct for the bias introduced from the sloppy solves alone.
The stopping criterion for our solver is $10^{-10}$ and $10^{-4}$ for exact and sloppy solves, respectively.

The nucleon interpolating operators used in this work are given by
\begin{equation}
    N^{(s)}_\alpha (x, t) = \varepsilon_{abc} u^{(s)}_{a \alpha} (x, t) (u_b^{(s)} (x, t)^T C \gamma_5 d^{(s)}_c (x, t)) ,
\end{equation}
where $C = \gamma_t \gamma_y$ is the charge-conjugation matrix, and the superscript on the quark fields indicates whether the quarks are smeared ($s = S$) or not ($s = P$).
Additionally, we use momentum smearing~\cite{Bali:2016lva} in order to improve the overlaps with highly boosted hadron states.
Our choice for the boosted quarks is $\frac{2 \pi k_z}{L} \hat{z}$, where $k_z$ depends on the choice of the final momentum projection in our three-point functions $\pvec_f \equiv \frac{2 \pi n_z}{L} \hat{z}$.
Note that one might naively choose $k_z = n_z/3$,
as there are three valence quarks in a baryon.
However, it was shown in Ref.~\cite{Bali:2016lva} that a value of $k_z \approx 0.45 n_z$ was the optimal choice for the nucleon at their pion mass $m_\pi \approx 295$ MeV,
and we make similar choices for $k_z$ in this work.

Some of the more important details of our setup can be found in \Cref{tab:setup}.

\begin{table}
\centering
\begin{tabular}{c|c|c|c|c|c|c}
\hline
\hline
Ensembles & $m_\pi$ & $N_{\rm cfg}$ & $n_z$ & $k_z$ & $\ts/a$ & (\#ex,\#sl) \\
$a,L_t \times L_s^3$ & (GeV) & & & \\
\hline
$a=0.076$ fm     & 0.14 & 350 & 0 & 0 & 6         & (1, 16) \\
$64 \times 64^3$ &      &     & 0 & 0 & 8,10      & (1, 32) \\
                 &      &     & 0 & 0 & 12        & (2, 64) \\
                 &      &     & 1 & 0 & 6,8,10,12 & (1, 32) \\
                 &      &     & 4 & 2 & 6         & (1, 32) \\
                 &      &     & 4 & 2 & 8,10,12   & (4, 128) \\
                 &      &     & 6 & 3 & 6         & (1, 20) \\
                 &      &     & 6 & 3 & 8         & (4, 100) \\
                 &      &     & 6 & 3 & 10,12     & (5, 140) \\
\hline
\hline
\end{tabular}
\caption{Some details on the ensemble and the statistics gathered for our calculation.
         The $n_z$ give the momentum projection at the sink for the three-point functions in integer units of the lattice.
         The $k_z$ are the boost momentum for the smeared quark fields.
         The various sink-source separations are given by $t_{\rm sep}$.
         And, the number of sources used for exact and sloppy solves is given by \#ex and \#sl, respectively.}
\label{tab:setup}
\end{table}

\subsection{Two-point correlation functions}
\label{sec:two_point}

The two-point correlation functions we compute are given by
\begin{equation}
\begin{split}
    C^{\rm 2pt} & (\pvec, \ts; \xvec, t_0) = \\
    &\sum_{\yvec} e^{-i \pvec \cdot (\yvec - \xvec)} \mathcal{P}^{\rm 2pt}_{\alpha \beta} \braket{N^{(s)}_\alpha (\yvec, \ts + t_0) \overline{N}^{(s^\prime)}_\beta (\xvec, t_0)} ,
\end{split}
\end{equation}
where we project to the positive parity states with $\mathcal{P}^{\rm 2pt} = \frac{1}{2}(1 + \gamma_t)$.
In order to increase statistics, we average these correlators with the negative time correlators while changing the projector to $\mathcal{P}^{\rm 2pt} = \frac{1}{2}(1 - \gamma_t)$.
The change in parity projector is needed because the backward propagating antibaryon is not the antiparticle of the forward propagating baryon but of its parity partner.
We always use a smeared source ($s^\prime = S$)
but consider both smeared ($s=S$) and unsmeared ($s=P$) sinks which leads to SS and SP correlators respectively.
The SS and SP correlators share the same energy eigenstates, but differ in their overlaps onto these states allowing for a more reliable determination of the spectrum by looking for agreement in the spectrum determined from each.

For each value of $k_z$, we computed two-point correlators with momentum projections at the sink of $\pvec = \frac{2 \pi n_z}{L} \hat{z}$ for all $n_z \le 10$,
which is larger than the optimal $n_z$ even for our largest $k_z$.
All of these momenta are not strictly needed, as we only included a much smaller number of sink momentum projections in the three point functions.
However, the larger range of momentum for the two-point functions comes at a minimal extra cost and gives us more confidence in our analysis to clearly see a consistent momentum dependence for the spectrum.

\subsection{Three-point correlation functions}
\label{sec:three_point}

The three-point correlation functions we computed are of the form
\begin{equation}
\begin{split}
    C^{\rm 3pt}_\Gamma & (\pvec_f , \qvec , \ts, \ti, z ; \xvec, t_0) = \\
    & \sum_{\yvec, \zvec_0} e^{-i \pvec_f \cdot (\yvec - \xvec)} e^{-i \qvec \cdot (\xvec - \zvec_0)} \mathcal{P}^{\rm 3pt}_{\alpha \beta} \\
    &\times \braket{N_\alpha (\yvec, \ts + t_0) \mathcal{O}^\Gamma (\zvec_0 + z \hat{z}, \ti + t_0) \overline{N}_\beta (\xvec, t_0)} ,
\end{split}
\end{equation}
where $\qvec = \pvec_f - \pvec_i$, and the inserted operator is given by
\begin{equation}\label{eq.3pt}
\begin{split}
    \mathcal{O}&^\Gamma (\zvec_0 + z \hat{z}, \ti + t_0) = \overline{q} (\zvec_0, \ti + t_0) \Gamma \tau_3 \\
    &\times W(\zvec_0, \ti + t_0 ; \zvec_0 + z\hat{z} , \ti + t_0) q(\zvec_0 + z\hat{z}, \ti + t_0) ,
\end{split}
\end{equation}
where $q(\zvec, t)$ is an isospin doublet consisting of the light quark fields $u$ and $d$, 
$W$ is a Wilson line of length $z$ that connects the positions of the quark fields via a straight spatial path along the $z$-axis,
and $\tau_3$ results in the isovector combination (i.e. $\overline{u} \Gamma W u - \overline{d} \Gamma W d$) which leads to the cancellation of the disconnected diagrams.
The gauge links entering the Wilson line are the 1-step HYP-smeared gauge links.
The sink momenta considered here are all in the z-direction $\pvec_f = \frac{2 \pi}{L} n_z \hat{z}$ with $n_z \in \left\{0, 1, 4, 6 \right\}$ leading to momentum in physical units of $P_z = \left\{0, 0.25, 1.02, 1.53 \right\}$ GeV,
where $P_z \equiv \frac{2 \pi}{L} n_z$

In order to access the unpolarized PDF, we use $\mathcal{P}^{\rm 3pt} = \frac{1}{2}(1 + \gamma_t)$.
Additionally, we have the choice of $\Gamma = \gamma_t, \gamma_z$.
But, to avoid mixing under renormalization, we use $\Gamma = \gamma_t$~\cite{Constantinou:2017sej,Chen:2017mie}.

\section{Analysis of Correlation functions}
\label{sec:analysis}

Our goal is to obtain the bare ground-state matrix elements from the three-point functions.
However, there can be significant unwanted contributions to these correlators coming from excited states.
These contributions are especially important for the present work at the physical pion mass,
as the contamination from higher states becomes more severe as the pion mass is lowered.
Thus, in general we must include the effects of excited states in our analysis.
In practice, it is very difficult to reliably separate these effects from the desired ground state matrix elements directly from the three-point functions.
Fortunately, we can extract the lowest few energies from the two-point functions and use this information in our fits to the three-point functions.

\subsection{Two-point function analysis}

\begin{figure}
    \centering
    \includegraphics[width=\columnwidth]{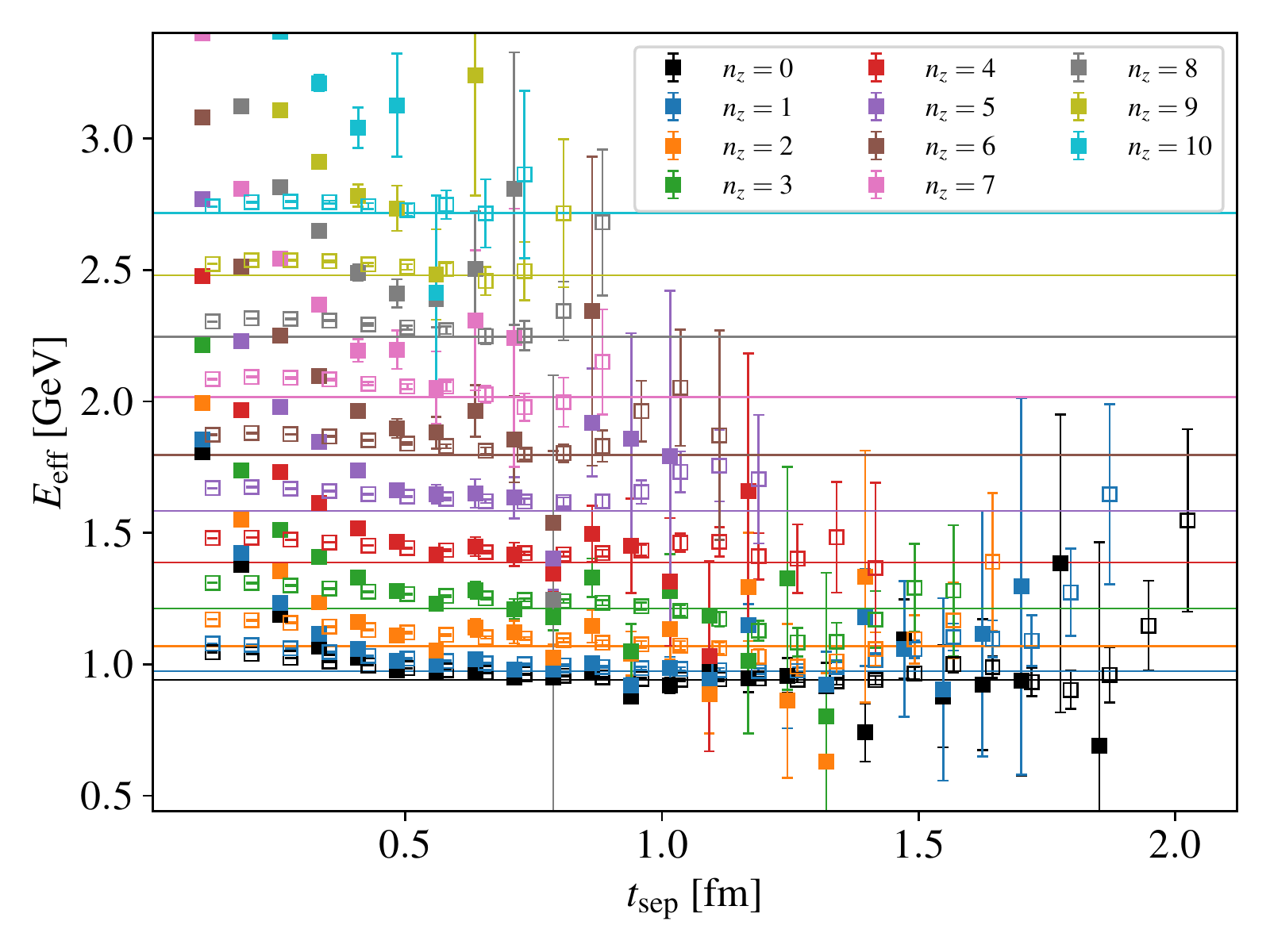}
    \caption{The effective energies from SS (filled) and SP (open) two-point correlators for all $n_z \le 10$.
             The horizontal lines correspond to predictions of the ground-state energies for each $n_z$ using
             the continuum dispersion relation with the estimate for the ground-state energy extracted
             from a one-state fit to the $n_z = 0$ SS two-point correlator.}
    \label{fig:effective_energies}
\end{figure}

Our aim in this subsection is twofold.
First, as already pointed out, we would like to determine the energies that contribute to the three-point functions.
And, second, we need to understand at what $\ts$ and $\ti$ various energies are no longer relevant.
In \Cref{fig:effective_energies}, we show the effective energies for all two-point correlators we have computed,
along with predictions for the ground-state energies for all $n_z \le 10$ using the ground-state energy for $n_z = 0$ with the continuum dispersion relation.

The fit form we use comes from the spectral decomposition for the two-point functions
\begin{equation}
    C^{\rm 2pt}(\pvec, \ts) = \sum_n A_\alpha^{(n)} (\pvec) A_\alpha^{(n)} (\pvec)^\ast e^{-E_n (\pvec) \ts} ,
\end{equation}
where $A_\alpha^{(n)} \equiv \bra{\Omega} N_\beta \mathcal{P}^{\rm 2pt}_{\beta \alpha} \ket{n, \pvec}$ ($\ket{\Omega}$ denotes the vacuum state), thermal effects are ignored, and $t_0$ is shifted to zero.
We then truncate the number of contributing states to $N$ and rearrange to find
\begin{equation}
    C^{\rm 2pt}_N (\pvec, \ts) = C_0 e^{-E_0 \ts} \Big[ 1 + \sum_{i=1}^{N - 1} R_i \prod_{j=1}^i e^{-\Delta_{j, j-1} \ts} \Big] ,
\end{equation}
where $\Delta_{j, j-1} \equiv E_j (\pvec) - E_{j-1} (\pvec)$, $R_i = C_i / C_0$, and $C_n = A^{(n)}_\alpha (\pvec) A^{(n)}_\alpha (\pvec)^\ast$.
This particular form is useful in that it guarantees a proper ordering of the states.

\begin{figure}
    \centering
    \includegraphics[width=0.49\columnwidth]{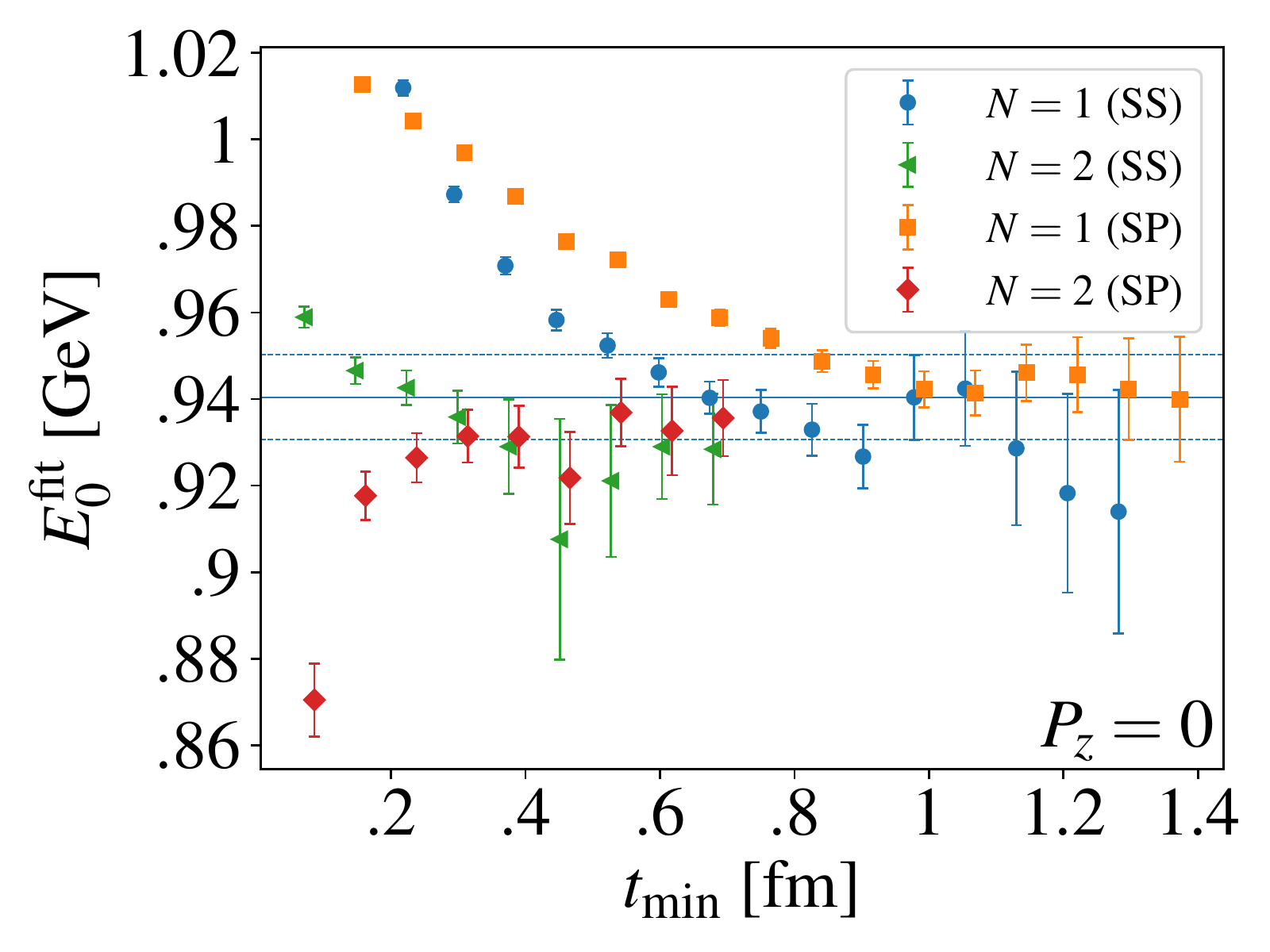}
    \includegraphics[width=0.49\columnwidth]{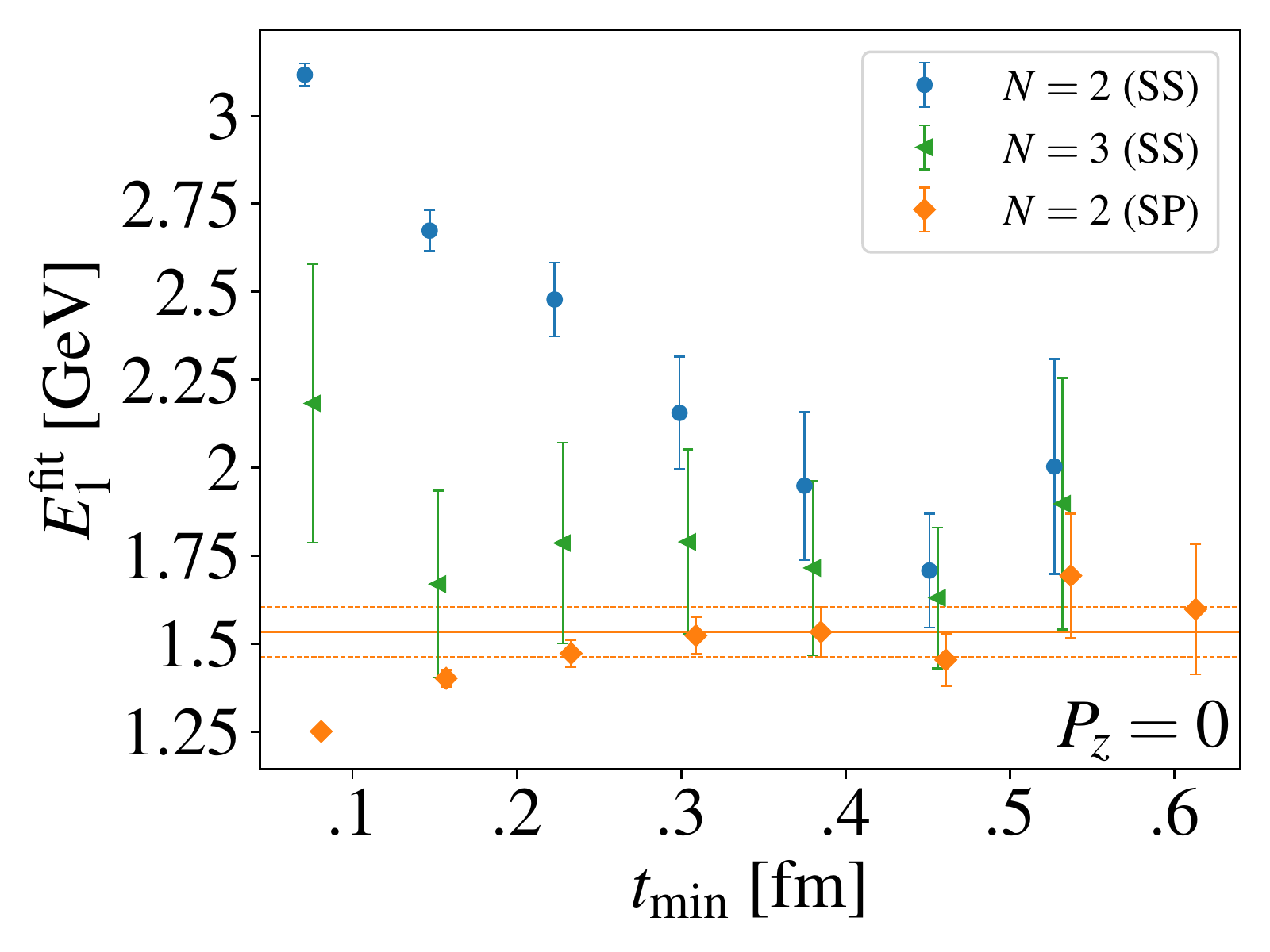}
    \includegraphics[width=0.49\columnwidth]{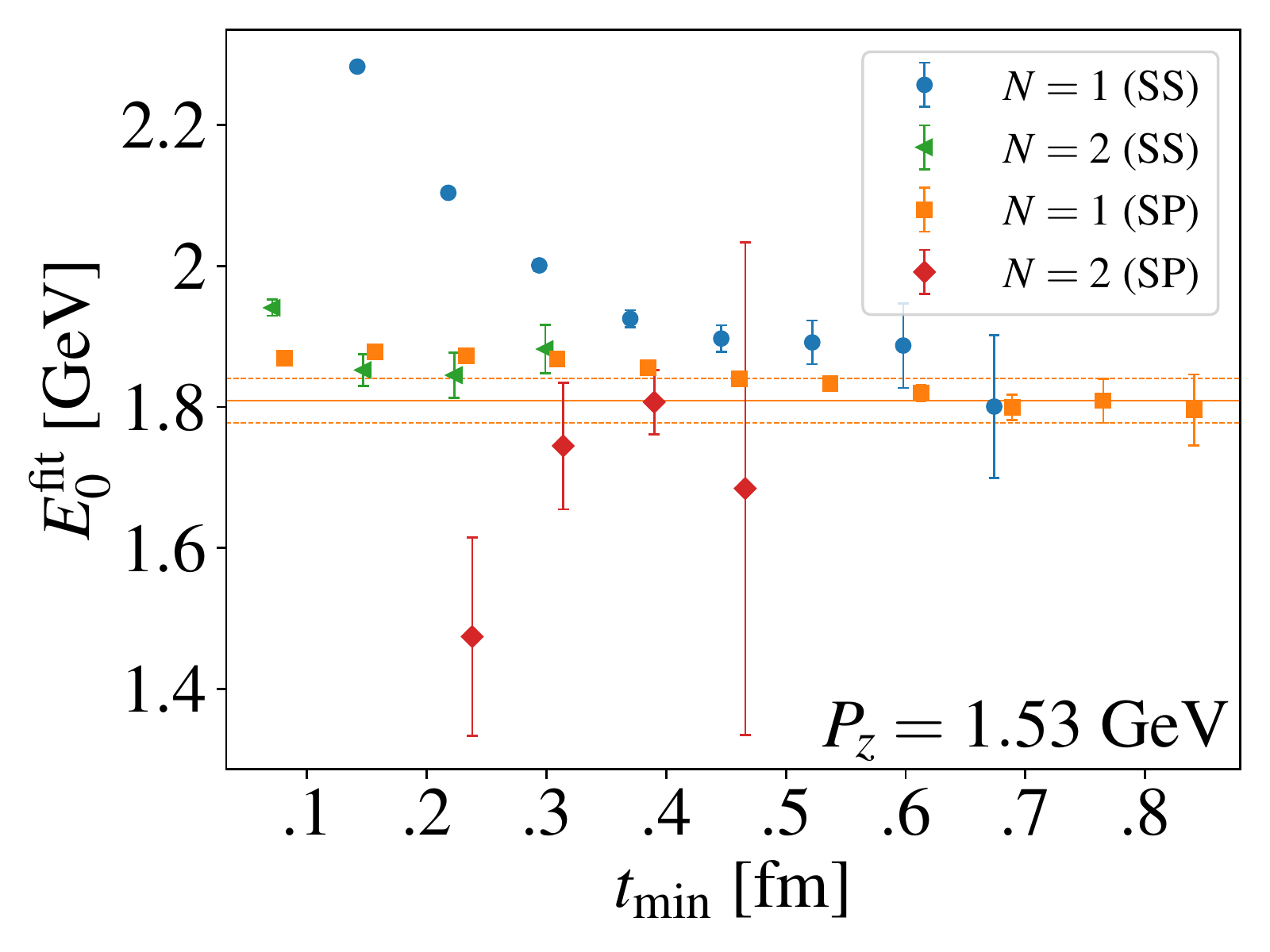}
    \includegraphics[width=0.49\columnwidth]{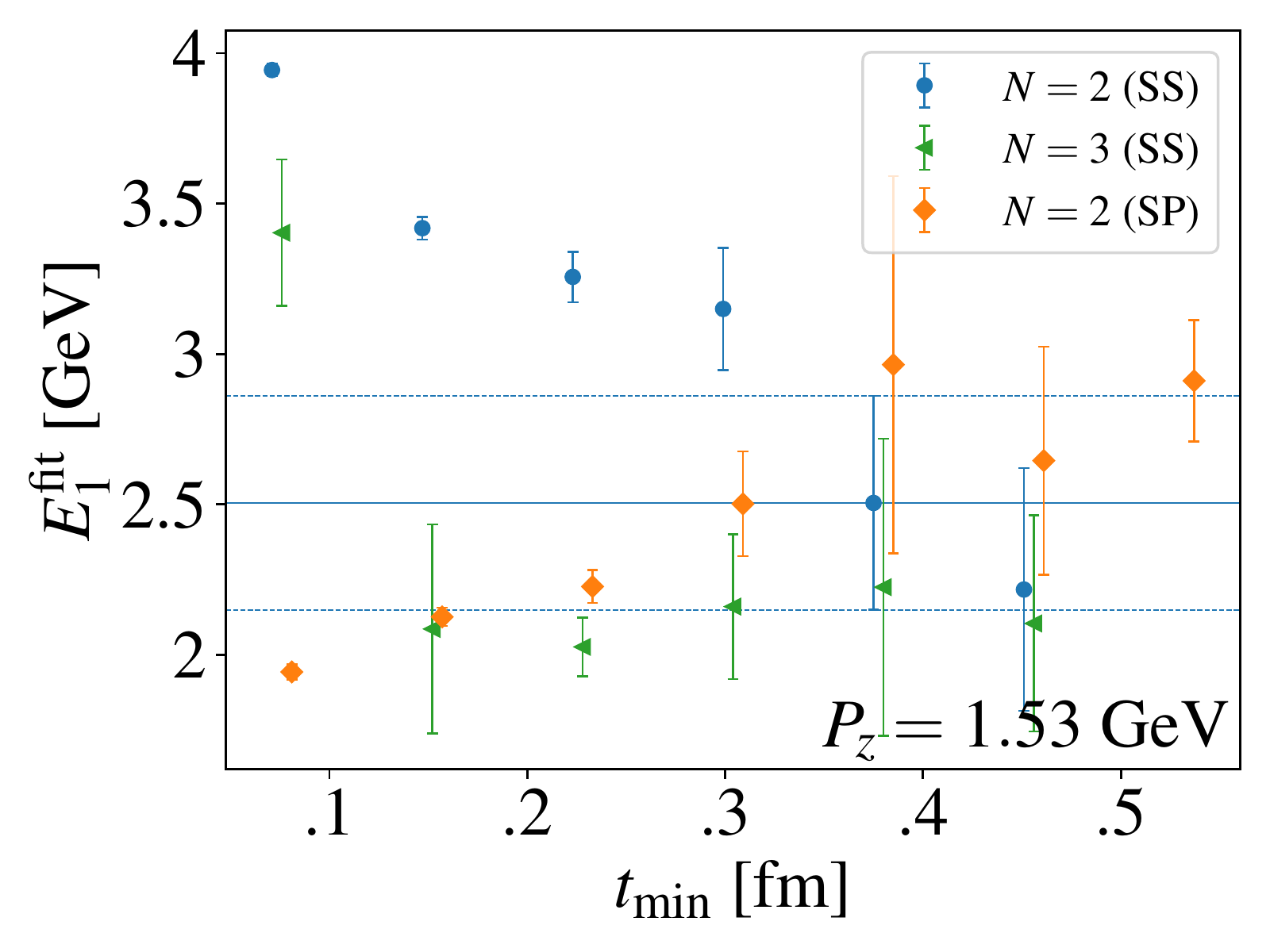}
    \caption{The $t_{\rm min}$ dependence of the (left) ground state and (right) first excited state for (upper)
             $P_z = 0$ and (lower) $P_z = 1.53$ GeV. The fits on the left-hand side are done without any 
             constraints, while the fits on the right-hand side are done with the ground-state energy
             constrained from the best one-state fit to the correlators with the same momentum.
             The legend indicates the number of states $N$ included in the fit function
             and the correlator being fit to (i.e. either SS or SP).}
    \label{fig:tmins}
\end{figure}

As a first step, we perform unconstrained one-state fits to all the two-point correlators,
and then pick the best fit for each momentum.
For each momentum, the different fits involve two choices: i) either SS or SP correlators, and ii) the minimum time separation $t_{\rm min}$ included in the fit.
The largest time separation $t_{\rm max}$ included in the fit is chosen to be the largest time separation such that $|C(t)| \geq 3 \, \delta C(t)$ for all $t \leq t_{\rm max}$, where $\delta C(t)$ is the error of $C(t)$.
The best fit is determined based on observing stability in the results from small changes in $t_{\rm min}$ in combination with a p-value $\gtrsim 0.1$.
If multiple fits satisfy these criteria, then fits to SS correlators are preferred and $t_{\rm min}$ is chosen to lie roughly in the middle of the region of stability.
In the left-hand side of \Cref{fig:tmins}, we show the dependence of the ground-state energy on the choice of $t_{\rm min}$ for the smallest and largest values of momentum used in the three-point functions.
For each momentum, we then perform two-state fits with a prior on the ground state energy coming from the best unconstrained one-state fit for the same momentum.
The right-hand side of \Cref{fig:tmins} shows the dependence of the first-excited state energy on the choice of $t_{\rm min}$ for the same momenta as in the ground state figures.
Our best fit estimates for the first two energies with $n_z = 0$ are
\begin{equation}
    E_0 = 0.9345(57) \, \text{GeV} , \; E_1 = 1.533(71) \, \text{GeV} .
\end{equation}

\begin{figure}
    \centering
    \includegraphics[width=\columnwidth]{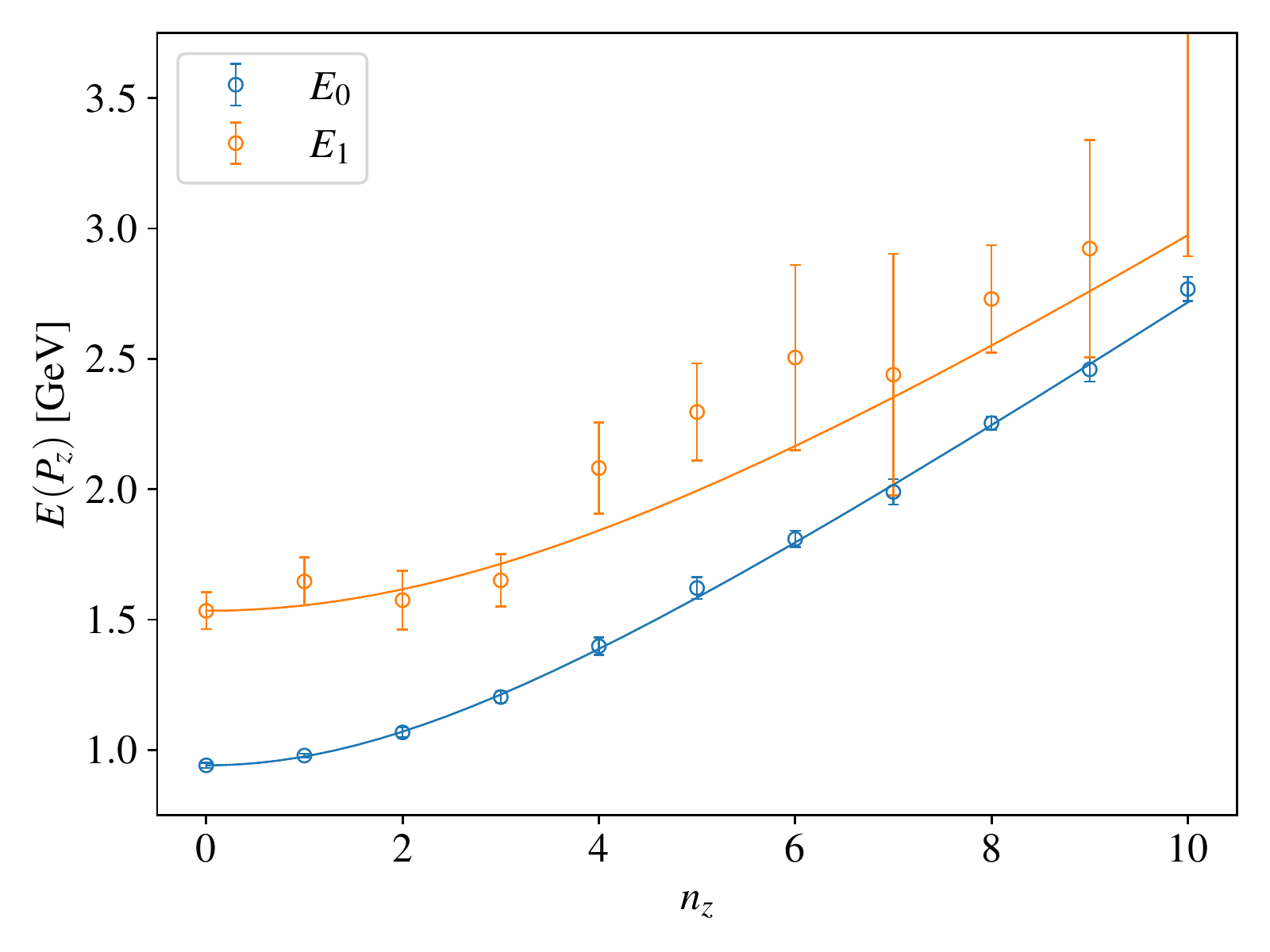}
    \caption{The ground (blue) and first excited (orange) states from one- and two-state fits, respectively,
             for each momentum with $n_z \le 10$. The curves come from predictions for the energies with $n_z \neq 0$ using the
             continuum dispersion relation with the mean value of the $n_z = 0$ energies.}
    \label{fig:dispersion}
\end{figure}

In \Cref{fig:dispersion}, the ground state and first excited state from one- and two-state fits, respectively, are plotted as a function of momentum.
The mean value of the energies for $n_z = 0$ are used to predict the energies for $n_z \neq 0$ using the continuum dispersion relation.
These predictions are shown as lines in \Cref{fig:dispersion}.
As expected, the ground state energies follow the continuum dispersion relation quite well, with a mass consistent with the proton, giving confidence in our control over excited states, even at large momentum.
More surprisingly, the first excited state appears to also follow a continuum dispersion relation with a mass that is a little larger than the Roper resonance, which is unexpected for calculations using the Wilson-clover action~\cite{xQCD:2019jke}.
The small disagreement with the Roper mass could also be from uncontrolled finite-volume effects, which are generally very important for resonances.
Further, we know that several multi-hadron states should lie below our estimate of the first-excited state, but our results suggest our operators must have very poor overlap onto those states.

\subsection{Three-point function analysis}
\label{subsec:c3pt_analysis}

The standard approach to extract the bare matrix elements is to form an appropriate ratio of the three-point functions to the two-point functions.
This is convenient in that in the limit as $\ts$ and $\ti$ become large, the ratio approaches the desired ground-state bare matrix element.
As an added benefit, the ratio leads to a cancellation of correlations, resulting in reduced statistical errors.

\begin{figure*}
    \centering
    \includegraphics[width=0.32\textwidth]{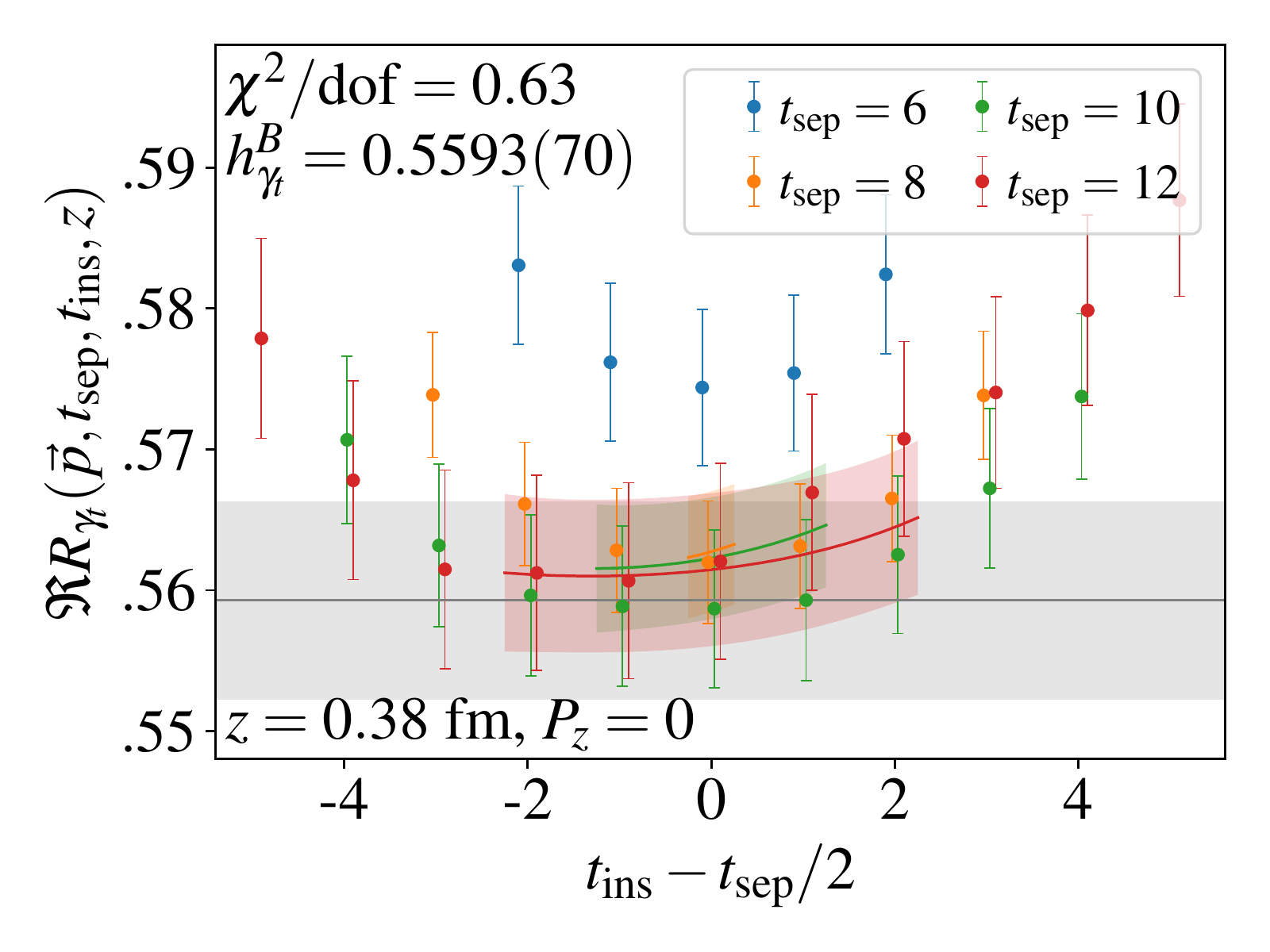}
    \includegraphics[width=0.32\textwidth]{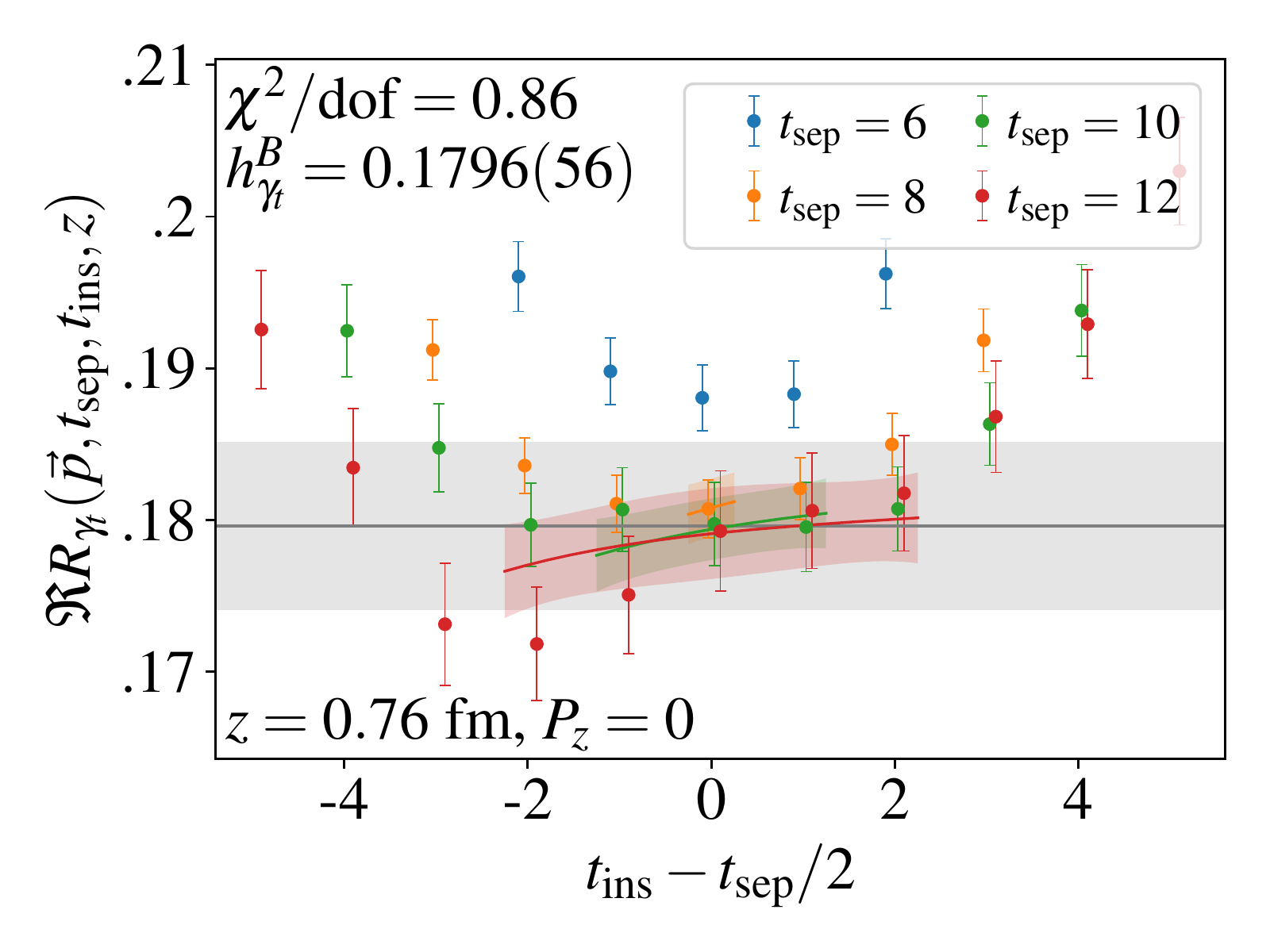}
    \includegraphics[width=0.32\textwidth]{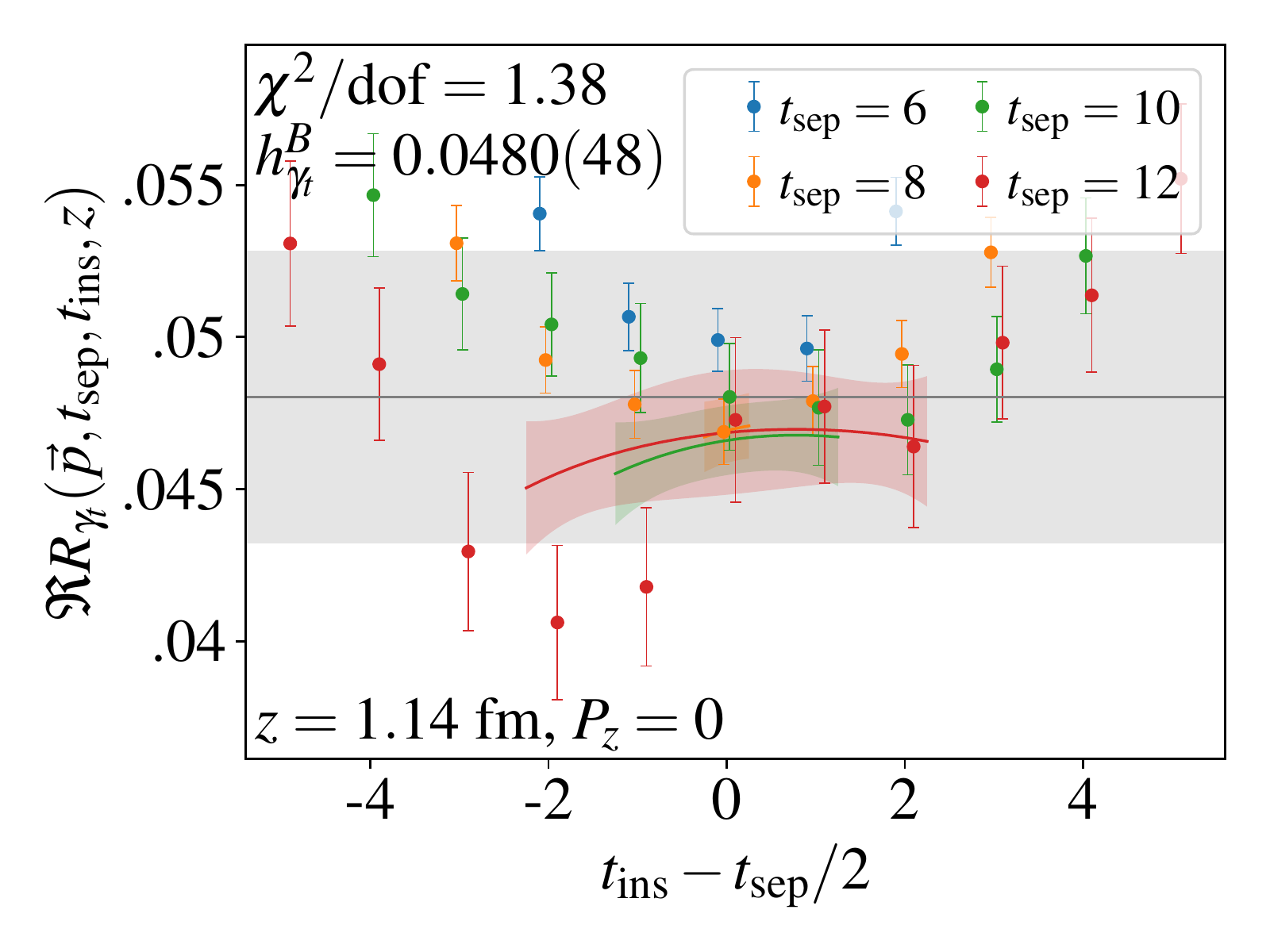}
    \includegraphics[width=0.32\textwidth]{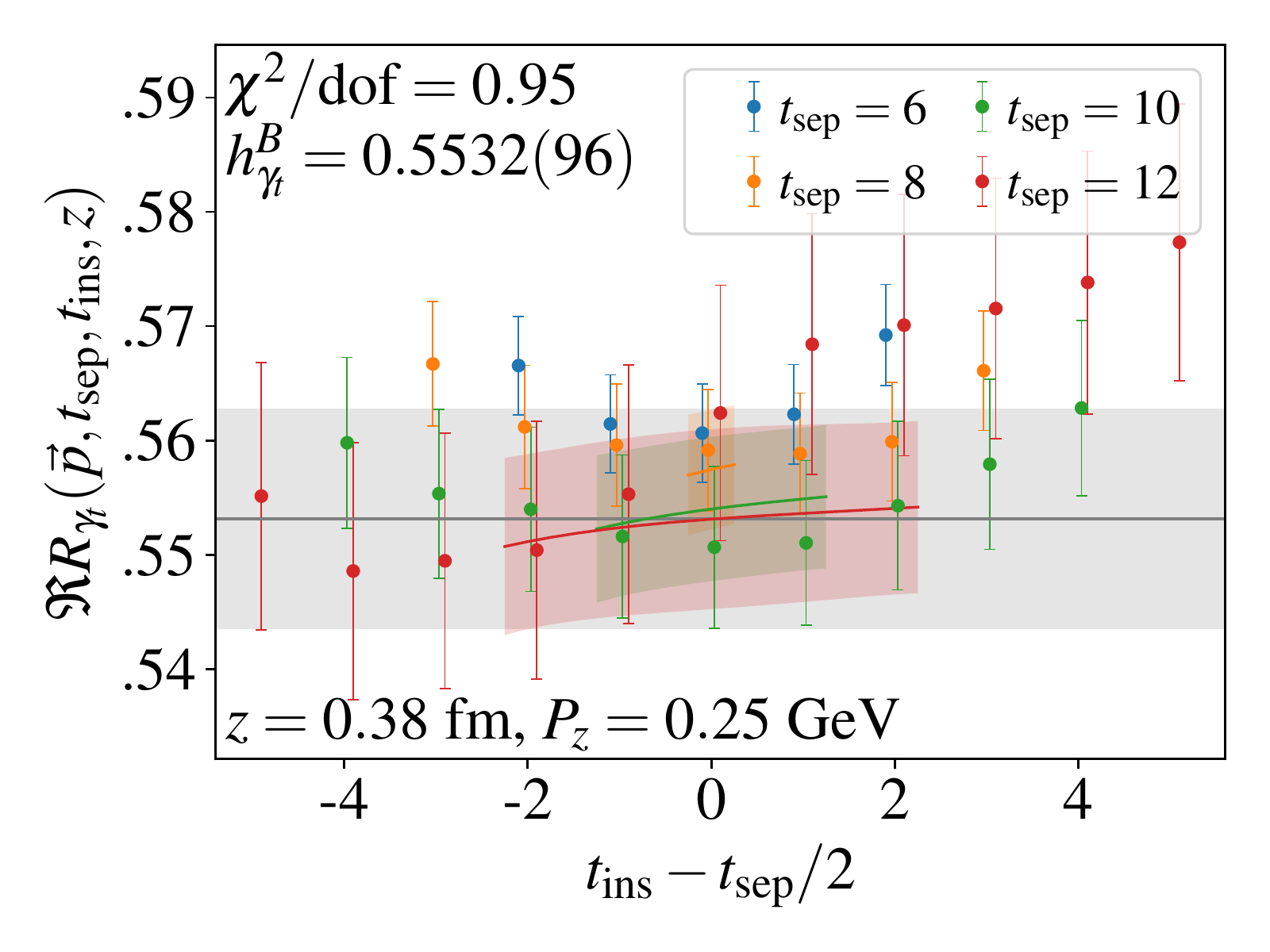}
    \includegraphics[width=0.32\textwidth]{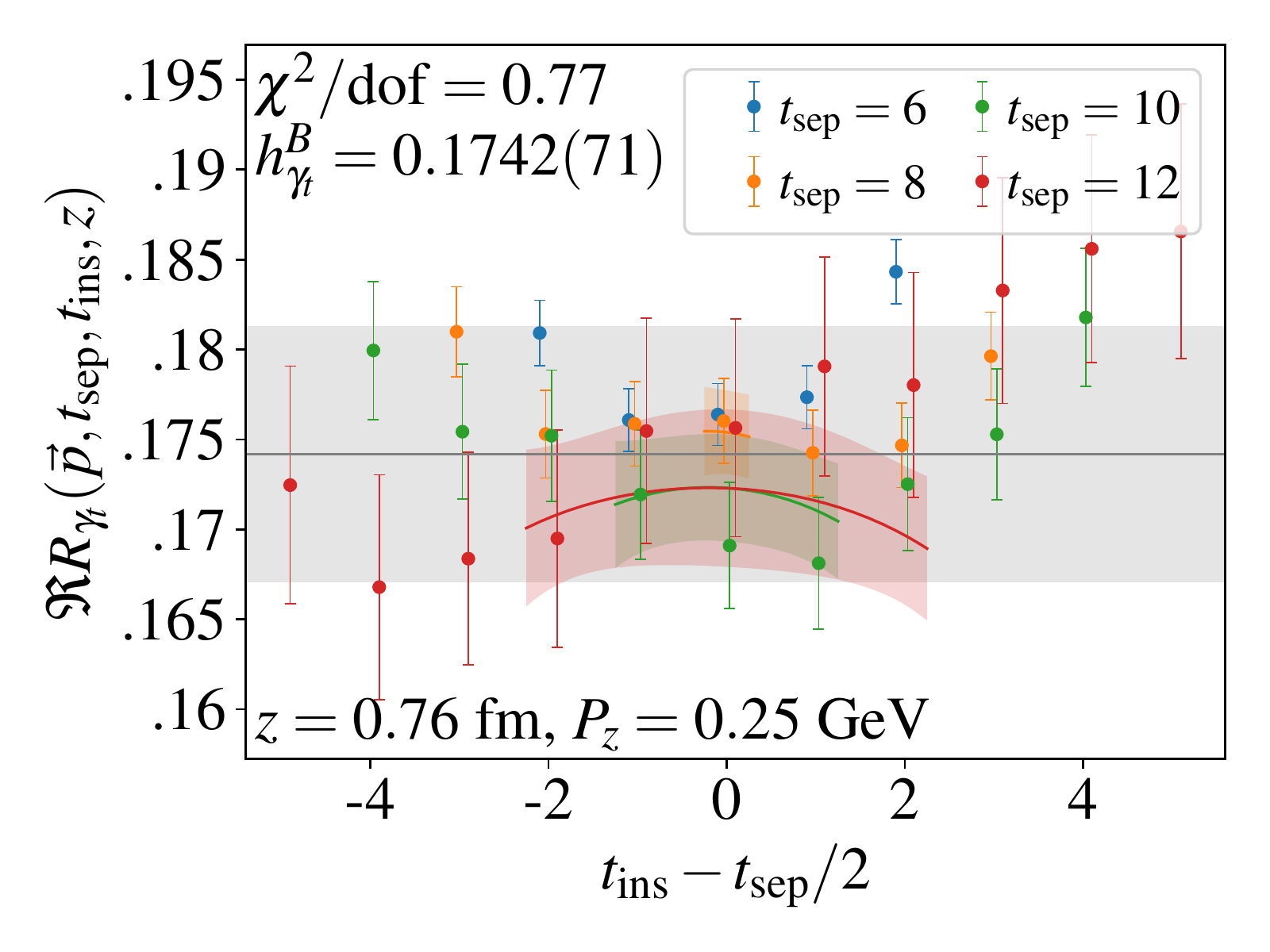}
    \includegraphics[width=0.32\textwidth]{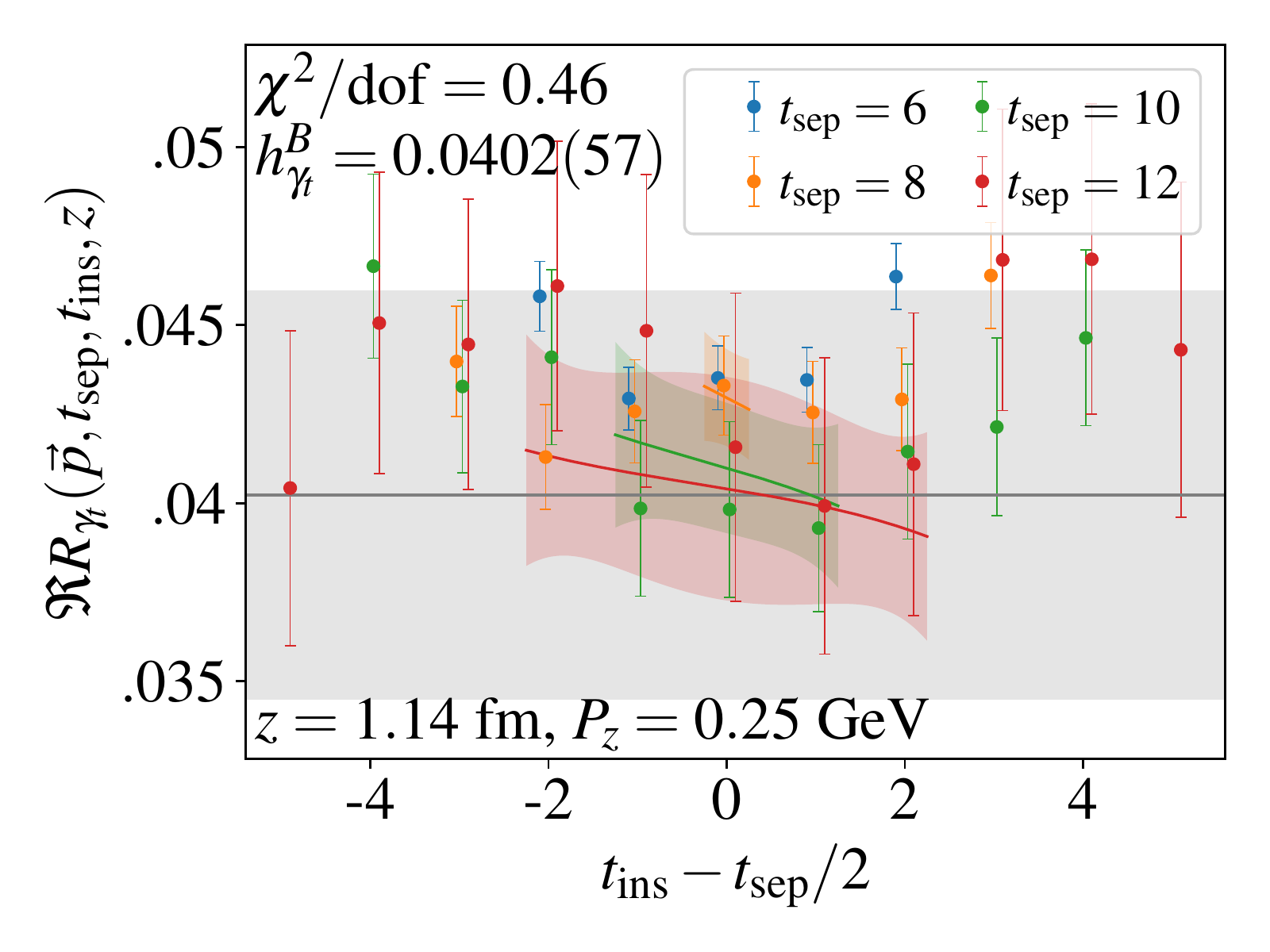}
    \includegraphics[width=0.32\textwidth]{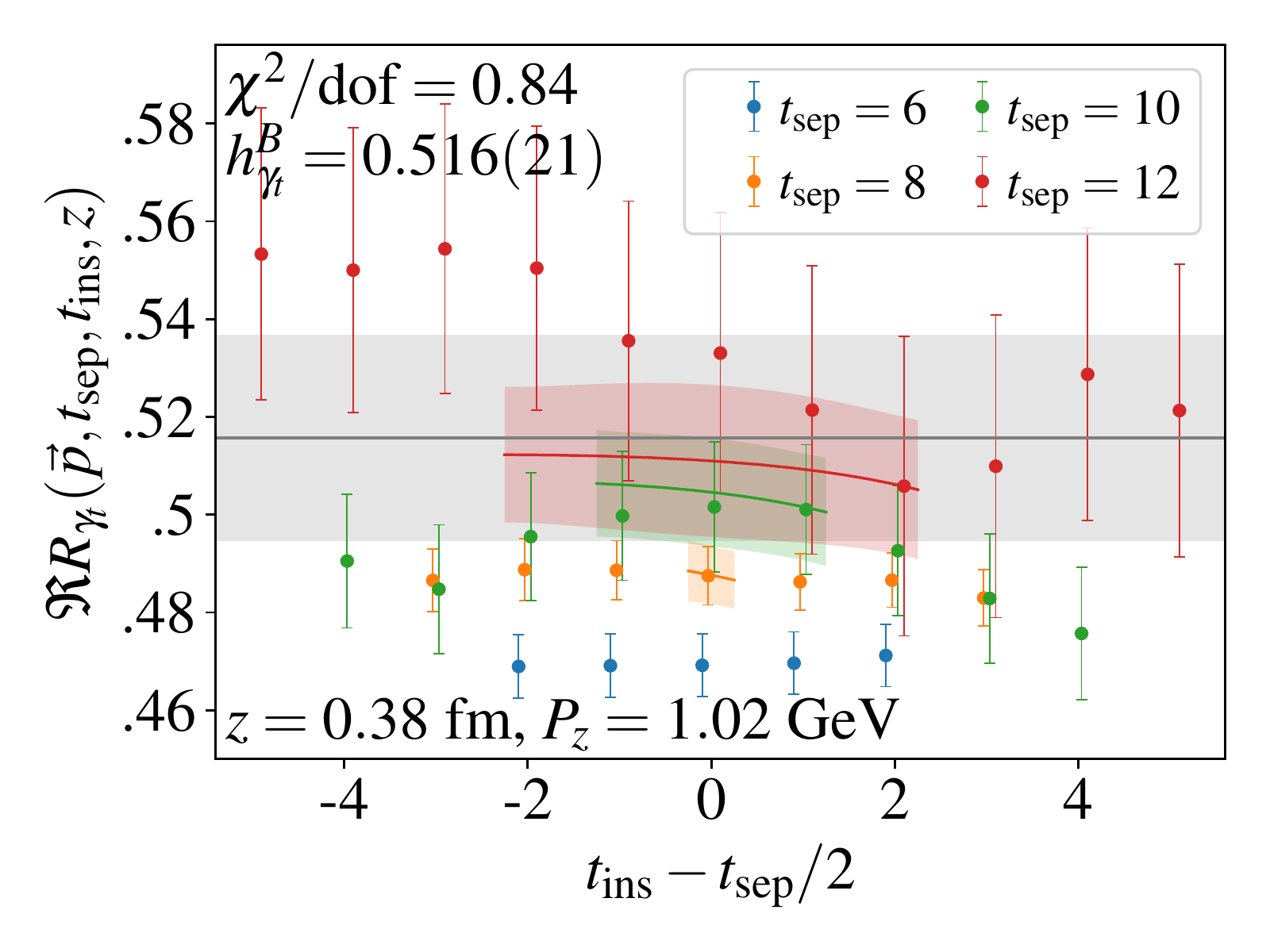}
    \includegraphics[width=0.32\textwidth]{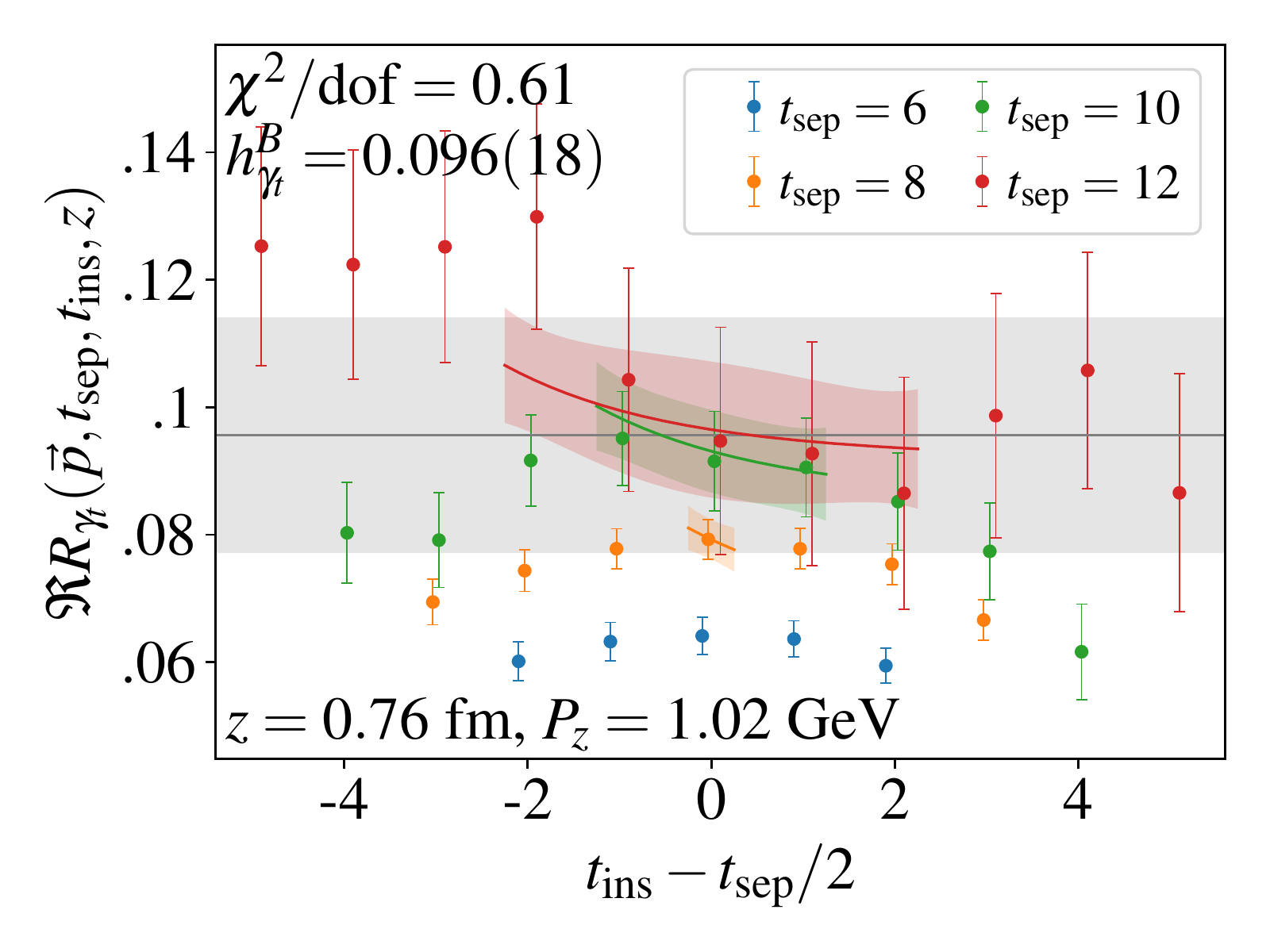}
    \includegraphics[width=0.32\textwidth]{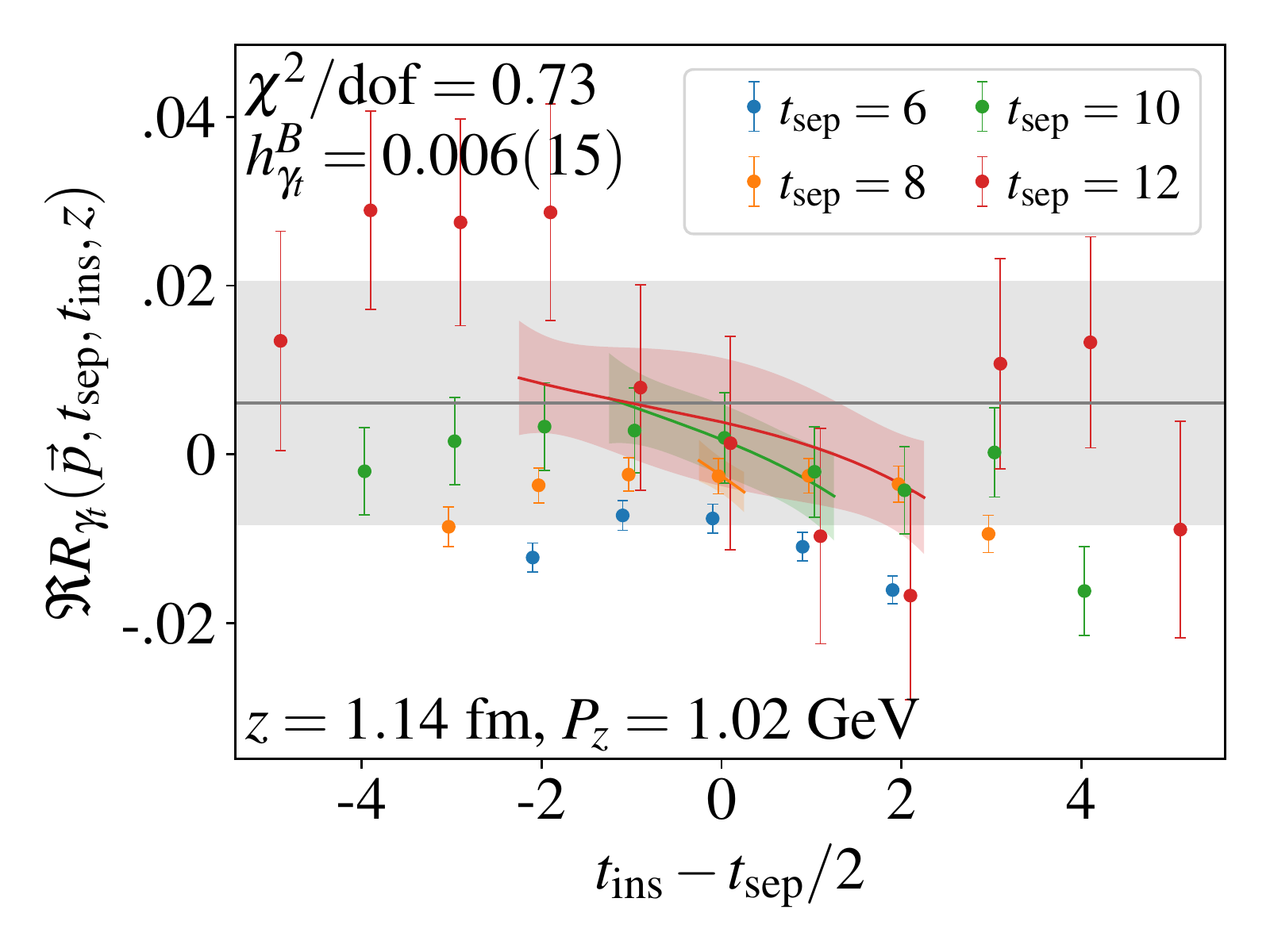}
    \includegraphics[width=0.32\textwidth]{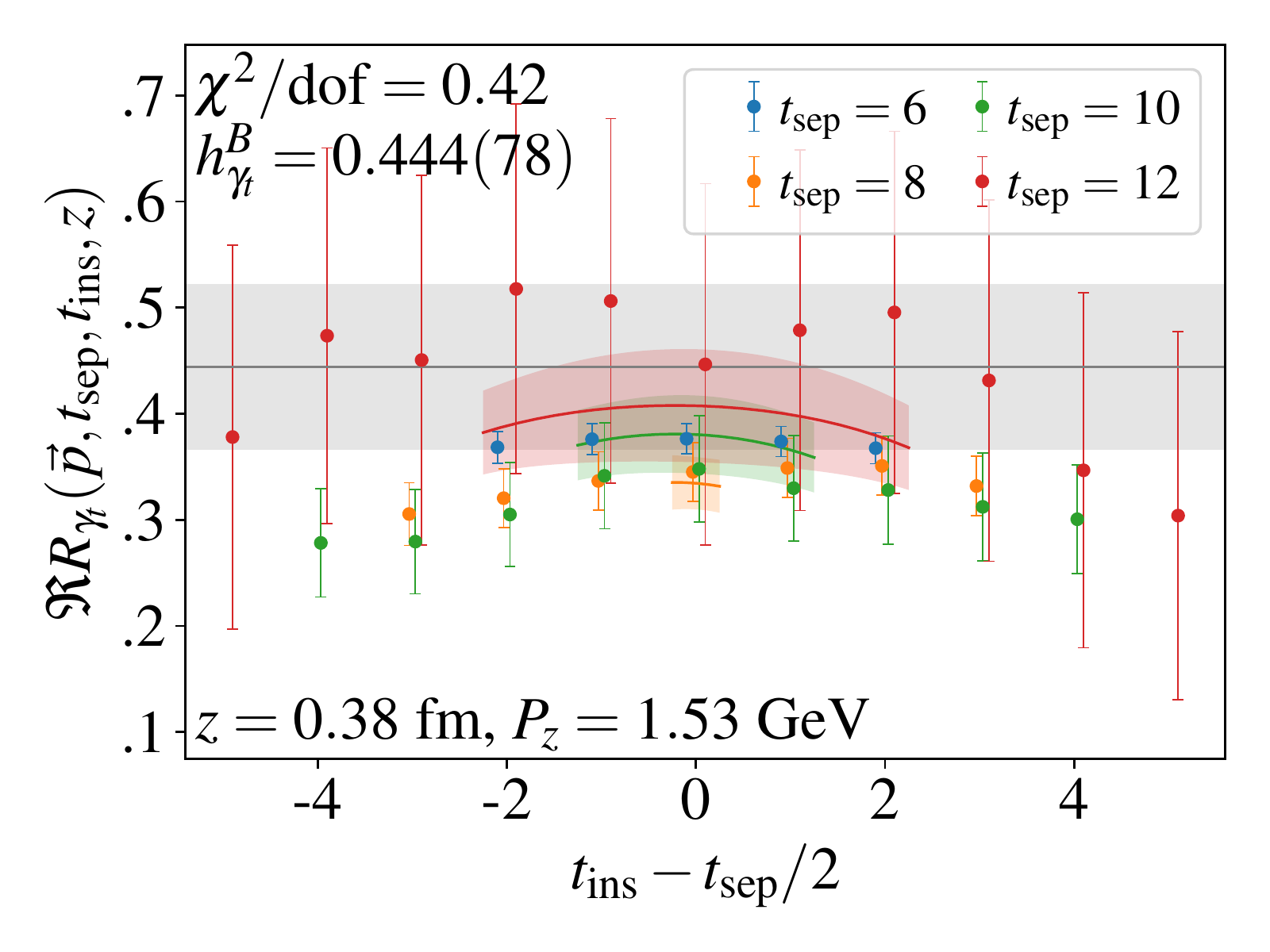}
    \includegraphics[width=0.32\textwidth]{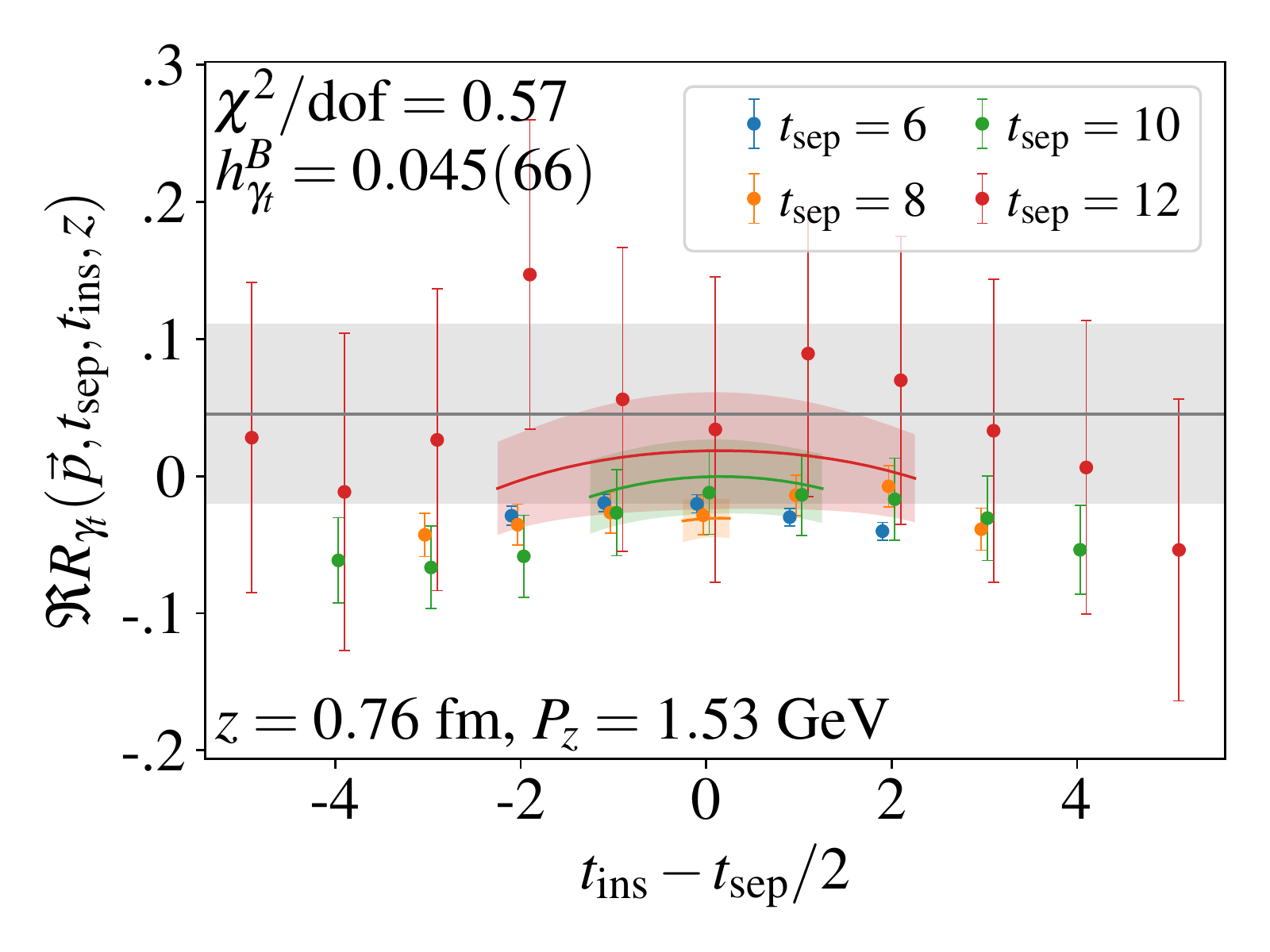}
    \includegraphics[width=0.32\textwidth]{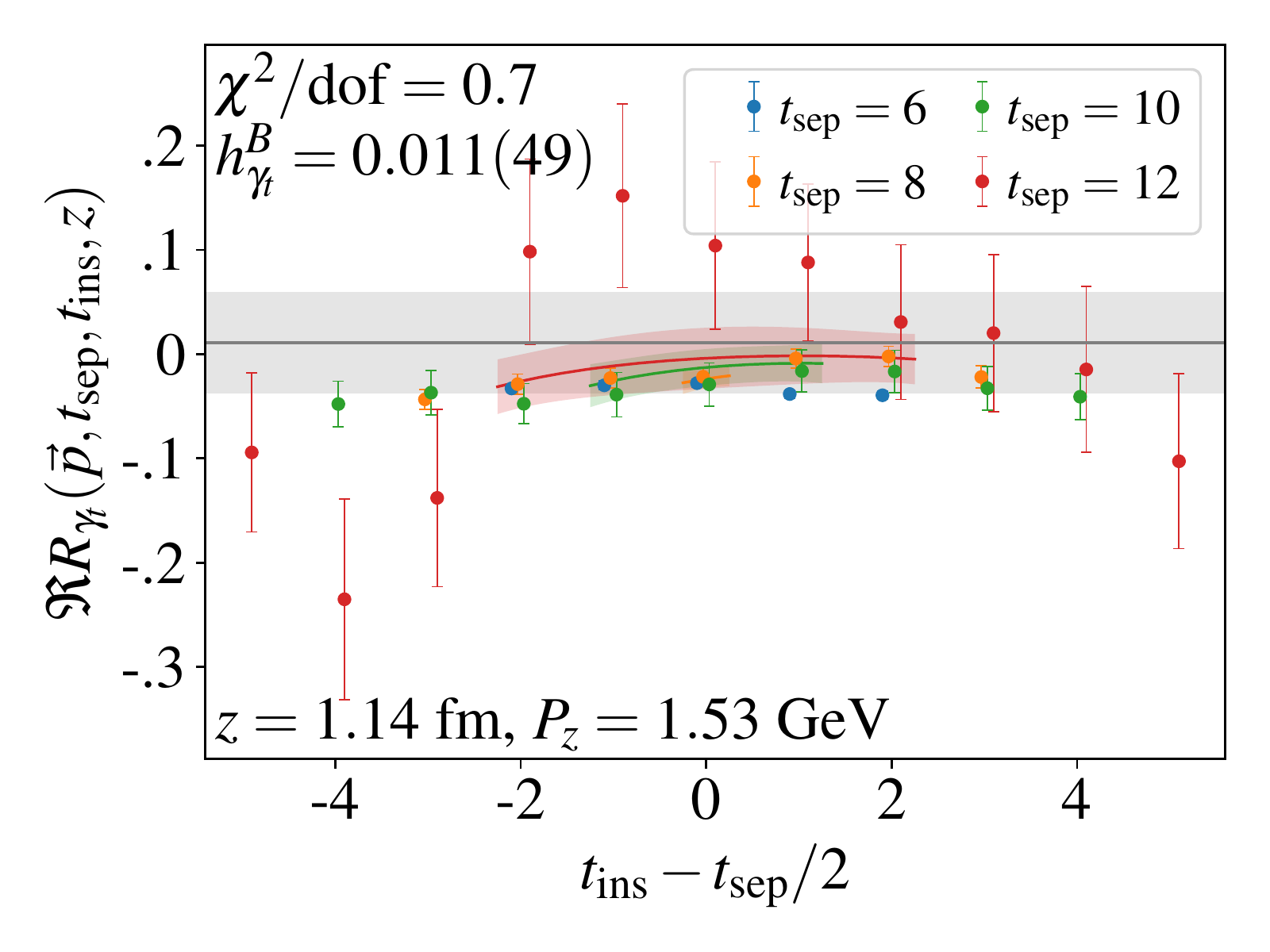}
    \caption{The real parts of the ratio of three-point to two-point functions for all values
             of momentum (one for each row) and a few representative values of Wilson-line
             length $z$ (one for each column). The $\chi^2/\textrm{dof}$ reported, estimate
             for the ground-state bare matrix element (also represented by a gray band),
             and $\ts$ fit bands come from the preferred fit strategy,
             i.e. the two-state fit to the ratio $R_{\gamma_t}$ with $n_{\rm exc} = 3$,
             where $n_{\rm exc}$ is the number of data points nearest both the source and sink that are not included in the fit.
             The range in $\ti$ of the $\ts$ bands
             covers the included data points in the fit.}
    \label{fig:c3pt_repr_fits_re}
\end{figure*}

\begin{figure*}
    \centering
    \includegraphics[width=0.32\textwidth]{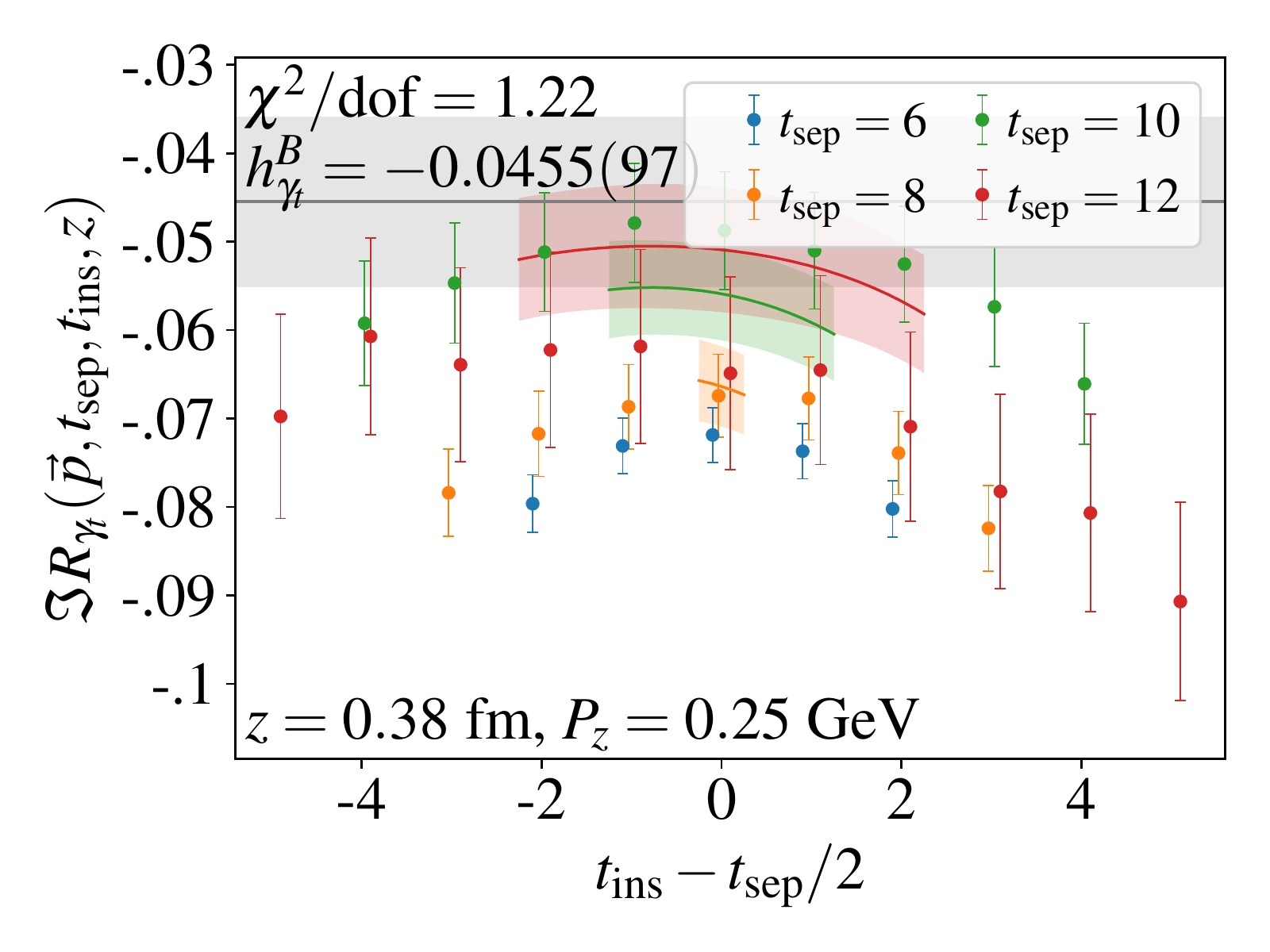}
    \includegraphics[width=0.32\textwidth]{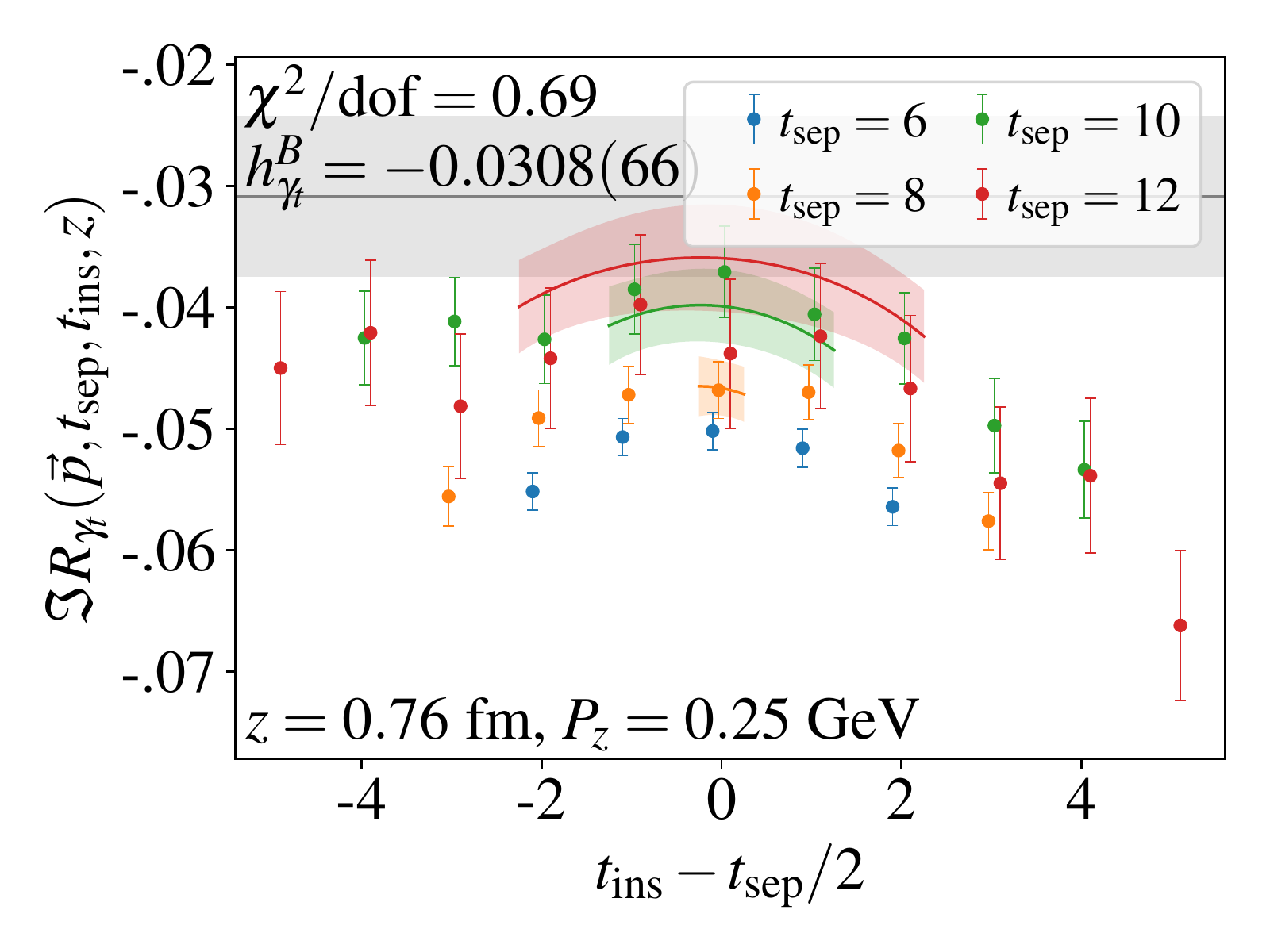}
    \includegraphics[width=0.32\textwidth]{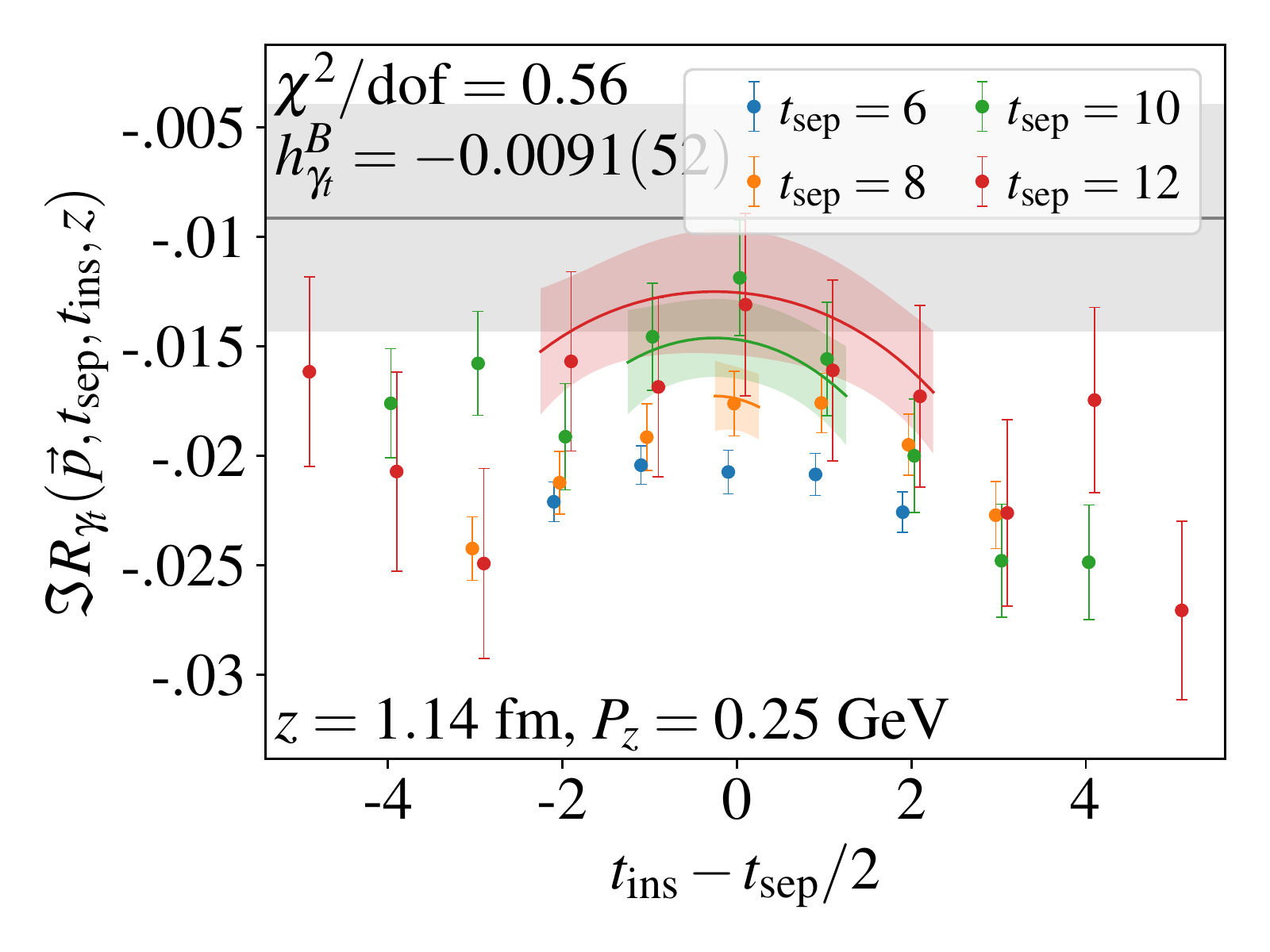}
    \includegraphics[width=0.32\textwidth]{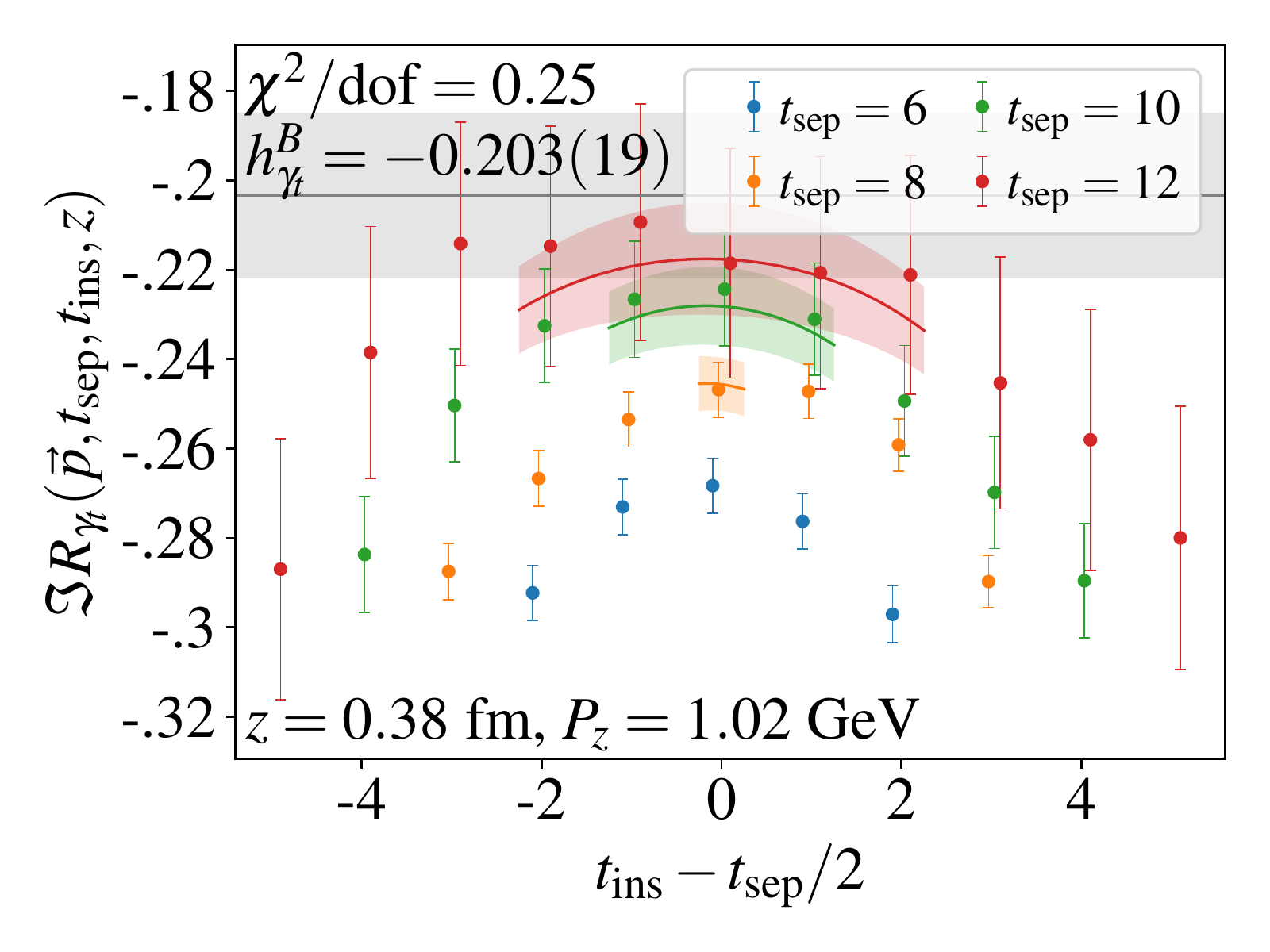}
    \includegraphics[width=0.32\textwidth]{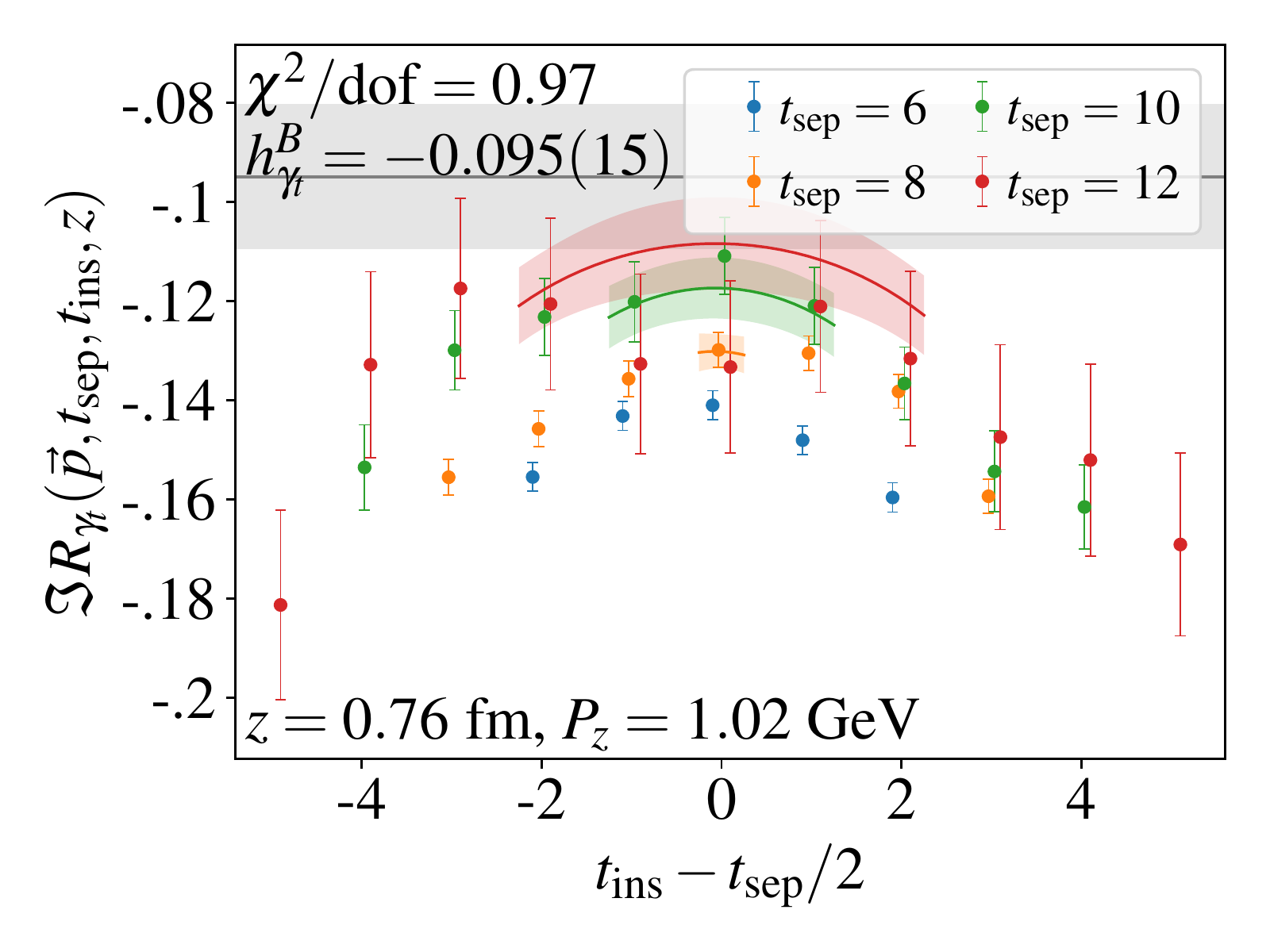}
    \includegraphics[width=0.32\textwidth]{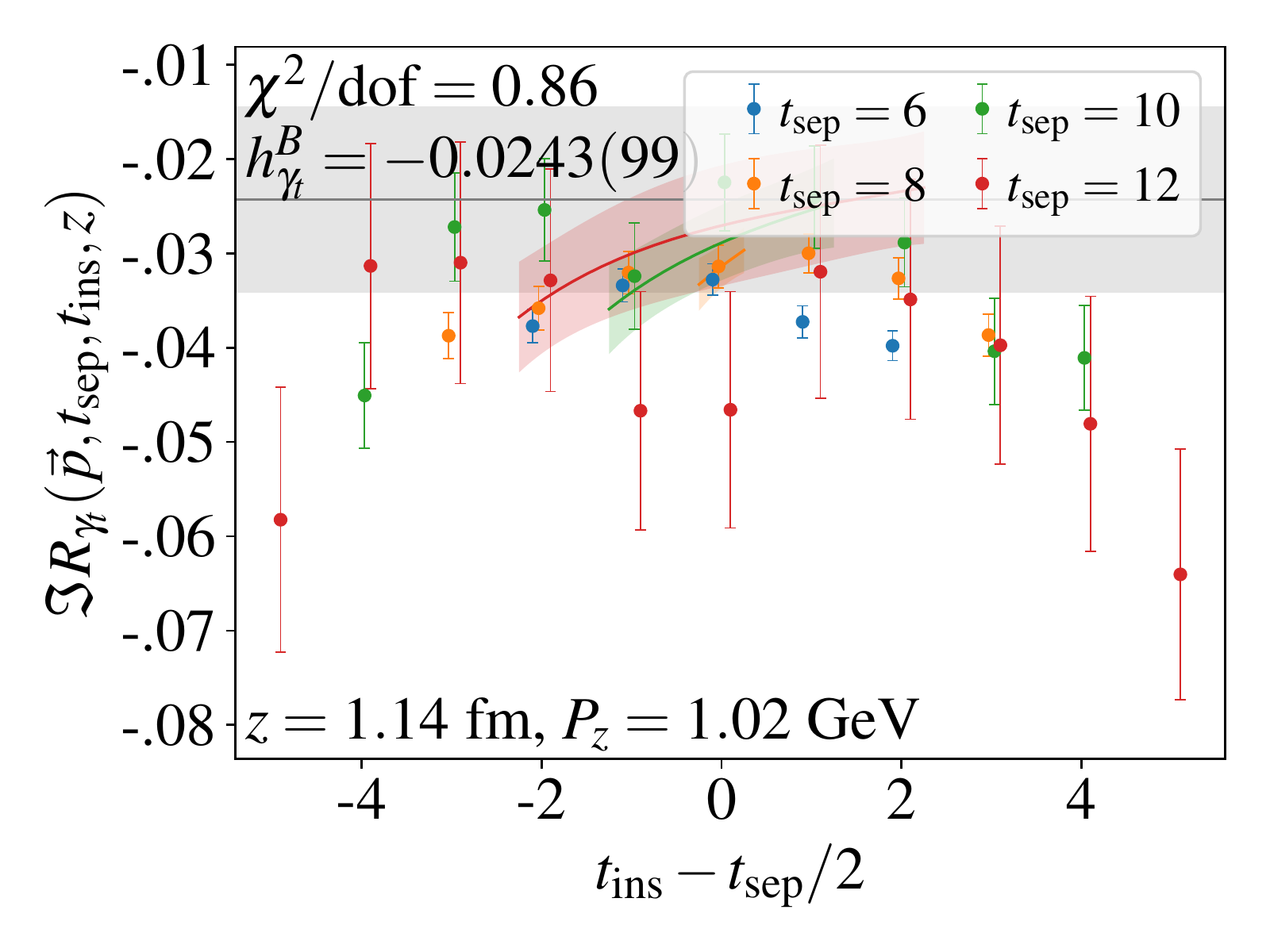}
    \includegraphics[width=0.32\textwidth]{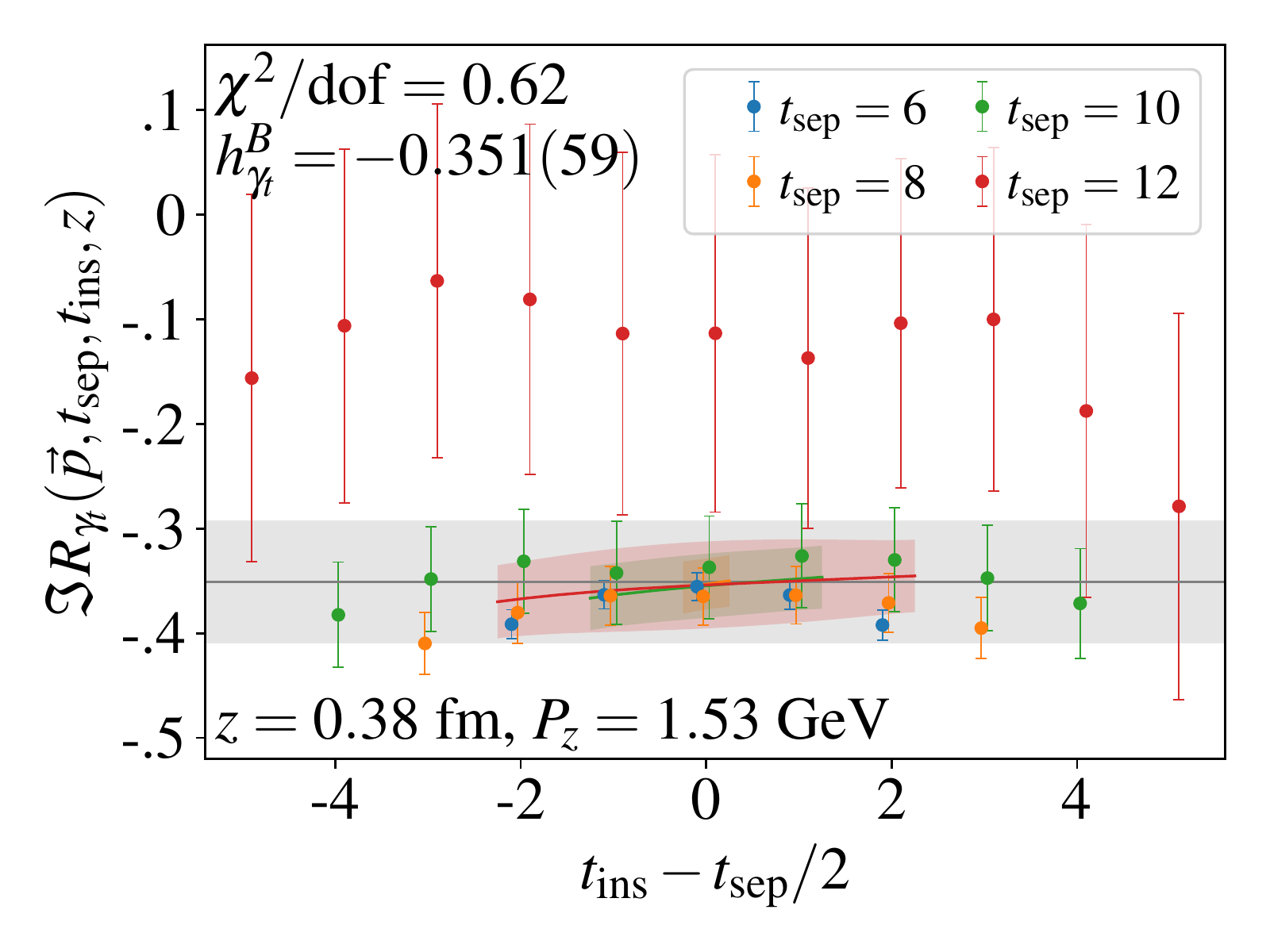}
    \includegraphics[width=0.32\textwidth]{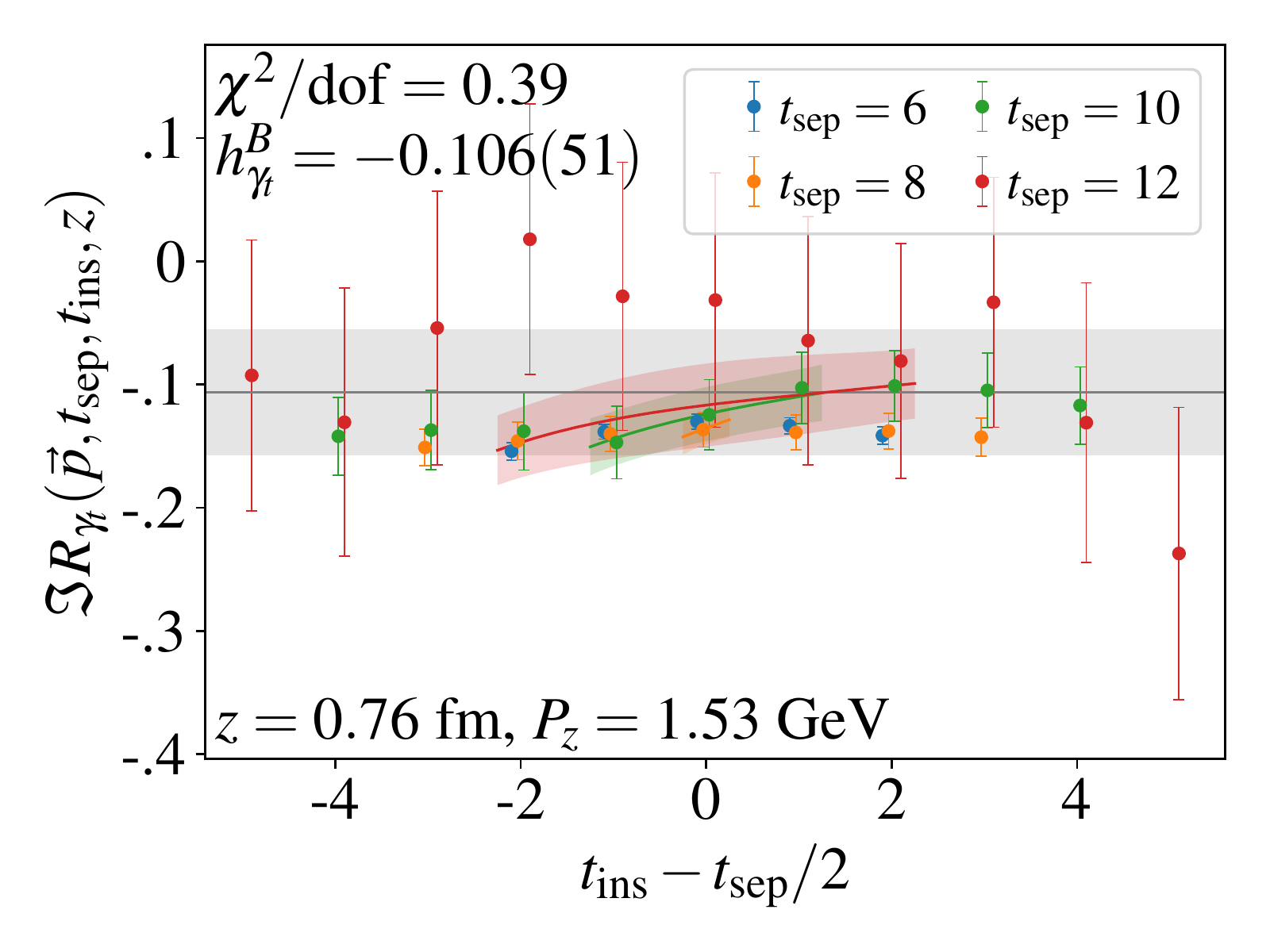}
    \includegraphics[width=0.32\textwidth]{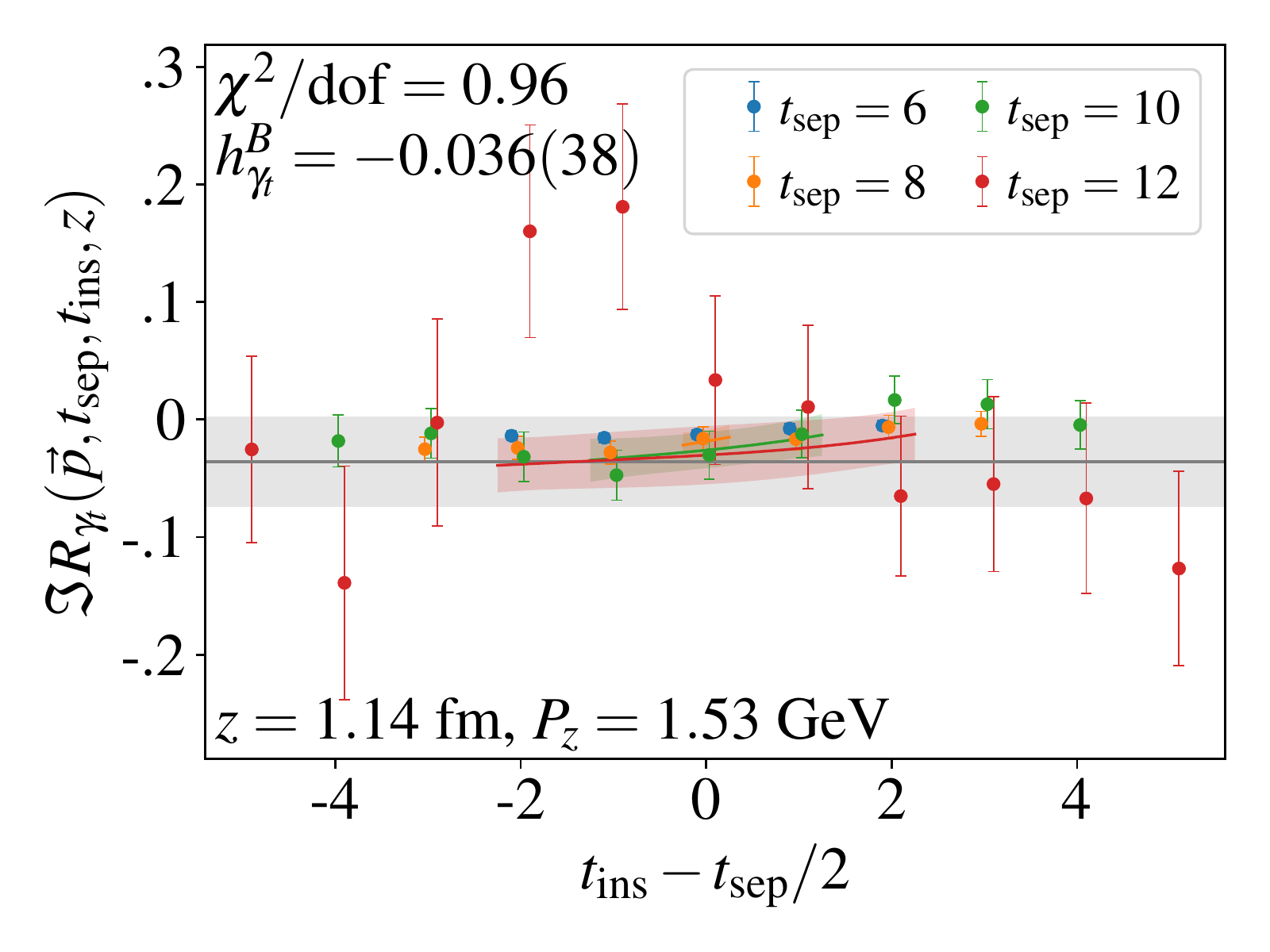}
    \caption{The same as \Cref{fig:c3pt_repr_fits_re}, but for the imaginary parts.
             Additionally, the zero-momentum matrix elements are not shown, as these are
             all consistent with zero (as they are expected to be).}
    \label{fig:c3pt_repr_fits_im}
\end{figure*}

To determine the needed ratio, it is beneficial to give the spectral decomposition of the three-point function
\begin{equation}
\begin{split}
    C^{\rm 3pt}_\Gamma & (\pvec_f, \pvec_i, \ts, \ti, z) = \\
    & \sum_{m,n} A^{(m)}_\alpha (\pvec_f) A^{(n)}_\alpha (\pvec_i)^\ast h^B_\Gamma (\pvec_f, m ; \pvec_i, n ; z) \\
    &\times e^{-E_m (\pvec_f) (\ts - \ti)} e^{-E_n (\pvec_i) \ti} ,
\end{split}
\label{eq:c3pt_spectral}
\end{equation}
where $A_\alpha^{(n)} \equiv \bra{\Omega} N_\beta \mathcal{P}^{\rm 3pt}_{\beta \alpha} \ket{n, \pvec}$,
$h^B_\Gamma (\pvec_f, m ; \pvec_i, n ; z)$ are the bare matrix elements, thermal effects are ignored, and it is assumed that $t_0$ has been shifted to zero.
The energies entering the spectral decomposition are the same for both the three-point and two-point functions.
However, the overlap factors are only the same if $\mathcal{P}^{\rm 2pt}$ and $\mathcal{P}^{\rm 3pt}$ are chosen appropriately.
In our case, this is trivially true since we consider $\mathcal{P}^{\rm 2pt} = \mathcal{P}^{\rm 3pt} = \frac{1}{2}(1 + \gamma_t)$.
Therefore, in the case where $\pvec \equiv \pvec_f = \pvec_i$, used in this work, we form
\begin{equation}
    R_{\gamma_t}(\pvec, \ts, \ti, z) \equiv \frac{C^{\rm 3pt}_{\gamma_t} (\pvec, \pvec, \ts, \ti, z)}{C^{\rm 2pt} (\pvec, \ts)} ,
    \label{eq:c3pt_ratio}
\end{equation}
where we now indicate the use of $\Gamma = \gamma_t$ (and continue to do so throughout the remainder of the paper).
This ratio can easily be seen to obey
\begin{equation}
    \lim_{\ti,\ts \to \infty} R_{\gamma_t} (\pvec, \ts, \ti, z) = h_{\gamma_t}^B (\pvec, 0 ; \pvec, 0; z) .
\end{equation}

The fit functions we consider are based on truncating the spectral decomposition in both the numerator and denominator of $R_{\gamma_t} (\pvec, \ts, \ti, z)$ to the same number of states $N$
\begin{equation}
    \begin{split}
    R^N_{\gamma_t} & (\pvec, \ts, \ti, z) = \\
     &\frac{\sum_{m,n=0}^{N-1} h^{\prime B}_{\gamma_t} \prod_{l,k,r=1}^{m} e^{-\Delta_{l,l-1}\ts} e^{(\Delta_{k,k-1} - \Delta_{r,r-1}) \ti}}{1 + \sum_{i=1}^{N-1} R_i \prod_{j=1}^{i} e^{-\Delta_{j,j-1}\ts}} ,
    \end{split}
\end{equation}
where $\Delta_{i,i-1}$, $R_i$, and $h^{\prime B}_{\gamma_t}$ are the fit parameters, and we have suppressed the function arguments.
For all of our fits, we prior the $\Delta_{i,i-1}$ and $R_i$ from the corresponding two-point function fit.
Note that the $h^{\prime B}_{\gamma_t} (\pvec, z; m, n)$ can be written in terms of the original matrix elements found in \Cref{eq:c3pt_spectral} as
\begin{equation}
    h^{\prime B}_{\gamma_t} (\pvec, z; m, n) \equiv \frac{A^{(m)}_\alpha (\pvec) A^{(n)}_\alpha (\pvec)^\ast h^B_{\gamma_t} (\pvec, m; \pvec, n ; z)}{A^{(0)}_\alpha (\pvec) A^{(0)}_\alpha (\pvec)^\ast} .
\end{equation}
For convenience, in what follows, we use $h^B_{\gamma_t} (z, P_z, a)$ to denote the ground-state bare matrix element,
where $a$ is the lattice spacing, as these are the only matrix elements used in the subsequent analysis.

\begin{figure*}
    \centering
    \includegraphics[width=0.32\textwidth]{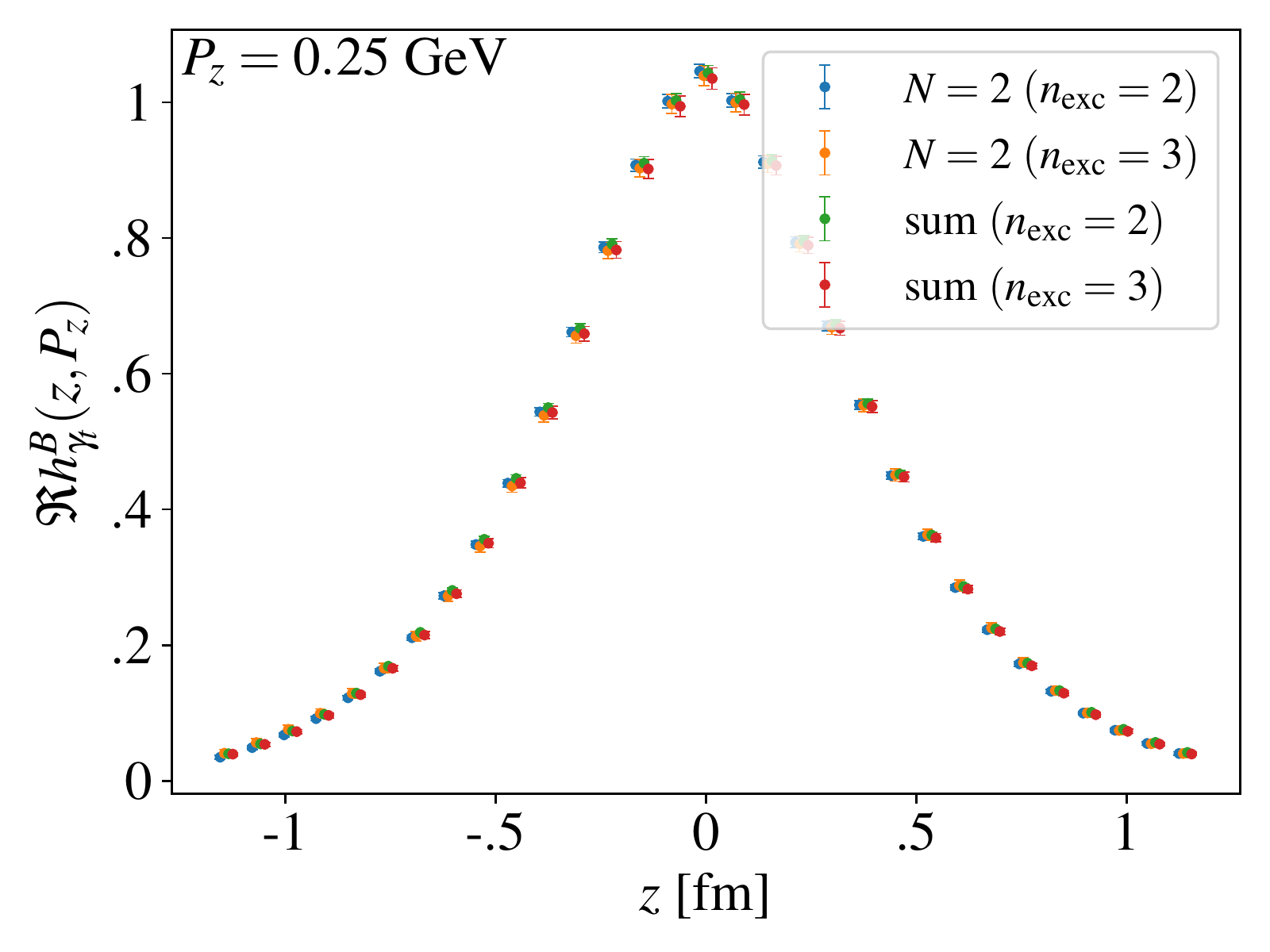}
    \includegraphics[width=0.32\textwidth]{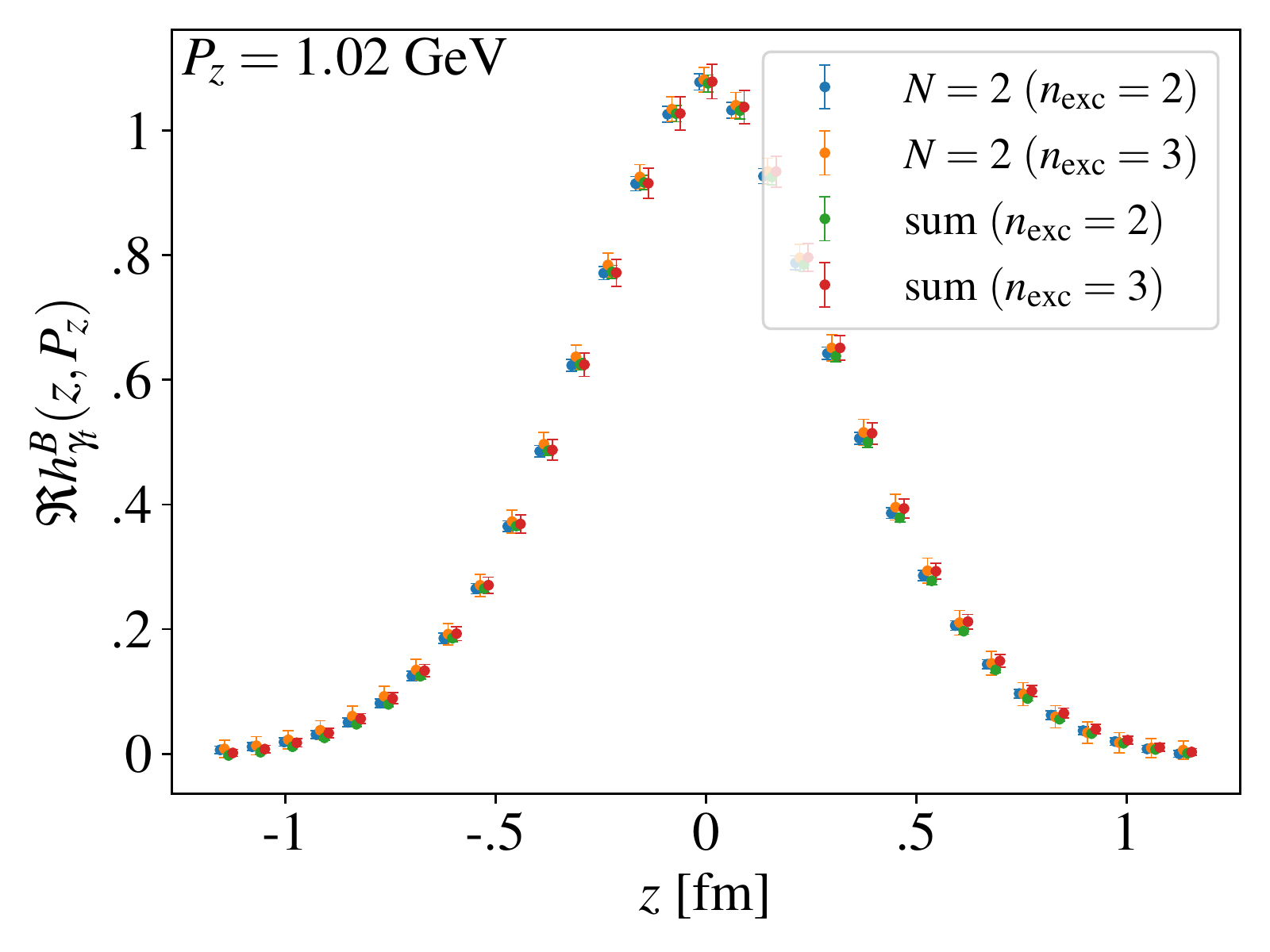}
    \includegraphics[width=0.32\textwidth]{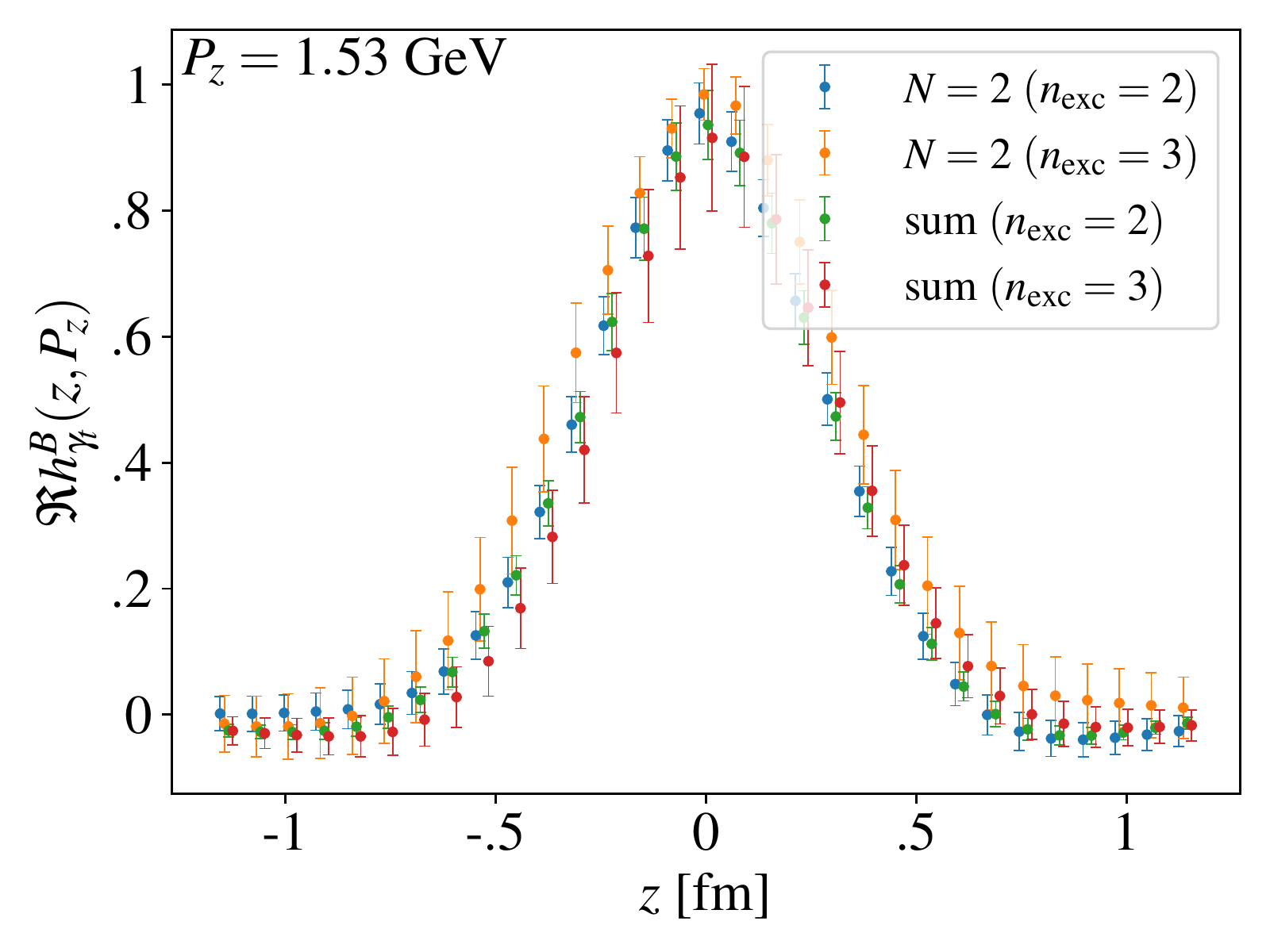}
    \includegraphics[width=0.32\textwidth]{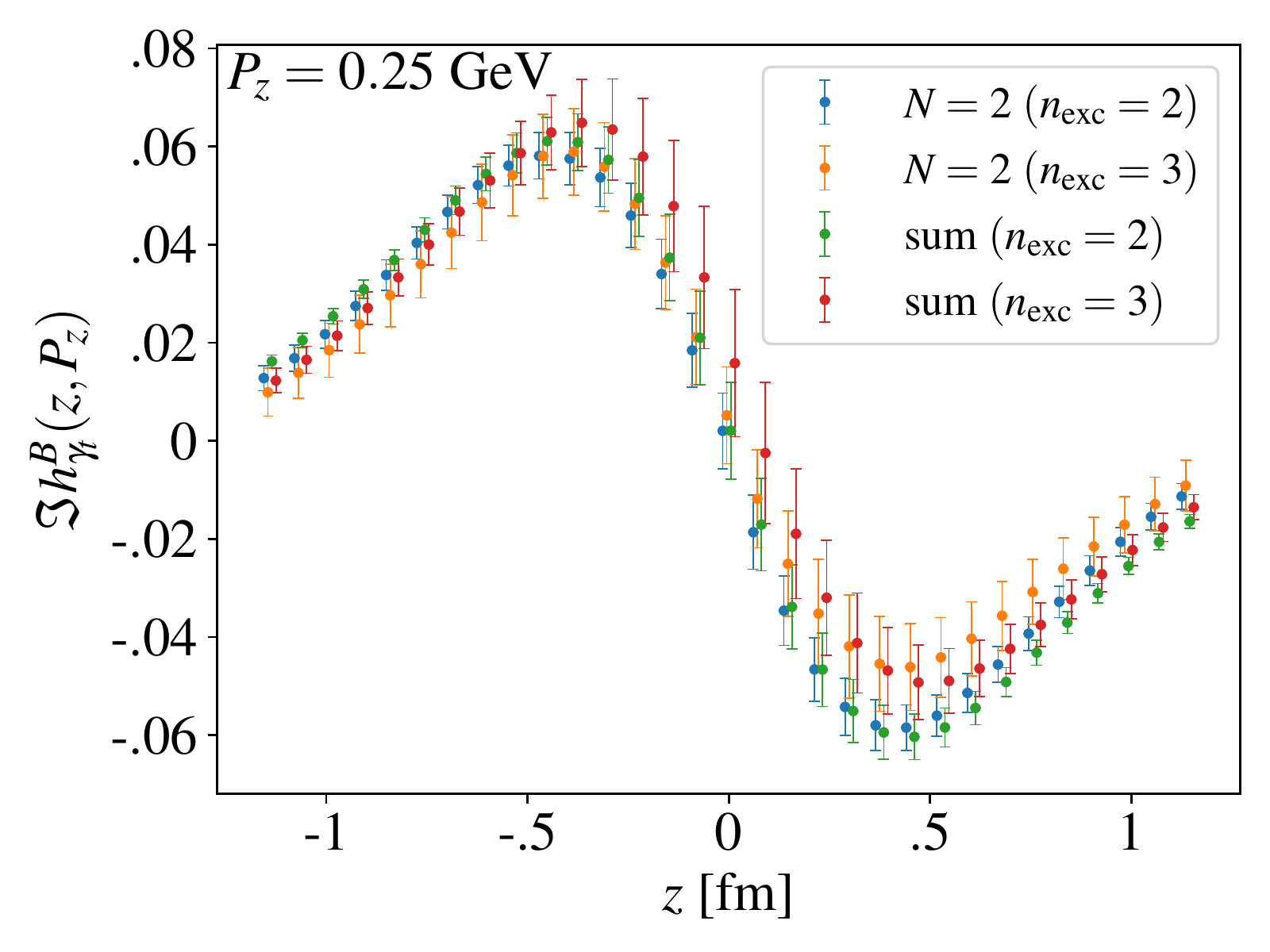}
    \includegraphics[width=0.32\textwidth]{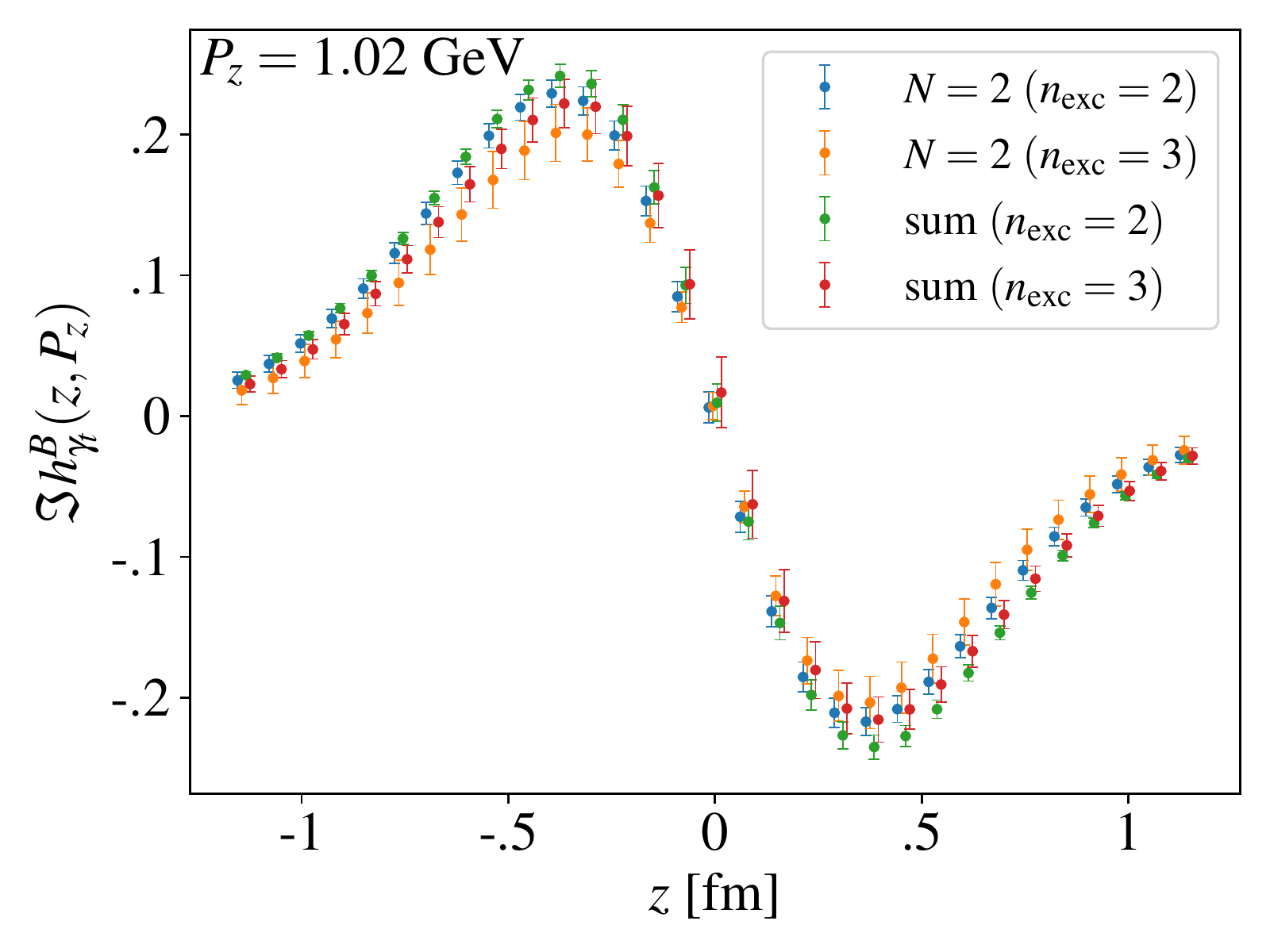}
    \includegraphics[width=0.32\textwidth]{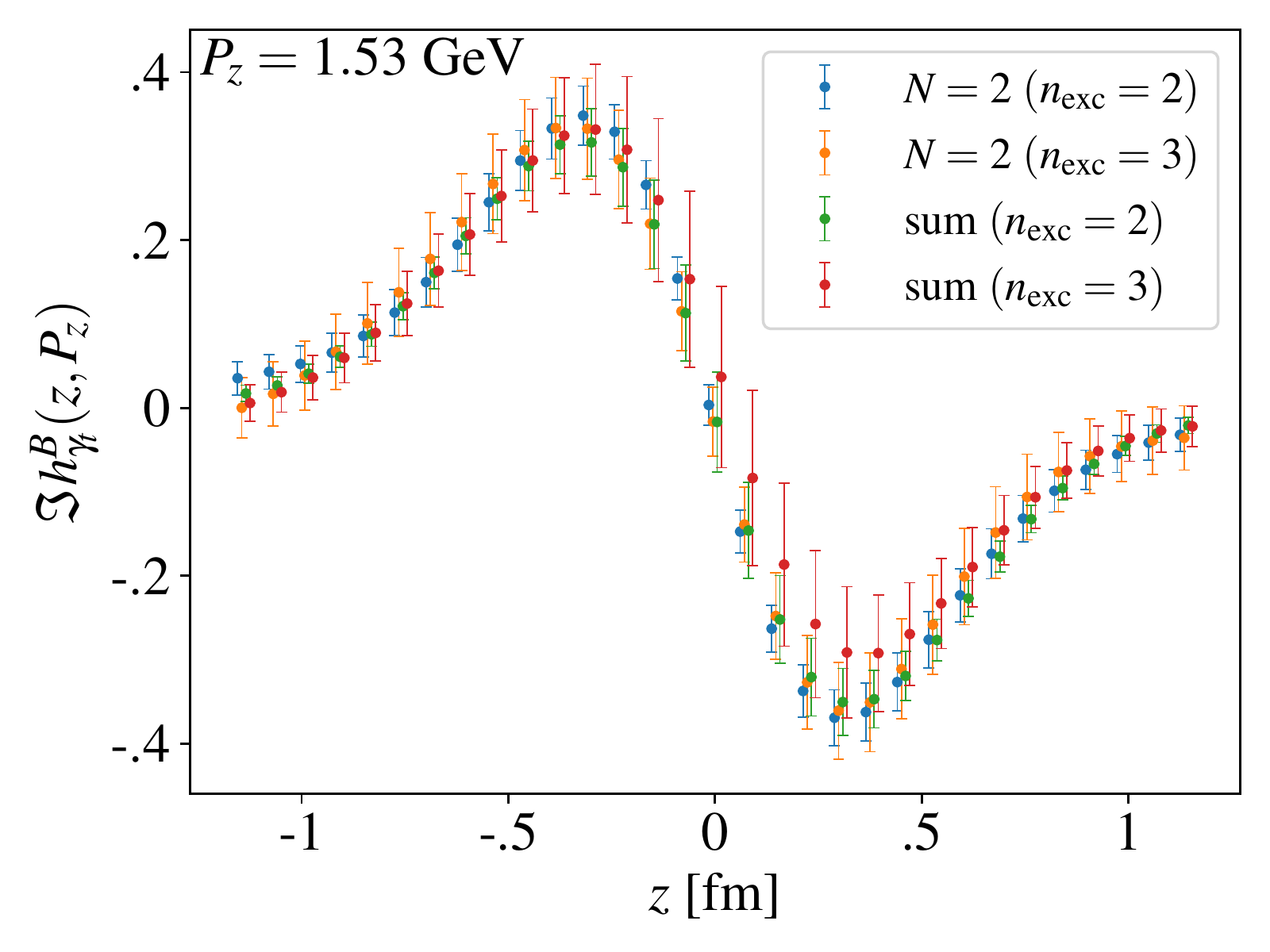}    
    \caption{The Wilson-line length dependence of the (upper) real and (lower) imaginary parts of the
             ground-state bare matrix elements from two-state and summation fits with $n_{\rm exc} = 2,3$
             for the three nonzero values of momentum considered (one for each column).}
    \label{fig:bare_mat_comp}
\end{figure*}

Further, we also consider the summation method as an alternative strategy, which sums $R_{\gamma_t}(\pvec, \ts, \ti, z)$
over a subset of $\ti \in [n_{\rm exc}+1, \ts - n_{\rm exc} - 1]$
\begin{equation}
    S^{n_{\rm exc}}_{\gamma_t} (\pvec, \ts, z) \equiv \sum_{\ti = n_{\rm exc}+1}^{\ts - n_{\rm exc} - 1} R_{\gamma_t} (\pvec, \ts, \ti, z) ,
\end{equation}
which in turn reduces the leading excited-state contamination to be $\mathcal{O} (e^{-\Delta_{1,0} \ts})$,
as opposed to $\mathcal{O} (e^{-\Delta_{1,0} \ts / 2})$ present in the ratio of \Cref{eq:c3pt_ratio}.
The summed ratio can then be fit to a linear function
\begin{equation}
    S^{n_{\rm exc}}_{\gamma_t} (\pvec, \ts, z) = B_0 + \ts h^B_{\gamma_t} (z, P_z, a) .
\end{equation}
We will use the summation fits as a consistency check for our multi-state fits directly to the ratio of three-point to two-point functions.

For our limited number of sink-source separations and statistics, we find the three-state fits to be unreliable,
therefore we only consider fits that include up to two states in the ratio of \Cref{eq:c3pt_ratio}.
Further, to reduce the effects of excited states as much as possible, we remove some of the insertion times near the source/sink symmetrically.
The number of excluded times $n_{\rm exc}$ is given such that the insertion times included in the fit are $\ti \in [n_{\rm exc}+1, \ts - n_{\rm exc} - 1]$.
In order to retain enough data for our fits, we only consider $n_{\rm exc} \le 3$ meaning that all fits will include insertion times in which more than two states are contributing, as evidenced from significant $t_{\rm min}$ dependence still present in the two-state fits shown in \Cref{fig:tmins} for $\ts \le n_{\rm exc} + 1$.
We must therefore use an effective value for the gap $\Delta_{1,0}$ that mocks up the effects of all higher excited states present for the smallest/largest insertion time included in the fits.
To this end, we simply use the extracted value of $\Delta_{1,0}$ coming from a two-state fit to the appropriate SS two-point function with $t_{\rm min} = n_{\rm exc} + 1$.

In order to reduce the effects of excited states as much as possible, our preferred fit is the two-state fit to the ratio \Cref{eq:c3pt_ratio} with $n_{\rm exc} = 3$, which completely excludes $\ts = 6$ from the fit.
In \Cref{fig:c3pt_repr_fits_re,fig:c3pt_repr_fits_im}, we show the results using this fit strategy for all computed values of momentum and a few representative values of Wilson-line length $z$.
Then, in \Cref{fig:bare_mat_comp}, we compare the extracted ground-state bare matrix elements as a function of the Wilson-line length $z$ from two-state fits and the summation method with $n_{\rm exc} = 2, 3$.
There is reasonable consistency among the various fits, but there is still some tension in a handful of cases,
which further motivates our use of the more conservative fits with $n_{\rm exc} = 3$.
\section{Mellin Moments from the Leading-twist OPE}
\label{sec:leading_twist_OPE}

The bare matrix elements $h^B_{\gamma_t} (z, P_z, a)$ are multiplicatively renormalizable,
and therefore we can cancel the renormalization factors, which only depend on the lattice spacing $a$ and the Wilson-line length $z$, by forming the ratio~\cite{Fan:2020nzz}
\begin{equation}
    \mathcal{M}_{\gamma_t} (\lambda, z^2; P_z^0) = \frac{h^B_{\gamma_t} (z, P_z, a)}{h^B_{\gamma_t} (z, P_z^0, a)} \big/ \frac{h^B_{\gamma_t} (0, P_z, a)}{h^B_{\gamma_t} (0, P_z^0, a)} ,
    \label{eq:ratio}
\end{equation}
where $\lambda \equiv z P_z$ is referred to as the Ioffe time and the ratio itself is the Ioffe time pseudo-distribution (pseudo-ITD).
This quantity is thus a renormalization-group invariant quantity.
The additional $z=0$ matrix elements appearing in the ratio are not strictly required but they enforce an exact normalization and further reduce correlations and systematics.
The lattice spacing dependence of the pseduo-ITD is suppressed for convenience.
The ratio for the specific case of $P_z^0 = 0$, used in what follows, is referred to as reduced pseudo-ITD~\cite{Orginos:2017kos,Joo:2019jct,Joo:2020spy,Bhat:2020ktg,Karpie:2021pap,Egerer:2021ymv,Bhat:2022zrw},
and will be denoted by $\mathcal{M}^0_{\gamma_t}(\lambda, z^2)$.

Using the OPE for the $\gamma_t$ matrix elements, we can extract the first few Mellin moments by fitting the pseudo-ITD to
\begin{equation}
\begin{split}
    \mathcal{M}_{\gamma_t} & (\lambda, z^2; P_z^0) = \\
    & \frac{\sum_{n=0} C_n (\mu^2 z^2) \frac{(-i\lambda)^n}{n!} \braket{x^n} (\mu) + \mathcal{O}(\Lambda^2_{\rm QCD} z^2)}{\sum_{n=0} C_n (\mu^2 z^2) \frac{(-i\lambda^0)^n}{n!} \braket{x^n} (\mu) + \mathcal{O}(\Lambda^2_{\rm QCD} z^2)} ,
\end{split}
\label{eq:fitrpITD}
\end{equation}
where $C_n(\mu^2 z^2)$ are Wilson coefficients which have been computed up to next-to-next-to-leading order (NNLO)~\cite{Chen:2020ody,Li:2020xml}, $\braket{x^n} (\mu)$ are the Mellin moments at a factorization scale $\mu$ defined by
\begin{equation}\label{eq.MellinMoments}
    \braket{x^n} (\mu) = \int_{-1}^{1} \mathrm{d}x \; x^n q^{u-d}(x, \mu) ,
\end{equation}
and
\begin{equation}
    q^{u-d}(x,\mu) \equiv
    \begin{cases}
        q^u(x,\mu) - q^d(x,\mu) , & x \geq 0 \\
        q^{\overline{u}}(-x,\mu) - q^{\overline{d}}(-x,\mu) , & x < 0
    \end{cases}
    \label{eq:isovector_pdf}
\end{equation}
where $q^{f}(x,\mu)$ and $q^{\overline{f}}(x,\mu)$ are the PDFs for the quark and antiquark of flavor $f$ defined for $x \in [0,1]$.
Additionally, although the effects are small, we include the target mass corrections by making the following substitution
\begin{equation}
    \braket{x^n} \to \braket{x^n} \sum_{k=0}^{n/2} \frac{(n-k)!}{k! (n-2k)!} \bigg(\frac{m^2_N}{4 P^2_z} \bigg)^k .
\end{equation}
The Wilson coefficients depend on the strong coupling constant $\alpha_s (\mu)$,
and we use the same estimates as in Ref.~\cite{Gao:2022iex}, which gives $\alpha_s (\mu = 2$ GeV$) = 0.2930$.

There are a few things to note regarding our fits to the ratio of OPEs.
First, we consider the leading-twist approximation where the higher-twist corrections that come in as $\mathcal{O}(\Lambda^2_{\rm QCD} z^2)$ are ignored.
Therefore, we must empirically determine at what value of $z^2$ this approximation breaks down, which can be done by looking for a strong dependence of our results on $z^2$.
Second, since we have chosen $P_z^0 = 0$, the denominator in \Cref{eq:fitrpITD} becomes unity.
With these simplifications, it can easily be seen that the real and imaginary parts of the ratio of OPEs in \Cref{eq:fitrpITD} correspond to even and odd moments, respectively.
Therefore, we separately fit the real and imaginary parts of the reduced pseudo-ITD to
\begin{equation}
\begin{split}
    \Re \mathcal{M}^0_{\gamma_t}(\lambda, z^2) &= \sum_{n=0}^{\lfloor N_{\rm max}/2 \rfloor} C_{2n} (\mu^2 z^2) \frac{(-i \lambda)^{2n}}{(2n)!} \braket{x^{2n}} , \\
    \Im \mathcal{M}^0_{\gamma_t}(\lambda, z^2) &= \sum_{n=1}^{\lceil N_{\rm max}/2 \rceil} C_{2n-1} (\mu^2 z^2) \frac{(-i \lambda)^{2n-1}}{(2n-1)!} \braket{x^{2n-1}} ,
\end{split}
\end{equation}
respectively, where the moments $\braket{x^n}$ are the fit parameters (except $\braket{x^0}$, which is fixed to one) and $N_{\rm max}$ is the largest moment considered.

\begin{figure}
    \centering
    \includegraphics[width=\columnwidth]{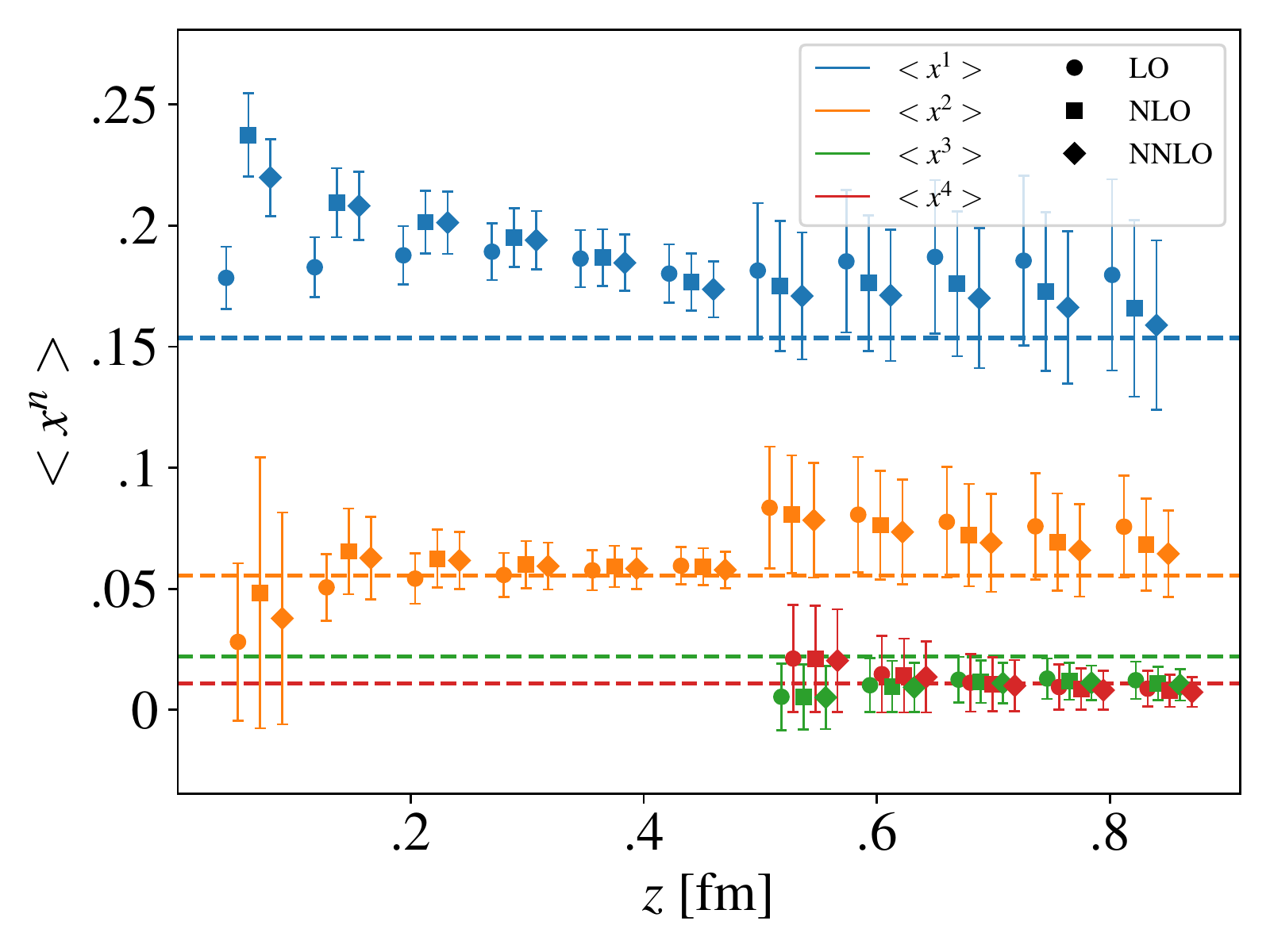}
    \caption{Results for the lowest four Mellin moments as a function of $z$
             from fits of the
             reduced pseudo-ITD at fixed $z$ with $n_z \in [1,4,6]$ to
             the leading-twist OPE using LO, NLO, and NNLO
             Wilson coefficients evaluated at $\mu = 2$ GeV.
             Only the first two
             moments are extracted for $z \leq 6a$.
             The horizontal dashed lines correspond to the central values
             of the moments extracted from the global analysis of NNPDF4.0~\cite{NNPDF:2021njg} defined at a scale $\mu = 2$ GeV.}
    \label{fig:fixed_z}
\end{figure}

As a first test of the validity of the leading-twist approximation, we perform fits including data at a fixed value of $z^2$ only.
As the higher-twist effects enter as $\mathcal{O}(\Lambda_{\rm QCD}^2 z^2)$, observation of a dependence in the extracted moments from a fixed-$z^2$ analysis on $z$ would likely indicate non-negligible higher-twist effects, invalidating the leading-twist approximation in the region of $z$ where this dependence is observed.
However, it should be noted that additional systematics from discretization effects and large logs (see Appendix B in Ref.~\cite{Gao:2022iex}) can lead to dependence of the moments on $z$ in the small-$z$ region (i.e. $z \lesssim 0.2$ fm) as well.
Additionally, the fixed-$z^2$ analysis also gives us an opportunity to determine the dependence on the perturbative order of the Wilson coefficients.
The results are shown in \Cref{fig:fixed_z}, with a comparison to the moments determined from the global analysis of NNPDF4.0~\cite{NNPDF:2021njg} shown as dashed lines.
The fits use a value of $\mu = 2$ GeV when evaluating the Wilson coefficients, and the NNPDF4.0 results are also defined at the scale $\mu = 2$ GeV.
We found that for values of $z \leq 6a \sim 0.456$ fm, the data is only sensitive to the lowest Mellin moment.
Therefore, we only include a higher moment in the fits when $z > 6a$.
This can be understood by evaluating the reduced pseudo-ITD using \Cref{eq:fitrpITD} with the moments extracted from the NNPDF4.0 global analysis~\cite{NNPDF:2021njg}.
From this, we can determine the dependence on the number of included moments in the leading-twist OPE.
We find that for $z \lesssim 0.4$ fm, there are no significant effects for any $N_{\rm max} \geq 2$.
Therefore, a choice of $N_{\rm max}=2$ is sufficient in this region.
But, beyond this, including a higher moment begins to make a difference.

There are a few interesting observations from these fits.
First, the only dependence on $z$ or the perturbative order is for $\braket{x}$.
Further, the perturbative order of the Wilson coefficients only seems to matter at very small values of $z$, where there is some mild $z$-dependence for $\braket{x}$ beyond leading order.
Following this, the results are rather independent of $z$ for  $z \gtrsim 0.4$ fm.
These effects are likely some combination of discretization effects and the need for resummation of large logs which was done in~\cite{Gao:2021hxl,Su:2022fiu,Gao:2022iex}.

\begin{figure}
    \centering
    \includegraphics[width=\columnwidth]{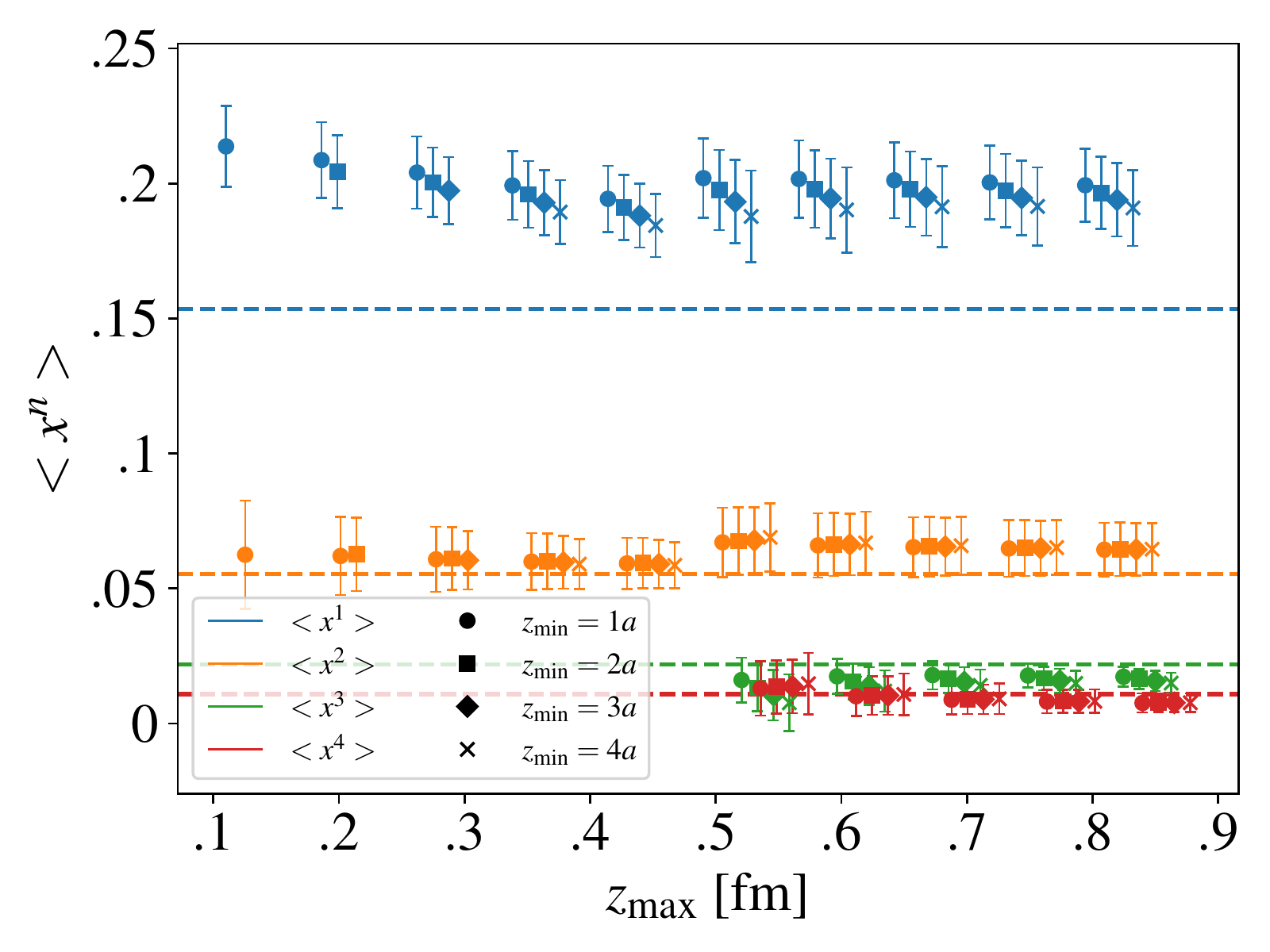}
    \caption{Results for the lowest four Mellin moments from uncorrelated fits of the reduced pseudo-ITD to
             the leading-twist OPE as a function of $z_{\rm max}$, with
             $z \in [z_{\rm min}, z_{\rm max}]$ and $n_z \in [1, 4, 6]$.
             The results use the NNLO Wilson coefficients evaluated at $\mu = 2$ GeV.
             Only the first two moments are considered for $z_{\rm max} \leq 6a$.
             The horizontal dashed lines are the same as in \Cref{fig:fixed_z}.}
    \label{fig:global_z}
\end{figure}

Next, we consider the inclusion of a range of $z$ for our fits,
where we still include all three values of nonzero momentum.
Up to this point, all fits have been fully correlated, however we found difficulty in obtaining reliable correlated fits when including multiple values of $z$ due to a high condition number for the covariance matrix.
For the cases in which a correlated fit was possible, the results are in agreement with a fully uncorrelated fit.
Therefore, we continue to use uncorrelated fits in what follows.
The results are shown in \Cref{fig:global_z}.

\begin{figure*}
    \centering
    \includegraphics[width=\columnwidth]{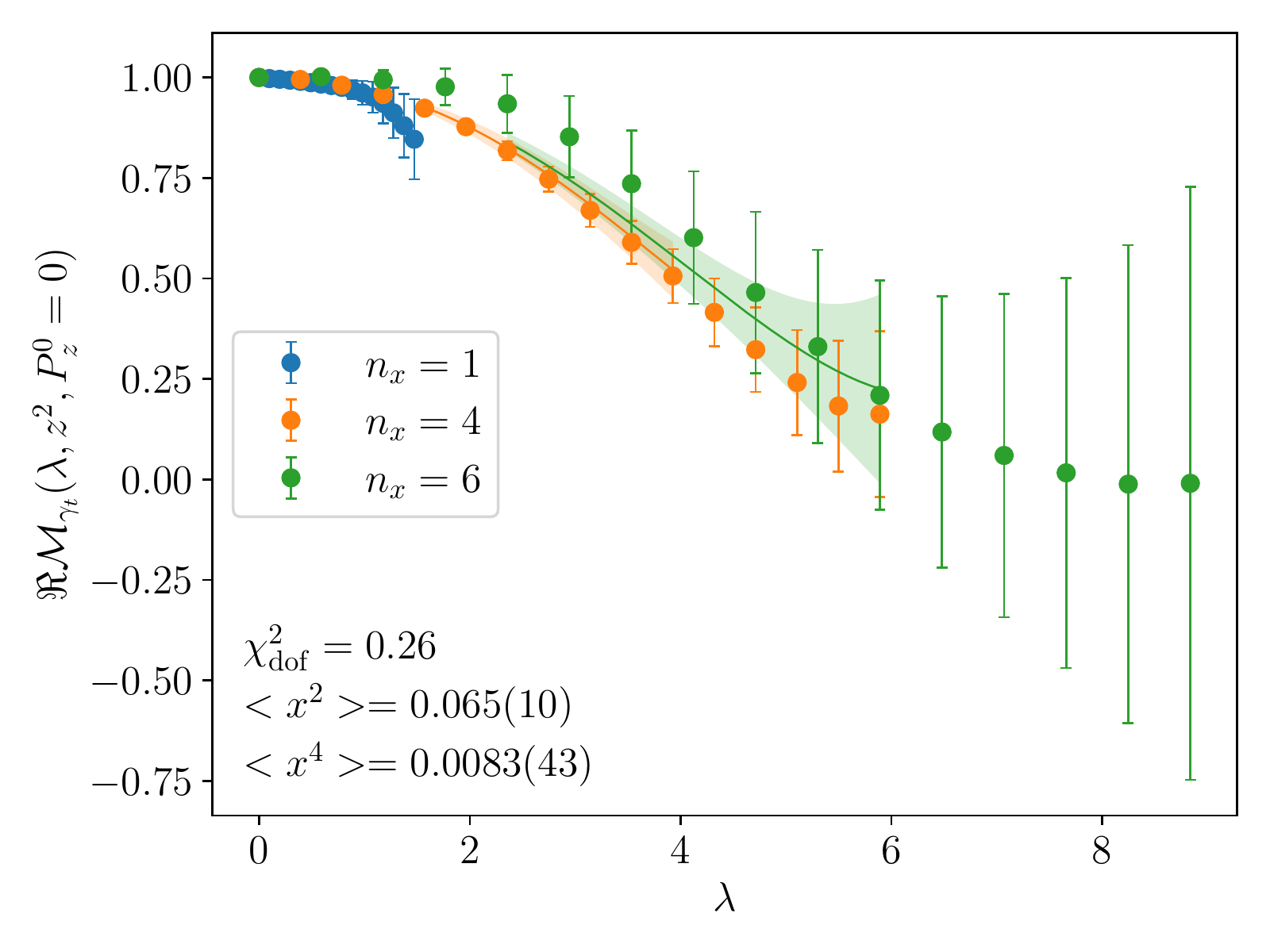}
    \includegraphics[width=\columnwidth]{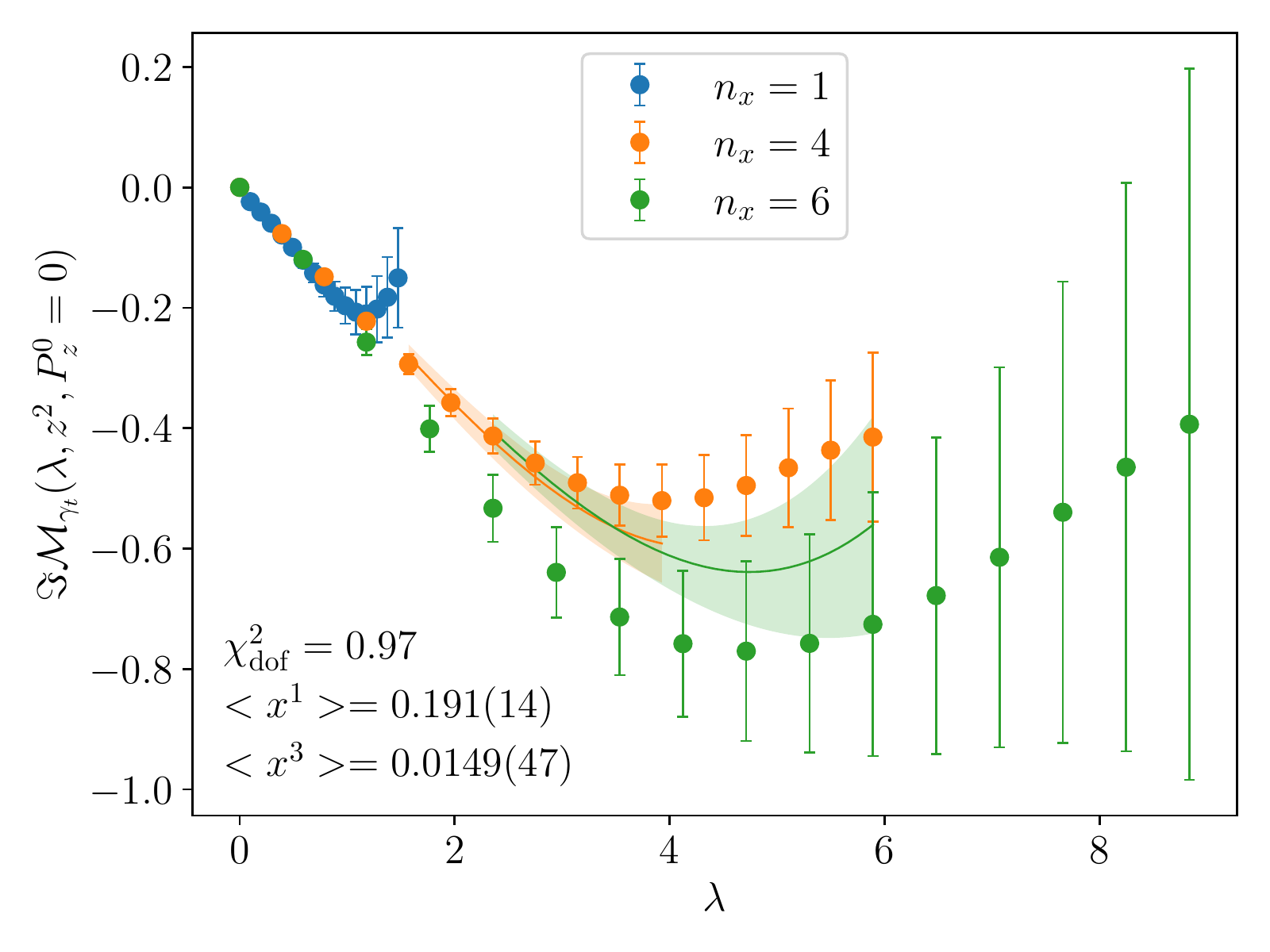}
    \caption{The (left) real and (right) imaginary parts of the reduced pseudo-ITD
             for the three momentum used in this work.
             The data come from our preferred fit strategy
             described in \Cref{subsec:c3pt_analysis}.
             The shaded bands correspond to the fits using the leading-twist
             OPE with NNLO Wilson coefficients evaluated at $\mu = 2$ GeV and
             including two moments in both the real and imaginary parts.}
    \label{fig:fit_ratio_matrix_elements}
\end{figure*}

As expected from the fixed-$z^2$ analysis, larger values of $z_{\rm min}$ tend to bring the value for $\braket{x}$ down closer to the global fit results.
But, there is generally good agreement among all fits.
As our final fit result, we choose $z_{\rm min} = 4a$ to avoid discretization effects and large logs, and $z_{\rm max} = 10a$ to avoid higher-twist effects.
The fit results using these choices for the real and imaginary parts of the reduced pseudo-ITD are shown in \Cref{fig:fit_ratio_matrix_elements}.
Our final results for the lowest four Mellin moments at $\mu = 2$ GeV are
\begin{equation}
\begin{split}
    \braket{x} = 0.191(14) , \; \braket{x^3} = 0.0149(47) , \; \chi^2_{\rm dof} = 0.97 , \\
    \braket{x^2} = 0.065(10) , \; \braket{x^4} = 0.0083(43) , \; \chi^2_{\rm dof} = 0.26 , \\
\end{split}
\end{equation}
which can be compared to the results from NNPDF4.0 also defined at $\mu = 2$ GeV
\begin{equation}
\begin{split}
    \braket{x}& = 0.15355(52) , \; \braket{x^3} = 0.02197(12) , \\
    \braket{x^2}& = 0.05536(22) , \; \braket{x^4} = 0.010846(78) .
\end{split}
\end{equation}

Some discussion is in order regarding our extracted value of $\braket{x}$, which is $2-3\sigma$ larger than the value obtained from phenomenological global fits.
There is a long history of lattice calculations consistently obtaining larger values as well.
First, there are several calculations that also utilized nonlocal operators to obtain the first few moments which see similar disagreements with the estimates from the global fits~\cite{Karpie:2018zaz,Joo:2019jct,Joo:2020spy,Bhat:2020ktg,Bhat:2022zrw}.
Further, these moments can also be extracted from local twist-2 operators, of which there are several lattice calculations which also see a value larger than expected from experiment~\cite{Bali:2014gha,Abdel-Rehim:2015owa,Alexandrou:2017oeh,Harris:2019bih}.
However, there was an older study consistent with the global fits that argued excited-state effects could cause larger extractions for $\braket{x}$~\cite{Green:2012ud}.
Finally, a recent study also obtained results consistent with the global phenomenological fits by improving their analysis strategy, focusing on excited state contamination, which also suggested that discretization effects could produce shifts of $\braket{x}$ upwards~\cite{Ottnad:2021tlx}.
More work is needed to fully resolve this issue.
\section{PDF from leading-twist OPE: DNN reconstruction}
\label{sec:dnn}

We now turn our attention to the extraction of the PDF itself.
In this section, we use the pseudo-distribution approach~\cite{Radyushkin:2017cyf,Orginos:2017kos},
which requires the solution of an ill-posed inverse problem that we solve via the use of a deep neural network (DNN).
We used this method in our previous work~\cite{Gao:2022iex},
and we repeat the pertinent details here.

Additionally, once we obtain the PDF, the moments can be extracted from it, allowing for a comparison to our extractions in the previous section.

\subsection{Method}\label{subsec:dnnMethod}

The leading-twist factorization formula for
the renormalized matrix element can be written as
\begin{equation}
    h^R(z,P_z,\mu) =\int_{-1}^1d\alpha\, \mathcal{C}(\alpha, \mu^2z^2)\, \int_{-1}^1 dy\, e^{-iy \alpha\lambda} q(y,\mu),
\end{equation}
where $q(y, \mu)$ is the light-cone PDF, and the matching kernel $\mathcal{C}(\alpha, \mu^2z^2)$ can be inferred from the Wilson coefficients $C_n(\mu^2 z^2)$ \cite{Radyushkin:2017cyf,Orginos:2017kos,Izubuchi:2018srq}, which connect the position-space matrix elements to the $x$-dependent PDFs. However, it has been shown that the ratio-scheme data $\mathcal{M}_{\gamma_t} (\lambda, z^2; P_z^0)$ only contains information on the first few moments of the PDFs~\cite{Gao:2020ito, Gao:2022vyh, Gao:2022iex}. To determine the $x$-dependence with limited information, priors have to be used. Conventionally one may choose certain models inspired by the end-point behavior of the PDFs like,
\begin{align}
    q(x)=Ax^\alpha(1-x)^\beta(1+\textup{subleading~terms}),
\end{align}
where the subleading terms can be modeled~\cite{Gao:2020ito} which however may lead to a bias. To reduce the model bias, more general functional forms like the Jacobi polynomial basis~\cite{Egerer:2021ymv, HadStruc:2021qdf, Karpie:2021pap} are proposed to make a more flexible PDF parametrization. In addition, the deep neural network technique, which in principle can approximate any function with enough complexity in a smooth and unbiased manner, is probably the most flexible method. The DNN has been proposed to parametrize the full PDF function~\cite{Karpie:2019eiq, DelDebbio:2020rgv} and the Ioffe-time distribution~\cite{Gao:2022iex}. In this work, instead we apply the DNN to only represent the subleading terms of the PDFs, which limits its contribution, in order to keep the end-point behavior and avoid any serious overfitting.

In our case, the real and imaginary parts of the ratio-scheme matrix elements $\mathcal{M}_{\gamma_t} (\lambda, z^2; P_z^0)$ with $P_z^0=0$ are related to $q^-(x)$ and $q^+(x)$, respectively, (e.g. see Ref.~\cite{Bhat:2022zrw})
\begin{align}\label{eq:isoVrelation}
\begin{split}
	&q^-(x)\equiv q^u(x)-q^d(x)-(q^{\bar u}(x)-q^{\bar d}(x)) ,\\
	&q^+(x)\equiv q^u(x)-q^d(x)+(q^{\bar u}(x)-q^{\bar d}(x)) ,
\end{split}
\end{align}
which are defined for $x \in [0,1]$ (see \Cref{eq:isovector_pdf} and the surrounding text).
Using the DNN we parametrize these functions as
\begin{align}\label{eq:DNNform}
\begin{split}
	q^-(x;\alpha^-,\beta^-,\boldsymbol{\theta}^-)\equiv A^-&x^{\alpha^-}(1-x)^{\beta^-} \\
        &\times [1+\delta \sin({f^-_{\rm DNN}(x,\boldsymbol{\theta}^-)})] \\
	q^+(x;\alpha^+,\beta^+,A^+,\boldsymbol{\theta}^+)\equiv A^+& x^{\alpha^+}(1-x)^{\beta^+} \\
        &\times [1+\delta \sin({f^+_{\rm DNN}(x,\boldsymbol{\theta}^+)})],
\end{split}
\end{align}
where $A^-$ is fixed by the normalization condition for $q^-$, i.e. $\int_0^1dx \; q^-(x;\alpha^-,\beta^-,\boldsymbol{\theta}^-)=1$, and $A^+$ is a free parameter. The $f^-_\textup{DNN}(x, \boldsymbol{\theta}^-)$ and $f^+_\textup{DNN}(x, \boldsymbol{\theta}^+)$ are DNN functions whose contributions are limited by $|\delta\sin(f^{\pm}_{\rm DNN})|\lesssim\delta$. With this setup, we can make sure the DNNs are only subleading contributions. In the case of $\delta=0$, contributions from the DNN are disabled and the forms in \Cref{eq:DNNform} simply reduce to the two ($q^-(x;\alpha^-,\beta^-)$) and three ($q^+(x;\alpha^+,\beta^+,A^+)$) parameter model fit. Although in this work we fix $\delta$ to be a constant, it could also be a function of $x$ if one assumes the contribution from subleading terms varies for different local $x$.

The DNN functions $f^{\pm}_\textup{DNN}(x, \boldsymbol{\theta}^{\pm})$ are composite multistep iterative functions, constructed layer by layer. The first layer is made up of a single node (i.e. $a_1^1$) and corresponds to the value of the input variable $x$. Then the hidden layers first perform a linear transformation,
\begin{align}
    z_{i}^{(l)} = b_{i}^{(l)}+\sum_jW_{ij}^{(l)}a_j^{(l-1)} ,
\end{align}
followed by the nonlinear activation $\sigma^{(l)}(z_{i}^{(l)})$ whose output gives the input to the next layer $a_i^{(l)}$. The particular activation function we used is the so-called exponential linear unit $\sigma_{\textsf{elu}}(z)=\theta(-z)(e^z-1)+\theta(z)z$.
Finally, the last layer produces the output $f^{\pm}_\text{DNN}(x,\boldsymbol{\theta}^{\pm})$, which is then used to evaluate $q^{\pm}(x;A^{\pm},\alpha^{\pm},\beta^{\pm},\mathbf{\theta}^{\pm})$.
The lower indices $i = 1,...,n^{(l)}$ label the particular node within the $l$th layer, where $n^{(l)}$ is the number of nodes in the $l$th layer.
The upper indices, in parenthesis, $l=1,...,N$ label the individual layers, where $N$ is the number of layers (i.e. the depth of the DNN).
The biases $b_{i}^{(l)}$ and weights $W_{ij}^{(l)}$, denoted by $\boldsymbol{\theta}^{\pm}$, are the DNN parameters to be optimized (trained) by minimizing the loss function,
\begin{align}
\begin{split}
J(\boldsymbol{\theta}^{\pm}) 
\equiv
    \frac{\eta}{2} \boldsymbol{\theta}^{\pm}\cdot\boldsymbol{\theta}^{\pm}+\frac{1}{2} \chi^2(\boldsymbol{\theta}^{\pm},\alpha^{\pm},\beta^{\pm},...),
\end{split}
\end{align}
where the first term is to prevent overfitting and makes sure the DNN-represented function is well behaved and smooth. The definition and details of the $\chi^2$ function can be found in \Cref{app:DNNchisq}. Given our low statistics, a simpler network structure like $\{1,16,16,1\}$ (where each entry gives the number of nodes in each layer) is good enough to approximate the $f^{\pm}_\text{DNN}(x,\boldsymbol{\theta}^{\pm})$ smoothly. Practically, we vary $\eta$ from $10^{0}$ to $10^{-2}$, and tried network structures of size $\{1,16,16,1\}$, $\{1,16,16,16,1\}$ and $\{1,32,32,1\}$. We found the results remain unchanged. We therefore chose $\eta=0.1$ and the DNN structure with four layers, including the input/output layer, to be $\{1,16,16,1\}$.

\subsection{DNN represented PDF}

\begin{table*}
\centering
\begin{tabular}{|c|c|c|c|c|c||c|c||}
\hline
\hline
&$z_{\rm max}$ [fm]&$\alpha^-$&$\beta^-$&$A^-$&$\chi^2/N_{data}$&$\langle x^2 \rangle$& $\langle x^4 \rangle$\cr
\hline
         & 0.61 & -0.47(0.75) & 1.38(2.38) & & 0.64&0.071(14)&0.022(18)\cr

$q^-(x)$ & 0.76 &  0.01(1.12) & 2.72(3.45) & & 0.52&0.070(12)&0.015(12)\cr

         & 0.92 &  0.16(1.18) & 3.49(3.76) & & 0.44&0.069(12)&0.014(9)\cr
\hline
\hline
&$z_{\rm max}$ [fm]&$\alpha^+$&$\beta^+$&$A^+$&$\chi^2/N_{data}$&$\langle x \rangle$& $\langle x^3 \rangle$\cr
\hline
         & 0.61 & -0.99(1.15) & 1.38(2.38) & 0.25(0.73) & 0.74&0.202(19)&0.027(16)\cr

$q^+(x)$ & 0.76 & -0.09(1.84) & 2.72(3.45) & 1.61(6.94) & 0.78&0.201(18)&0.030(12)\cr

         & 0.92 &  0.47(2.12) & 3.49(3.76) & 4.6(22.1) & 0.80&0.200(17)&0.031(10)\cr
\hline
\hline
\end{tabular}
\caption{The parameters from the DNN training with $\delta=0.1$ using the data with $z\in[2a, z_{\rm max}]$ are shown, where $N_{data}=21,27,33$ for $z_{\rm max} = 0.61,0.76,0.92$ fm. The moments inferred from the model fit results are also shown.}
\label{tb:model}\end{table*}

\begin{figure}
    \centering
    \includegraphics[width=0.45\textwidth]{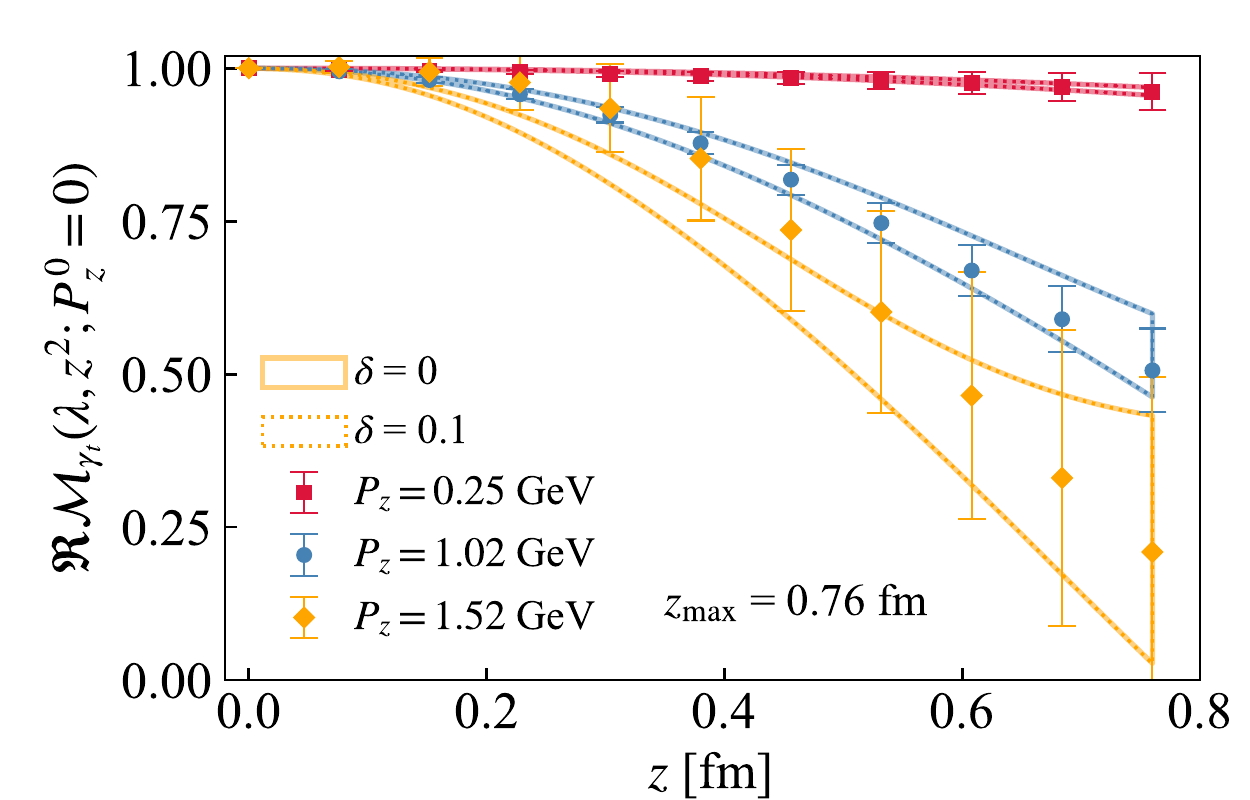}
    \includegraphics[width=0.45\textwidth]{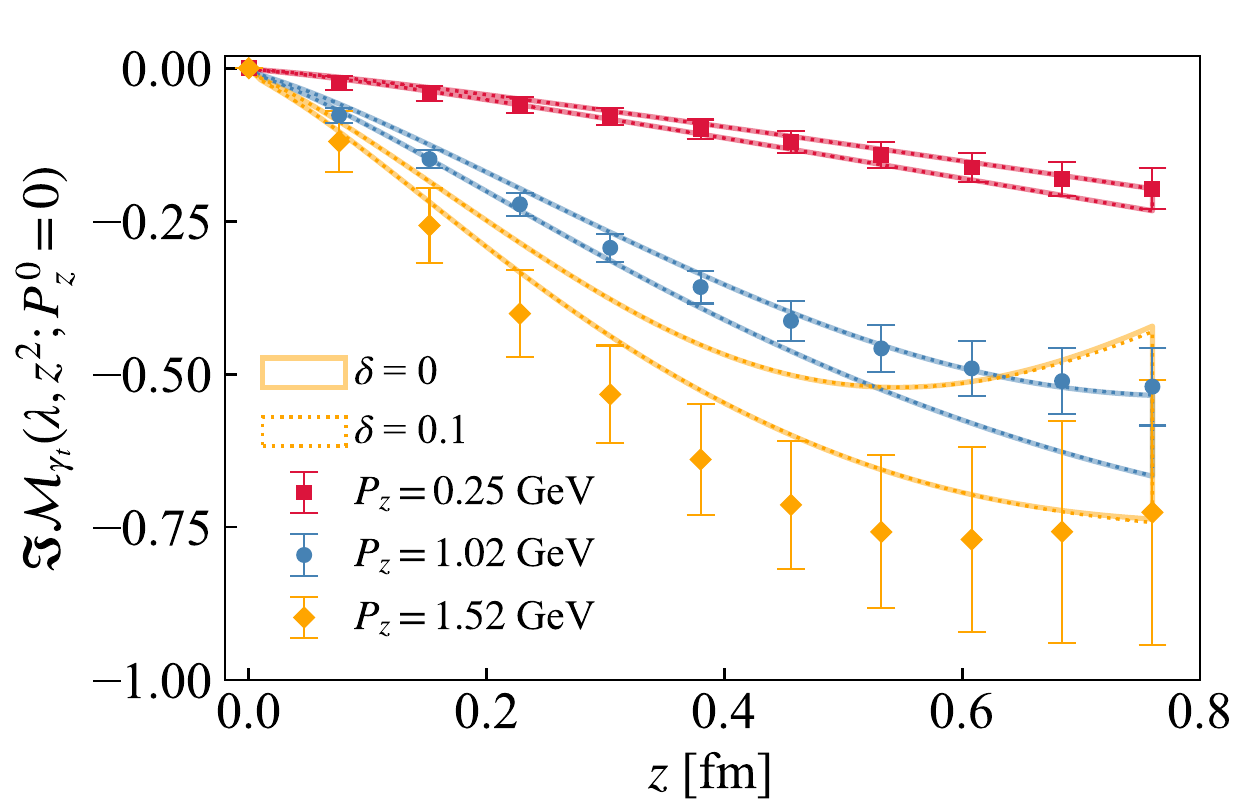}
    \caption{The DNN training results using the ratio-scheme renormalized matrix elements in the range $z\in[2a, 10a]$ (27 data points) for the real part (upper panel) and imaginary part (lower panel) are shown. The results using $\delta=0$ and $\delta=0.1$ are shown as the solid and dotted curves, respectively.}
    \label{fig:DNN_zmax76_rITD}
\end{figure}

\begin{figure}
    \centering
    \includegraphics[width=0.45\textwidth]{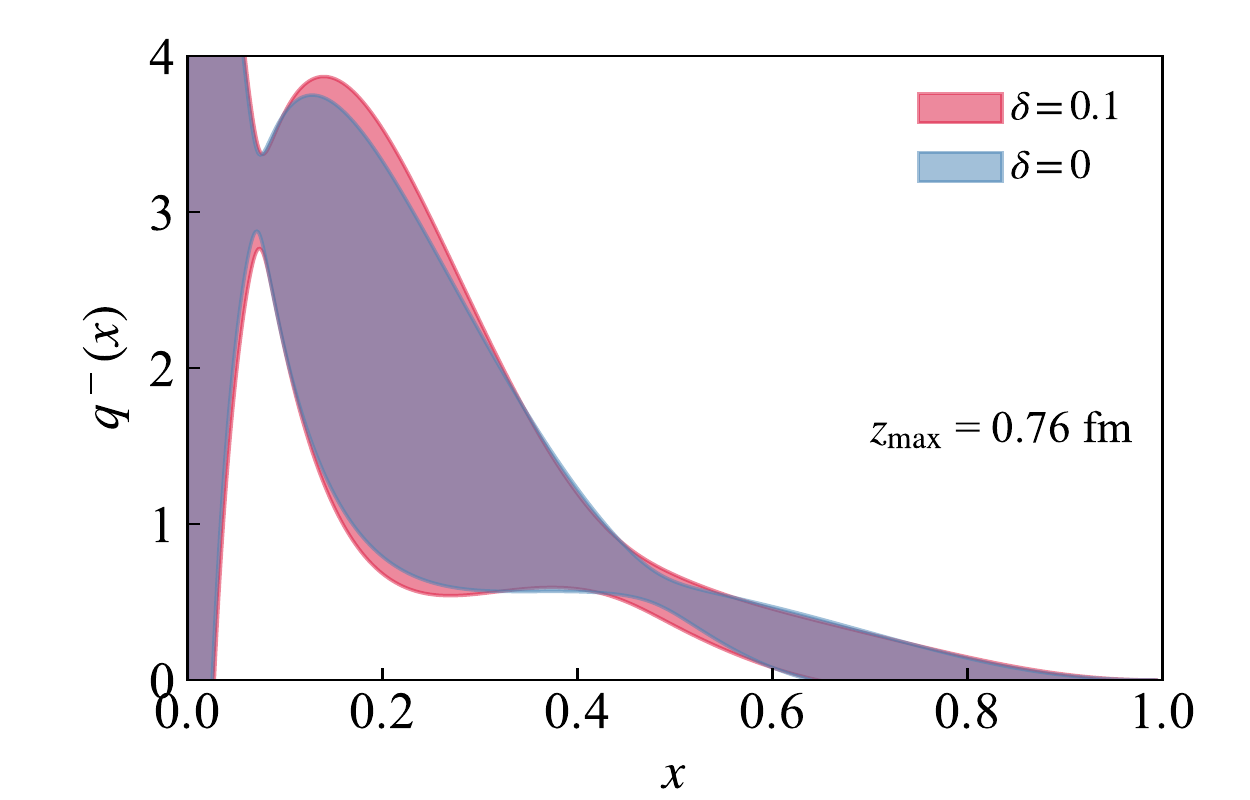}
    \includegraphics[width=0.45\textwidth]{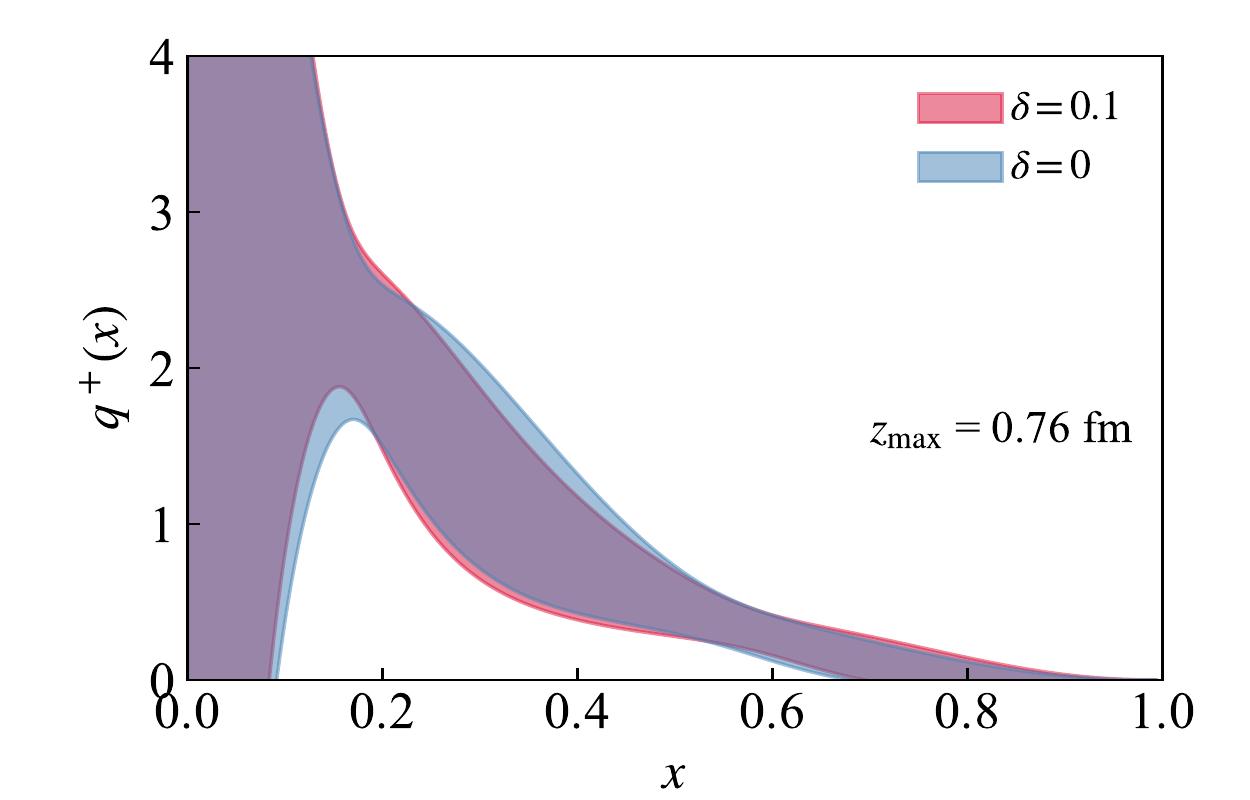}
    \caption{The DNN represented PDFs using the ratio-scheme renormalized matrix elements in the range $z\in[2a, 10a]$ (27 data points) for the real part (upper panel) and imaginary part (lower panel) are shown. The results with $\delta=0$ and $\delta=0.1$ are shown as the blue and red bands, respectively.}
    \label{fig:DNN_zmax76_PDF}
\end{figure}

We train $q^-(x)$ and $q^+(x)$ using the ratio-scheme data $\mathcal{M}_{\gamma_t} (\lambda, z^2; P_z^0)$ with $z\in[2a, z_{\rm max}]$, where we skip the first point $z=a$ to avoid the most serious discretization effects. In the upper panel of \Cref{fig:DNN_zmax76_rITD}, we show the results using $z\in[2a, 10a]$ (27 data points in total) for the real part with $\delta=0$ (solid curves) and $\delta=0.1$ (dotted curves). As one can see, the curves from the DNN go through the data points well, especially for the smaller momenta which are more precise. We vary $\delta$ to control the contributions from the DNN, where larger $\delta$ allows more flexibility of the PDF parametrization. However from $\delta=0$ to $\delta=0.1$, the fit results barely change, which is also reflected in the $\chi^2$ (total number of 27 data points) which only evolves from 14.078 to 14.077 between the different fits. The corresponding PDFs are shown in the upper panel of \Cref{fig:DNN_zmax76_PDF}, where the error bands increase for larger $\delta$. These observation suggest the DNN provides a very flexible parametrization but the data is very limited so that the simple two-parameter model $q^-(x;A^-,\alpha^-,\beta^-)$ can already describe the data well. It is  expected that for the imaginary part $q^+(x)$, the three-parameter ($A^+,\alpha^+,\beta^+$) fit will be extremely unstable. With the idea that antiquarks are supposed to have little contribution at large $x$ for the nucleon, where $q^+(x)$ and $q^-(x)$ are both dominated by $q^u(x)-q^d(x)$, we choose to prior $\beta^+$ in $q^+(x)$ using the result for $\beta^-$ in $q^-(x)$. The results for the the imaginary part of $\mathcal{M}_{\gamma_t} (\lambda, z^2; P_z^0)$ and for $q^+(x)$ are shown in the lower panels of \Cref{fig:DNN_zmax76_rITD} and \Cref{fig:DNN_zmax76_PDF}, respectively. Similarly, with a larger $\delta$, the fit results do not change much, with $\chi^2$ evolving from 21.28 to 21.09 and the error bands of the PDFs becoming slightly larger.
In what follows, we set $\delta=0.1$, but larger values of $\delta$ will play an important role for more precise data in the future.

\begin{figure}
    \centering
    \includegraphics[width=0.5\textwidth]{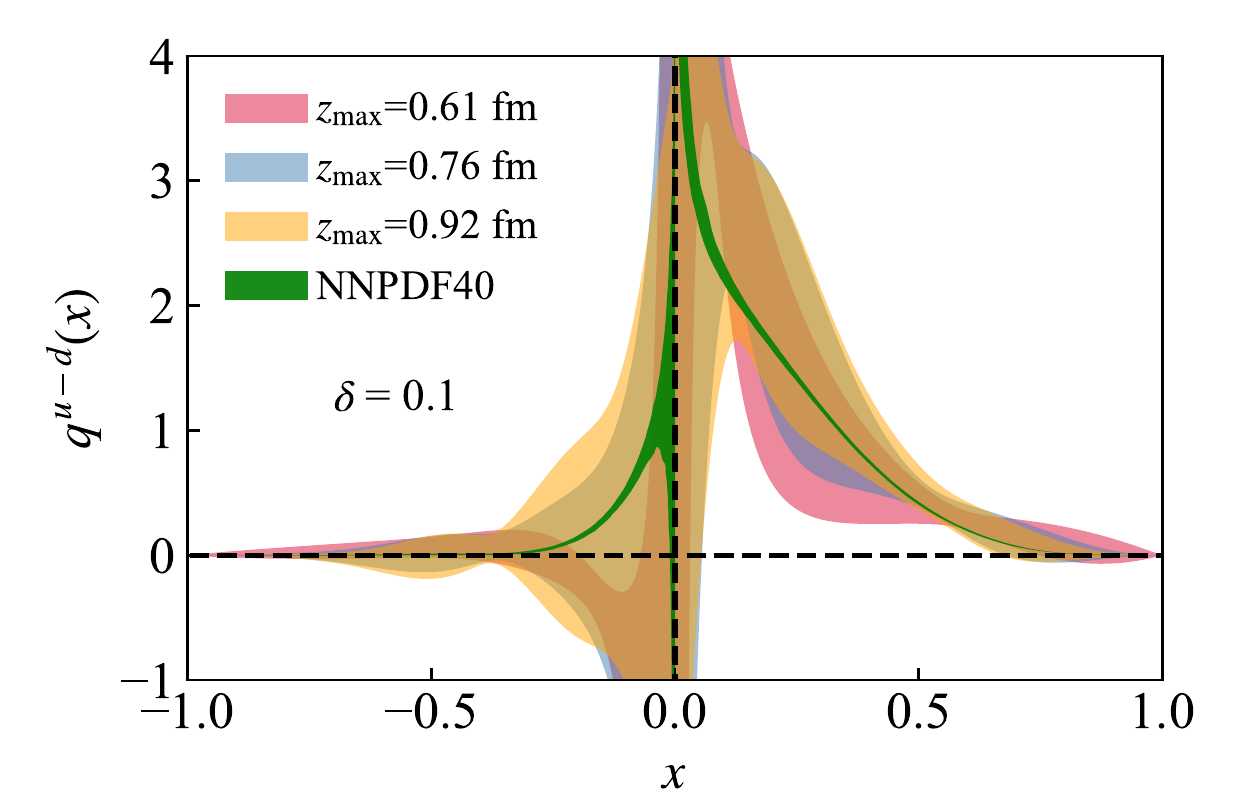}
    \caption{The isovector PDF of the nucleon using data in the range $z\in[2a,z_{\rm max}]$ and $\delta=0.1$ are shown. For comparison, we show the global analysis results from NNPDF4.0~\cite{NNPDF:2021njg}.}
    \label{fig:DNN_delta0_PDF}
\end{figure}

As higher-twist contamination can enter at large values of $z$, we train the PDFs using different $z_{\rm max}$ to study its dependence. The results are shown in \fig{DNN_delta0_PDF}, where we combine $q^-(x)$ and $q^+(x)$ to obtain the isovector PDF $q^{u-d}(x)$, defined for $x \in [-1,1]$, using the relations in \Cref{eq:isoVrelation}. It is observed that the $z_{\rm max}$ dependence is small relative to the large errors, suggesting that higher-twist effects are less important compared to the statistics for our data. Compared to the global analysis from NNPDF4.0~\cite{NNPDF:2021njg}, our results have significantly larger errors, leading to agreement with the global analysis results in most regions of $x$. We summarize the fit results and the $\chi^2$ in \Cref{tb:model}. As one can see, the errors on the fit parameters are quite large due to the limited statistics we were able to obtain. The first few moments inferred from the model fit results are also shown, which are consistent with the model-independent extraction in \Cref{sec:leading_twist_OPE}.
\section{\texorpdfstring{$x$}{x}-space Matching}
\label{sec:xspace}
This section demonstrates the calculation of the unpolarized proton PDF using the method of LaMET. This offers a consistency check with the previous method of leading-twist OPE combined with the deep neural network. In addition, it offers a more direct approach to the $x$-dependent PDF due to the ability to extrapolate to infinite distance and, thus, Fourier transform to momentum space.
\subsection{Method}
\label{sec:xspaceMethod}

The data is analyzed by the same method as that laid out in~\cite{Gao:2021dbh}. The process involves renormalizing the matrix elements in the hybrid-scheme; extrapolation of the renormalized matrix elements to infinite distance; Fourier transforming the matrix elements to momentum space and, finally, matching our data to the light-cone PDF.

\subsubsection{Renormalization}
\label{subsec:Renormalization}

The first step is to renormalize the bare matrix elements which are multiplicatively renormalizable~\cite{Ji:2017oey,Ishikawa:2017faj,Green:2017xeu} as
\begin{equation}\label{eq:op_renorm}
    h_{\gamma_t}^{\rm B}(z,P_z,a)=e^{-\delta m(a)|z|}Z_{\gamma_t}(a) h^{\rm R}_{\gamma_t}(z,P_z),
\end{equation}
where $Z_{\gamma_t}(a)$ contains the logarithmic ultraviolet (UV) divergences and is independent of $z$,
and $\delta m(a)$ includes the linear UV divergences coming from the Wilson-line self-energy.
Although the ratio scheme can be used at small values of $z$ where the OPE is valid, we need a method that does not introduce nonperturbative effects in the IR region at large values of $z$.
Thus, we use the recently developed hybrid scheme~\cite{Ji:2020brr} which explicitly includes the factor involving $\delta m(a)$ at large $z$.

An estimate for $\delta m(a)$ can be determined from the static quark-antiquark potential,
leading to a value of $a\delta m(a)=0.1597(16)$ taken from Refs.~\cite{HotQCD:2014kol,Bazavov:2016uvm,Bazavov:2017dsy,Bazavov:2018wmo,Petreczky:2021mef}.
However, the quantity $\delta m(a)$ has a scheme dependence, and in order to match to $\overline{\mathrm{MS}}$ we must determine the necessary shift $\overline{m}_0 \equiv \delta m^\prime - \delta m$ where $\delta m^\prime$ subtracts the linear divergence in the Wilson line and matches the result to the $\overline{\rm MS}$ scheme which includes an ${\cal O}(\Lambda_{\rm QCD})$ renormalon ambiguity~\cite{Braun:2018brg}. Following Ref.~\cite{Gao:2021dbh}, $\overline{m}_0$ can be estimated from fitting a ratio of the bare matrix elements, including factors of $e^{\delta m (a) z}$,
to a form motivated from the OPE
\begin{equation}
\label{eq:mass_sub_fit}
    e^{\delta m(a)(z-z_0)} \frac{h^B_{\gamma_t}(z,0,a)}{h^B_{\gamma_t}(z_0,0,a)}
    = e^{-\mzero(z-z_0)}\frac{C_0(\mu^2z^2)-\Lambda z^2}{C_0(\mu^2z_0^2)-\Lambda z_0^2} ,
\end{equation}
where $z_0 = 3a \sim 0.228$ fm, $C_0(\mu^2z^2)$ is computed at NNLO and is the only leading-twist Wilson coefficient contributing to the OPE at zero momentum, and the terms $(\mzero,\Lambda)$ are fit parameters. The value of $z_0$ is chosen large enough to neglect discretization effects and small enough to neglect higher-twist effects which become important for $z\gtrsim 0.2$ fm.
The $\Lambda$ term allows for the inclusion of larger values of $z$ in order to capture some of the higher-twist effects.

\begin{figure}
    \centering
    \includegraphics[width=\columnwidth]{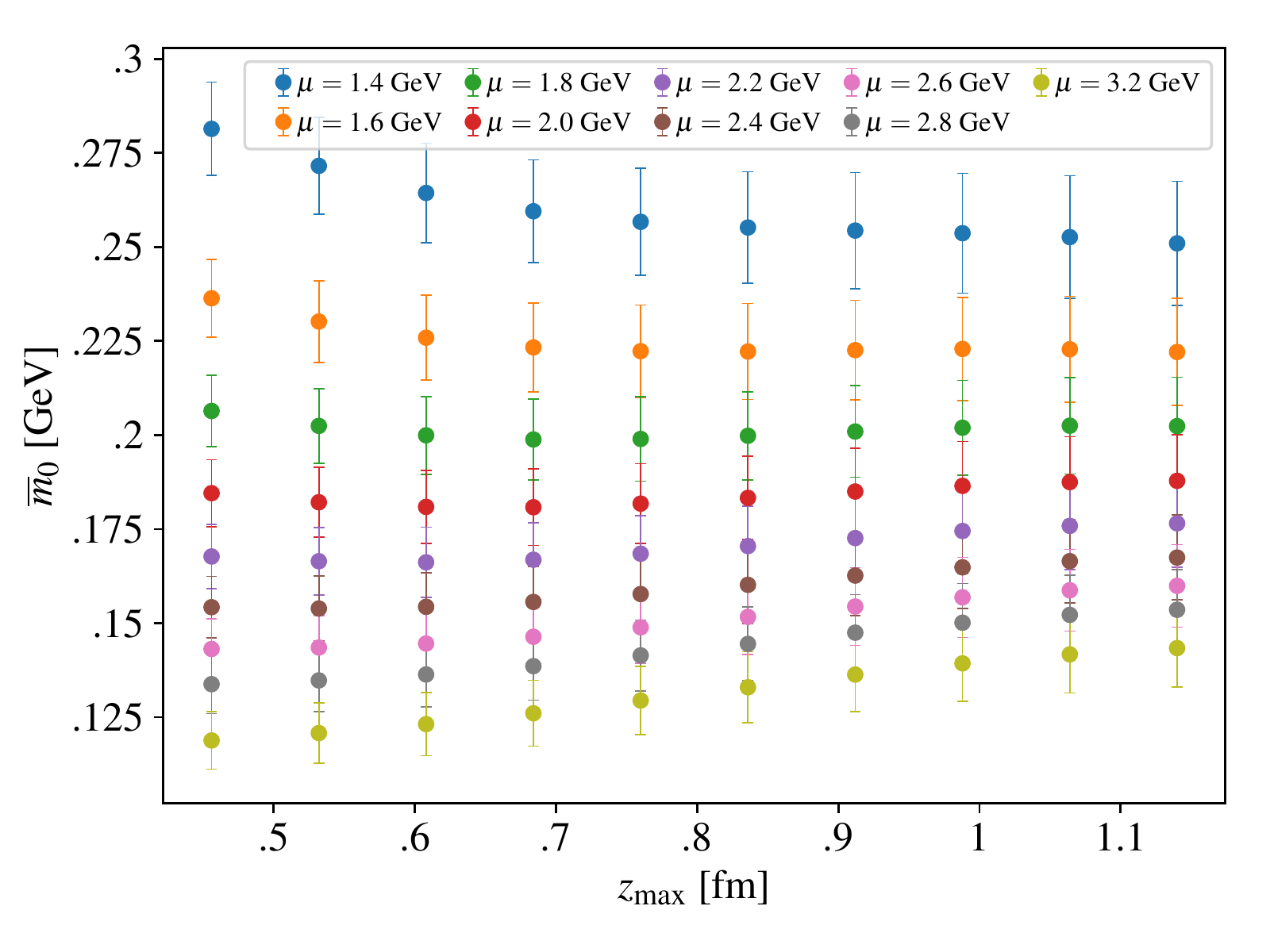}
    \includegraphics[width=\columnwidth]{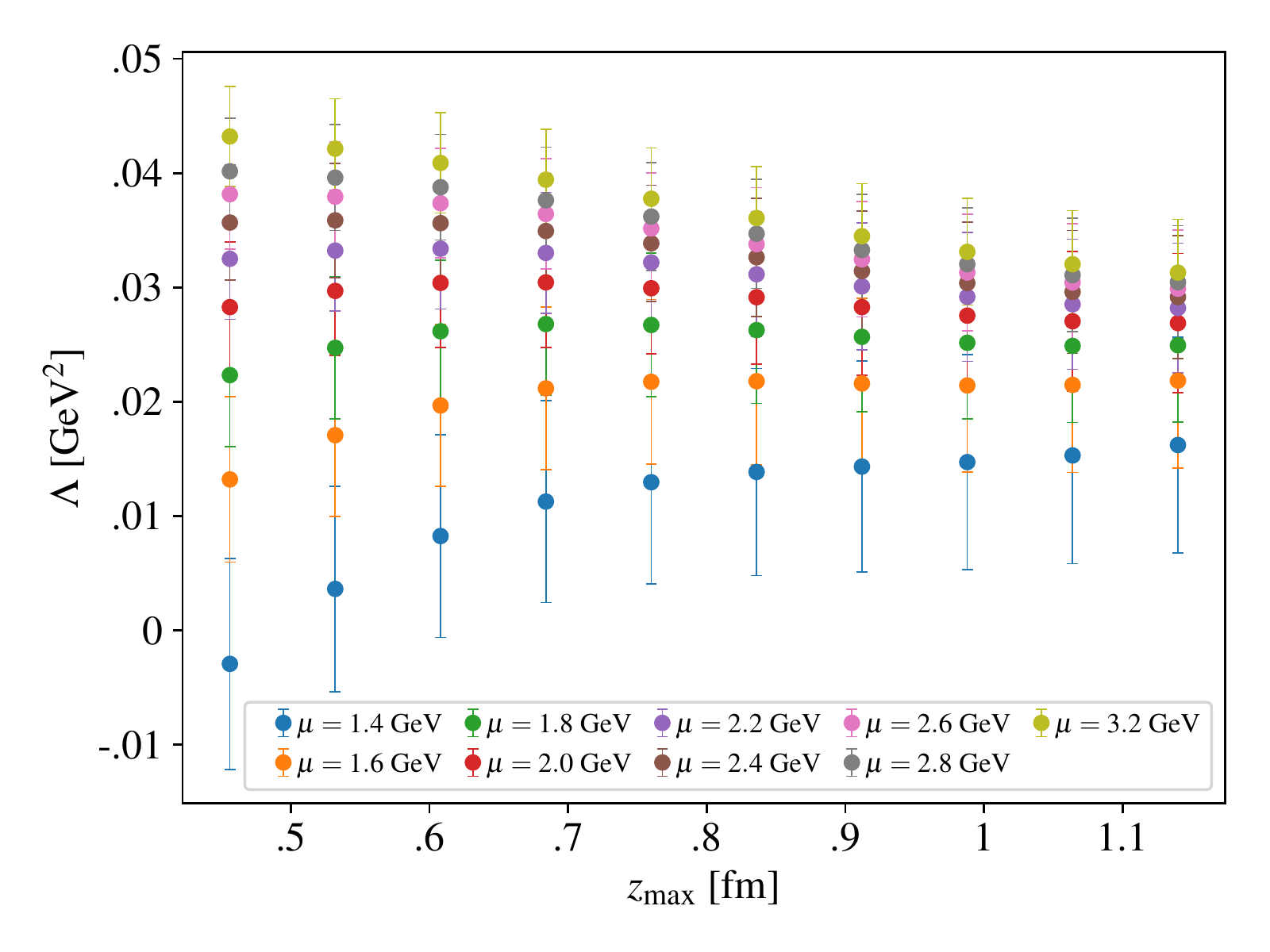}
    \caption{(Upper) $\mzero$ and (lower) $\Lambda$ values computed from fits using \Cref{eq:mass_sub_fit} at different energy scales $\mu$, $z_0 = 3a \sim 0.228$ fm, and up to different maximum $z$-values, $z_{\rm max}$.
    The Wilson coefficient $C_0(\mu^2 z^2)$ is computed at NNLO.}
    \label{fig:mass_sub_fits}
\end{figure}

In fixed order perturbation theory, the two parameters $(\mzero,\Lambda)$ depend on the renormalization scale $\mu$. As such, we must use a different set of fitting parameters for our calculation of the full PDF at different energy scales.
The two aforementioned parameters are independent of the external state.
We fit the parameters $\bar{m}_0$ and $\Lambda$ from both the nucleon and pion matrix elements, with the latter calculated in Ref.~\cite{Gao:2022iex}, and find any tension would lead to differences on the order of $3\%$ for the expected linear power corrections~\cite{Ji:2020brr} at the largest value of $P_z \sim 1.53$ GeV,
which is much less than the statistical uncertainty.
Therefore, we continue with the use of the parameters fitted from the proton matrix elements.
The $\mzero$ and $\Lambda$ values computed at different energy scales and up to different values of $z_{\rm max}$ are shown in \Cref{fig:mass_sub_fits}.

Finally, we can form the renormalized matrix elements in the hybrid scheme as
\begin{equation}
\label{eq:hybrid_renorm}
\begin{split}
    h^R_{\gamma_t} &(\lambda,\lambda_s,P_z,\mu) = \\
    \phantom{+} &N\frac{h^B_{\gamma_t}(z,P_z,a)}{h^B_{\gamma_t}(z,0,a)}\frac{C_0(\mu^2z^2)-\Lambda z^2}{C_0(\mu^2z^2)}\theta(z_s-z) \\
    + &Ne^{\delta m^\prime (z-z_s)}\frac{h^B_{\gamma_t}(z,P_z,a)}{h^B_{\gamma_t}(z_s,0,a)}\frac{C_0(\mu^2z_s^2)-\Lambda z_s^2}{C_0(\mu^2z_s^2)}\theta(z-z_s) ,
\end{split}
\end{equation}
where $N=h^B_{\gamma_t} (0,0,a)/h^B_{\gamma_t}(0,P_z,a)$ is a normalization, the correction $\Lambda z^2$ to $h^B_{\gamma_t}(z,0,a)$ is included for small $z$, we choose $z_s = 3a \sim 0.228$ fm, we have explicitly included $\delta m^\prime = \delta m + \bar{m}_0$ at large $z$, and the lattice spacing dependence for the renormalized matrix elements have been suppressed for convenience.
The form of the hybrid scheme comes from using the ratio scheme for $z<z_s$; for $z>z_s$, we use the matrix element with the linear divergence removed by the exponential involving $\delta m^\prime$. The additional factors on the side of $z>z_s$ are used to enforce continuity at $z=z_s$.

\subsubsection{Large-\texorpdfstring{$\lambda$}{lambda} extrapolation}
\label{subsec:largelambda}

In order to avoid unphysical oscillations in $x$-space, it is important that the Fourier transform not be truncated, which requires the matrix elements up to infinite $\lambda$. Of course, the lattice calculation can only produce values up to $\lambda_{\rm max} = P_z z_{\rm max}$.
Additionally, the signal can quickly deteriorate at large values of $\lambda$.
Therefore, we extrapolate to infinite $\lambda$ and perform the Fourier transform by a discrete sum over the data up to some maximum $\lambda_L = P_z z_L$ beyond which the extrapolation function is integrated to infinity. With sufficiently large $\lambda_{\rm max}$, the matrix elements falls close to zero at $\lambda_{\rm max}$, so the extrapolation will mainly affect the small $x$ region which is outside our prediction with LaMET.

We use the exponential decay model
\begin{equation}\label{eq:finalmodel}
    \frac{Ae^{-m_{\rm eff}\lambda/P_z}}{|\lambda|^d},
\end{equation}
where $(A, m_{\rm eff}, d)$ are fit parameters with the constraints $m_{\rm eff} > 0.1$ GeV, $A > 0$, and $d > 0$, as was done in Ref.~\cite{Gao:2021dbh}. The constraint $m_{\rm eff}>0.1$ GeV helps to ensure suppression at large values of $\lambda$, and does not noticeably affect the Fourier transform in the moderate-to-large $x$ region. This large-distance behavior can be derived under the auxiliary field formulation of the Wilson line~\cite{Green:2017xeu,Ji:2017oey}, where the nonlocal quark bilinear operator can be expressed as the product of two heavy-to-light currents. At large spacelike separation, this current product drops off exponentially with the decay constant given by the binding energy of the heavy-light meson.
From this behavior, the extrapolation model of Eq. (\ref{eq:finalmodel}) is derived. A more detailed derivation of the large-$\lambda$ model is provided in App. B of \cite{Gao:2021dbh}.

The real and imaginary parts of $h^R_{\gamma_t}(\lambda,\lambda_s,P_z,\mu)$ are fitted separately to \Cref{eq:finalmodel}.
Care must be taken in choosing the data points used in the fit to the extrapolation model. Data points at too large a value of $\lambda$ will have a poor signal-to-noise ratio and give large uncertainties in the extrapolation parameters. Data points at too small a value of $\lambda$ will not capture the exponential decay expected at finite momentum.
Our general guide is to select points for which $h^R_{\gamma_t}(\lambda,\lambda_s,P_z,\mu)$ becomes compatible with zero (if such points exist) or just before the point in which the matrix element begins to grow, contrary to the exponential decay expected from \Cref{eq:finalmodel}.

\begin{figure*}
    \centering
    \includegraphics[width=0.32\textwidth]{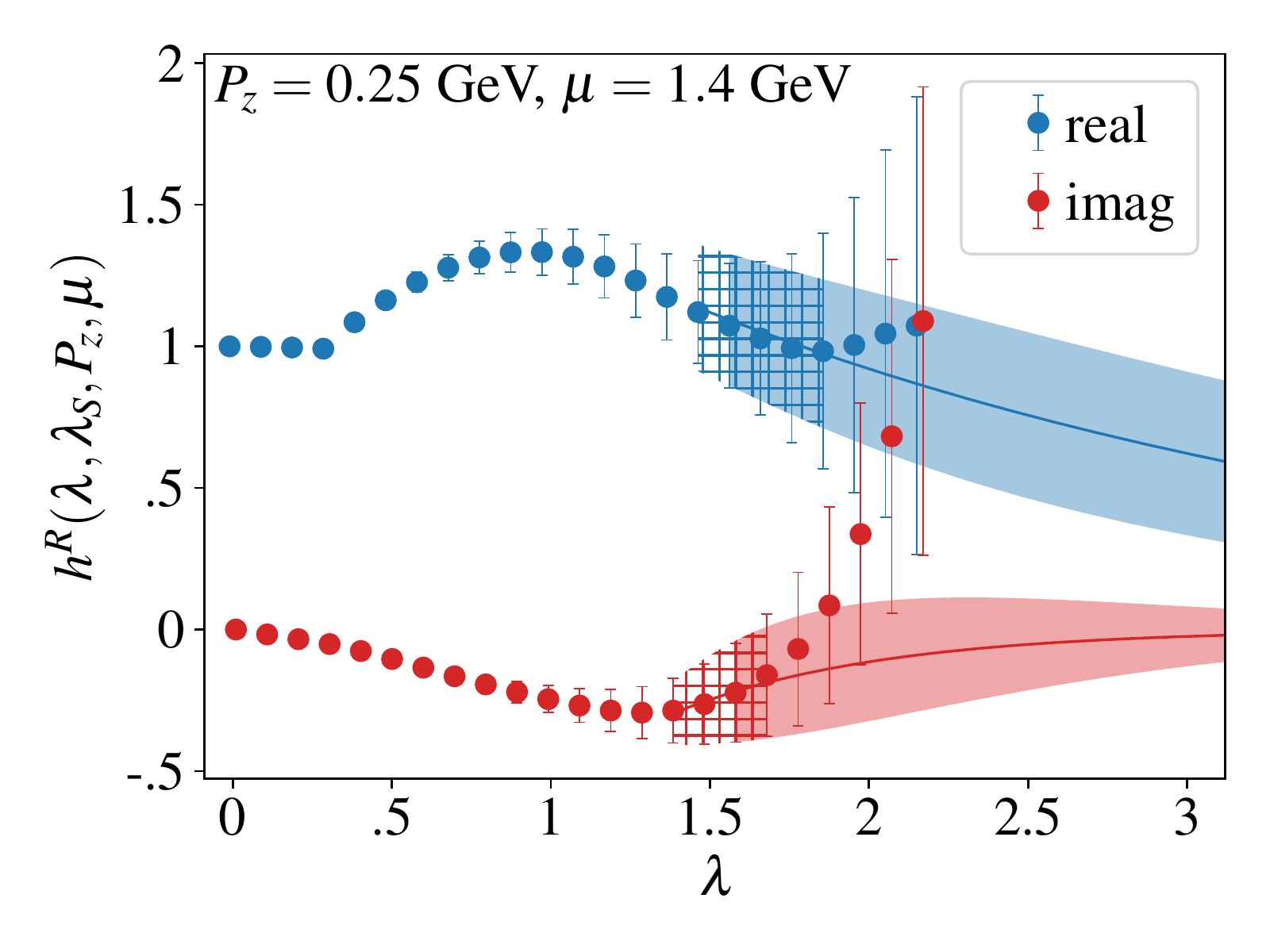}
    \includegraphics[width=0.32\textwidth]{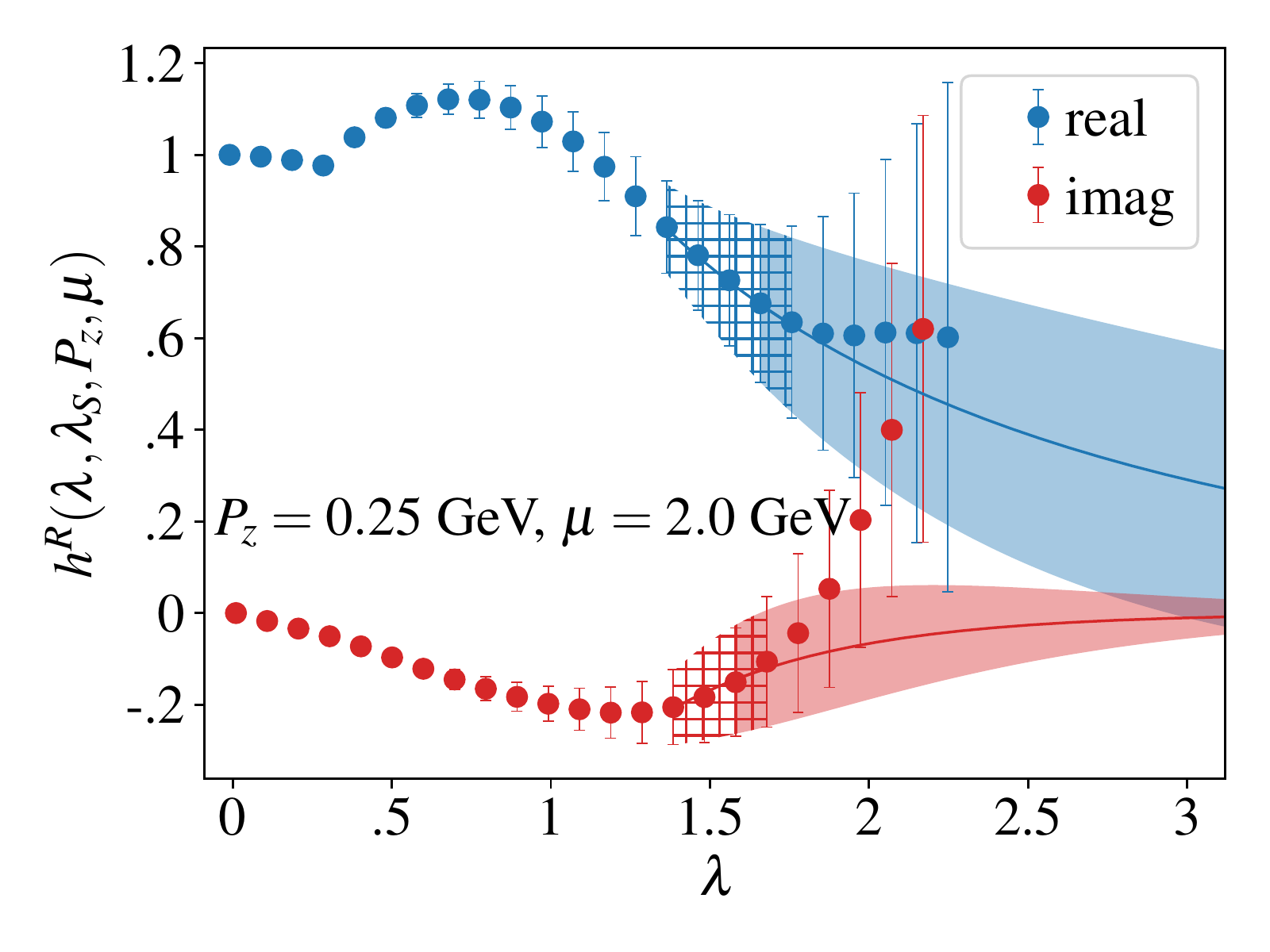}
    \includegraphics[width=0.32\textwidth]{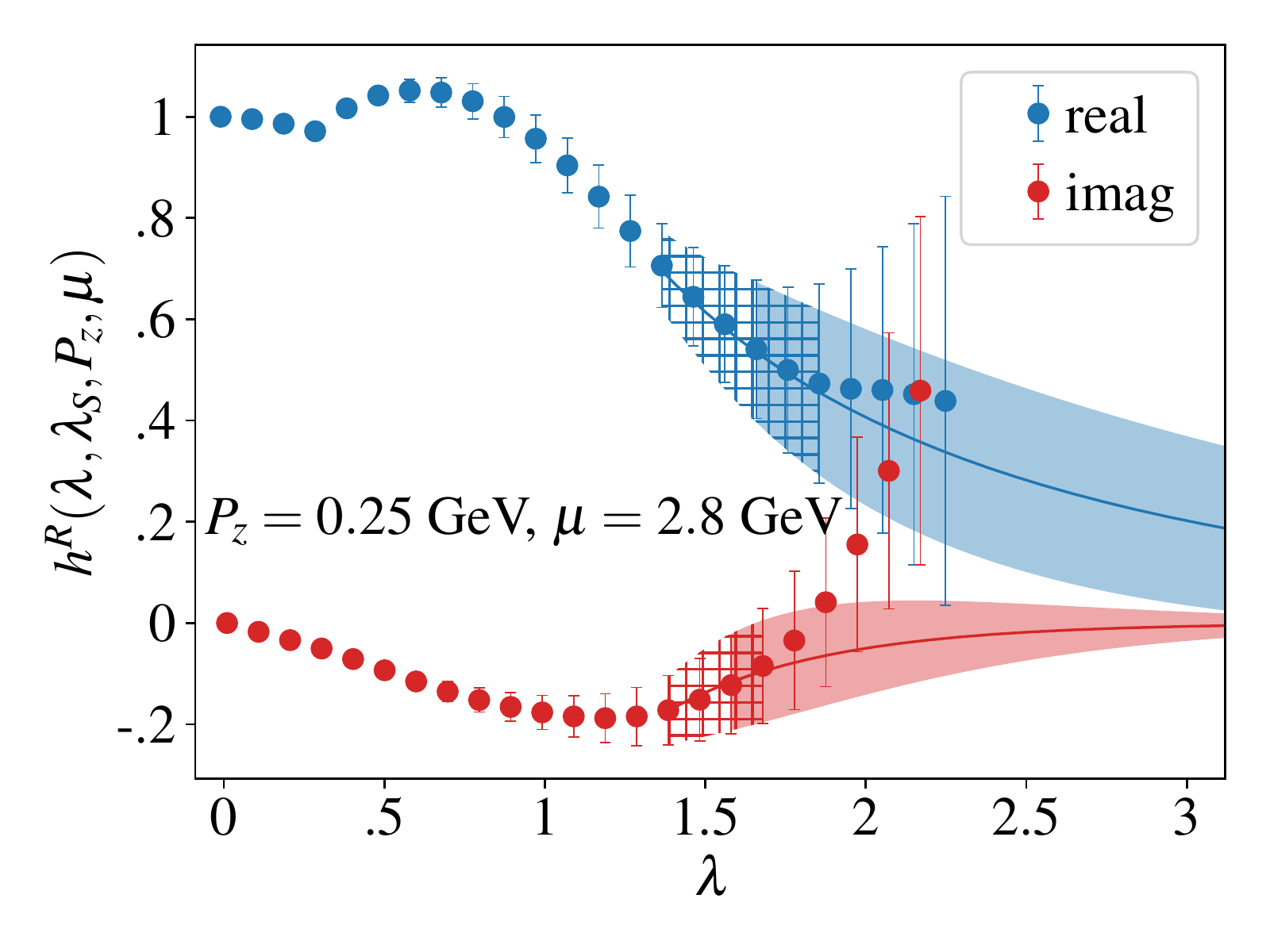}
    \includegraphics[width=0.32\textwidth]{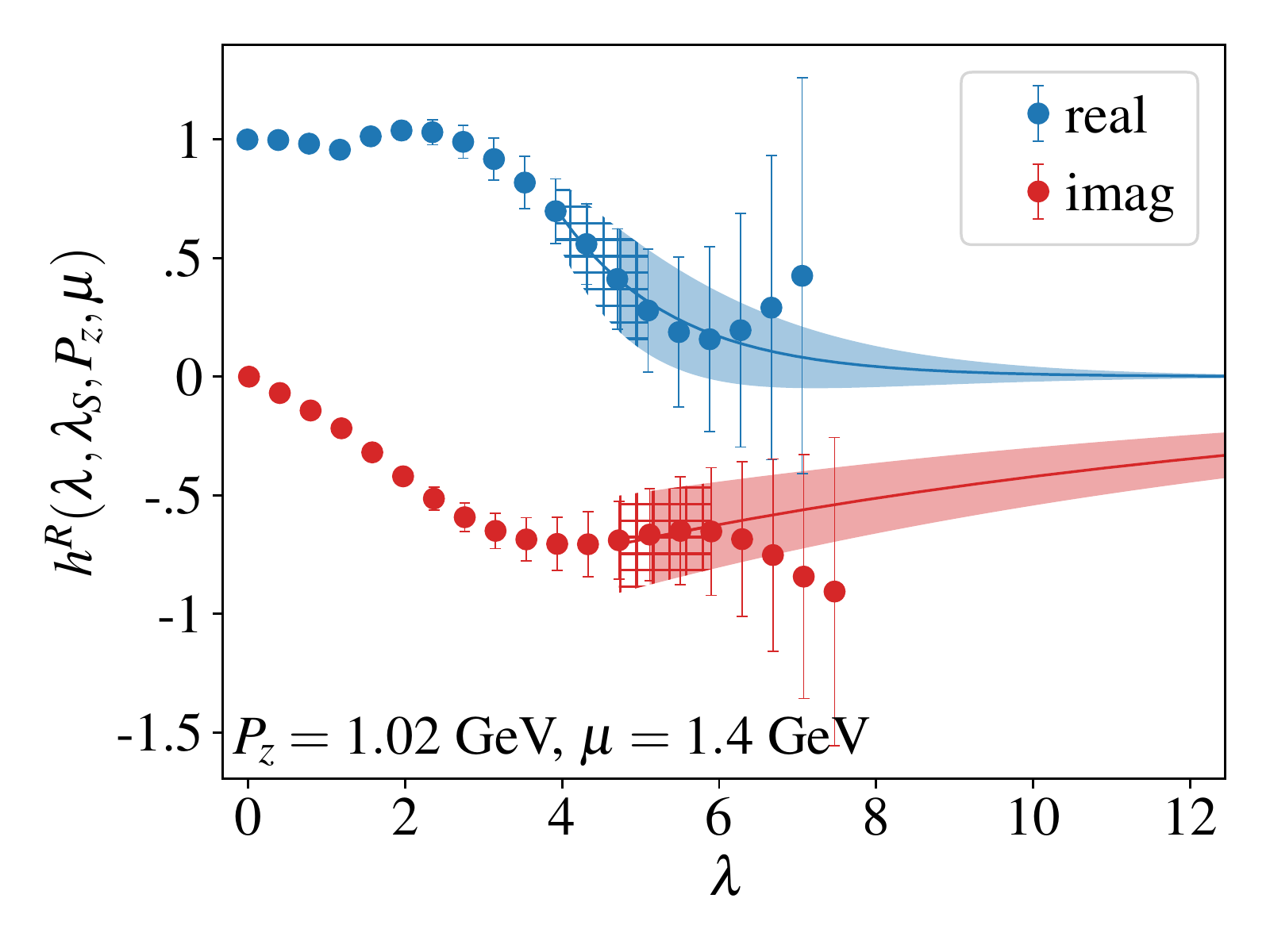}
    \includegraphics[width=0.32\textwidth]{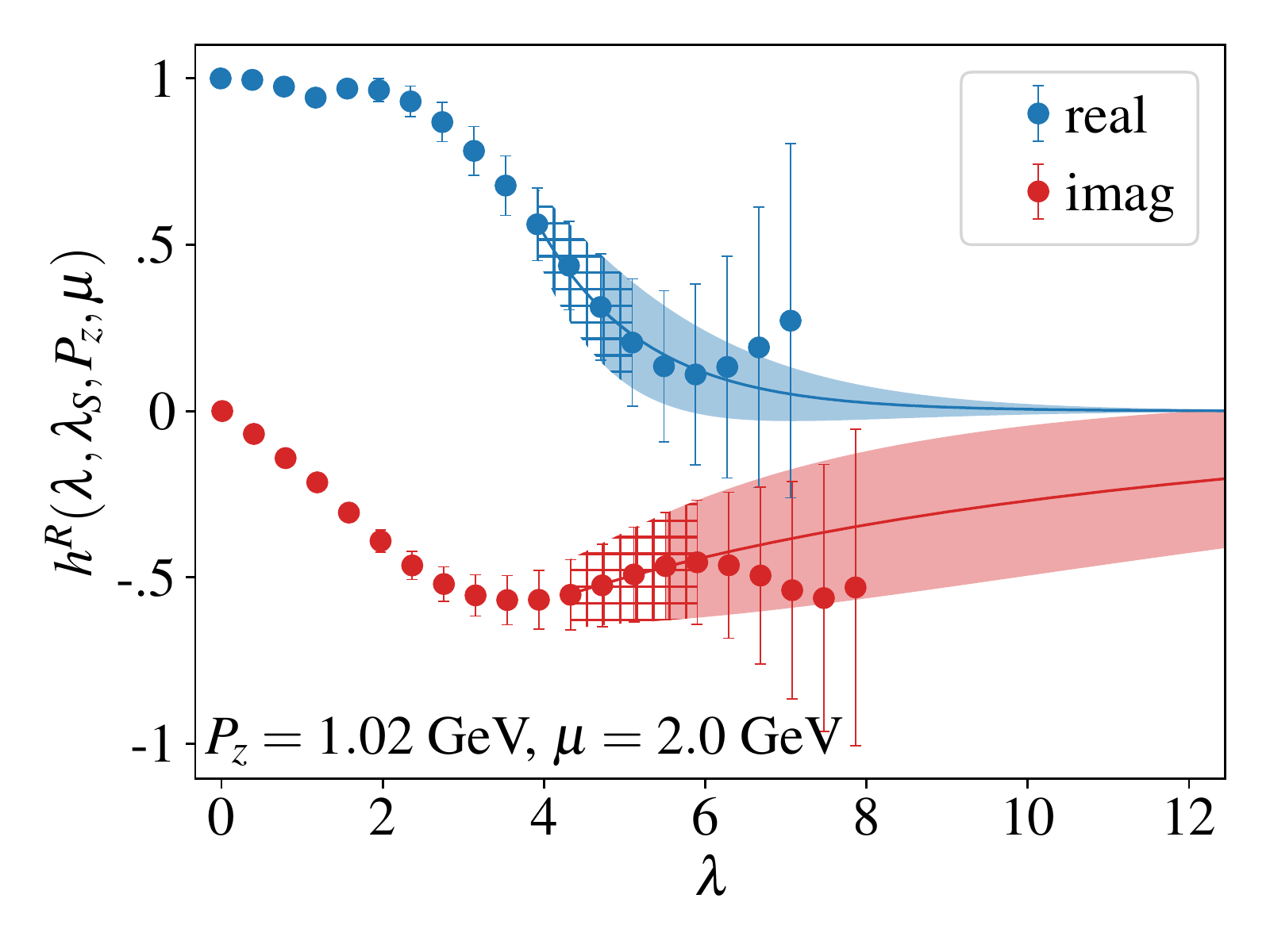}
    \includegraphics[width=0.32\textwidth]{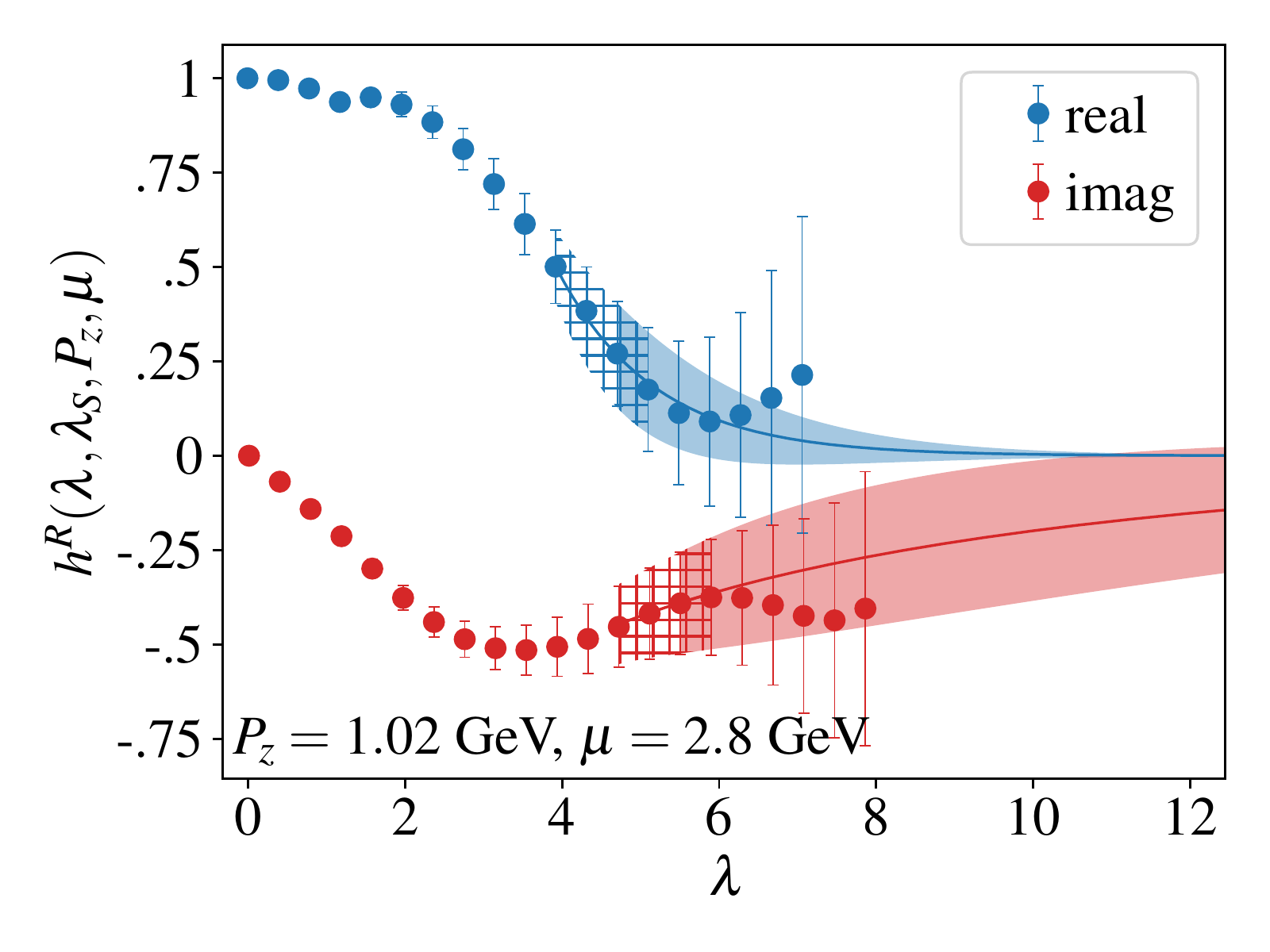}
    \includegraphics[width=0.32\textwidth]{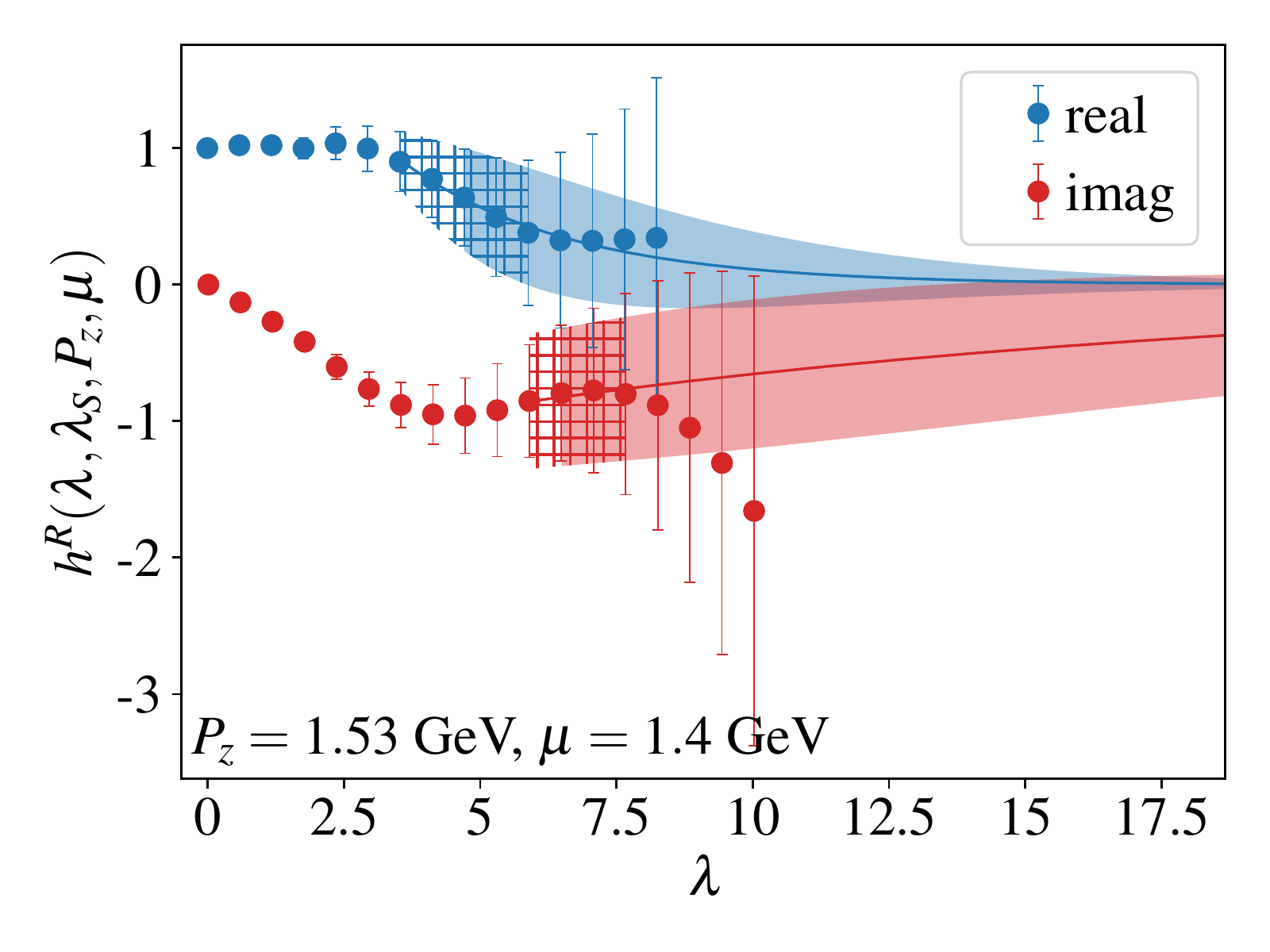}
    \includegraphics[width=0.32\textwidth]{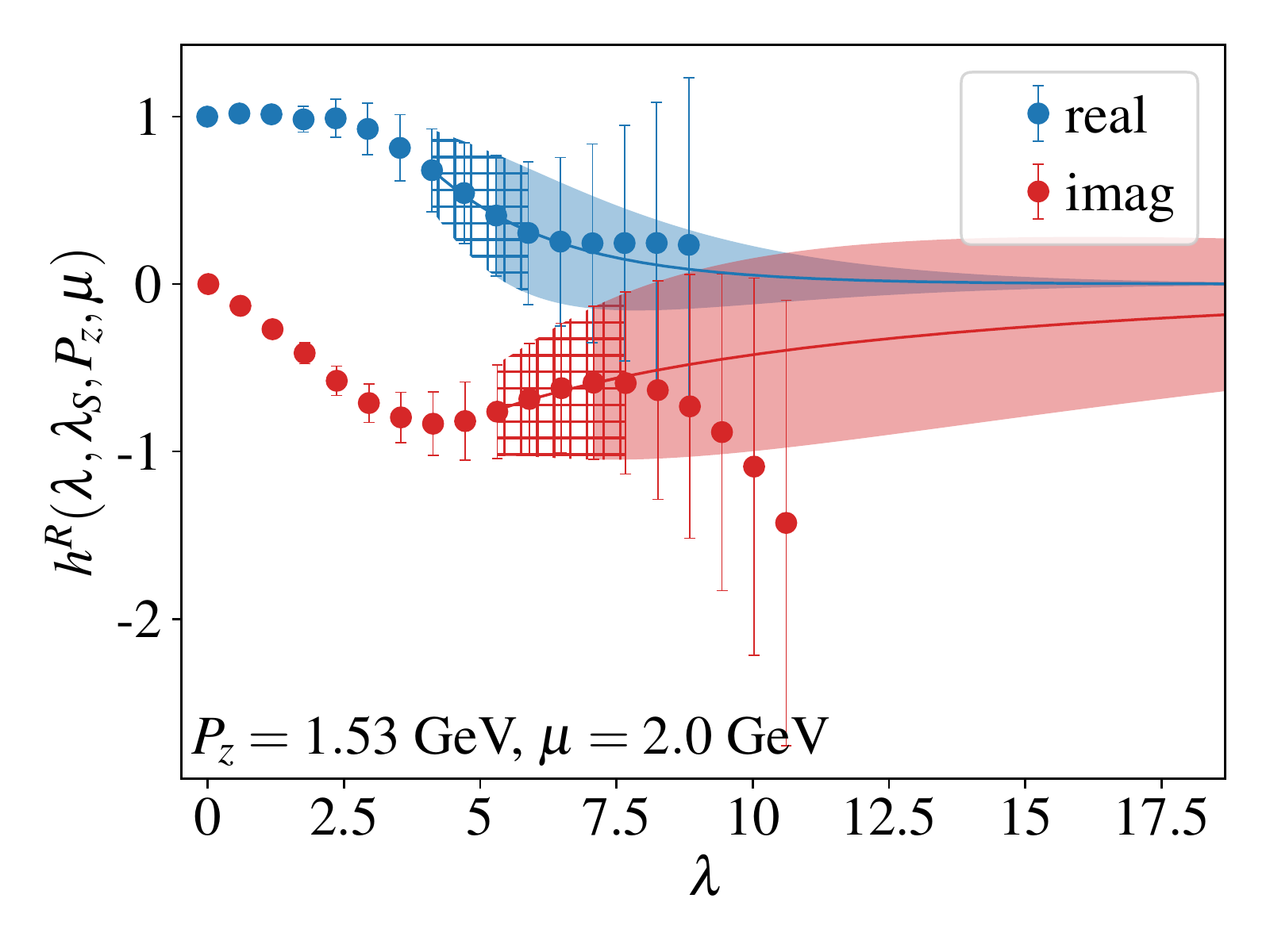}
    \includegraphics[width=0.32\textwidth]{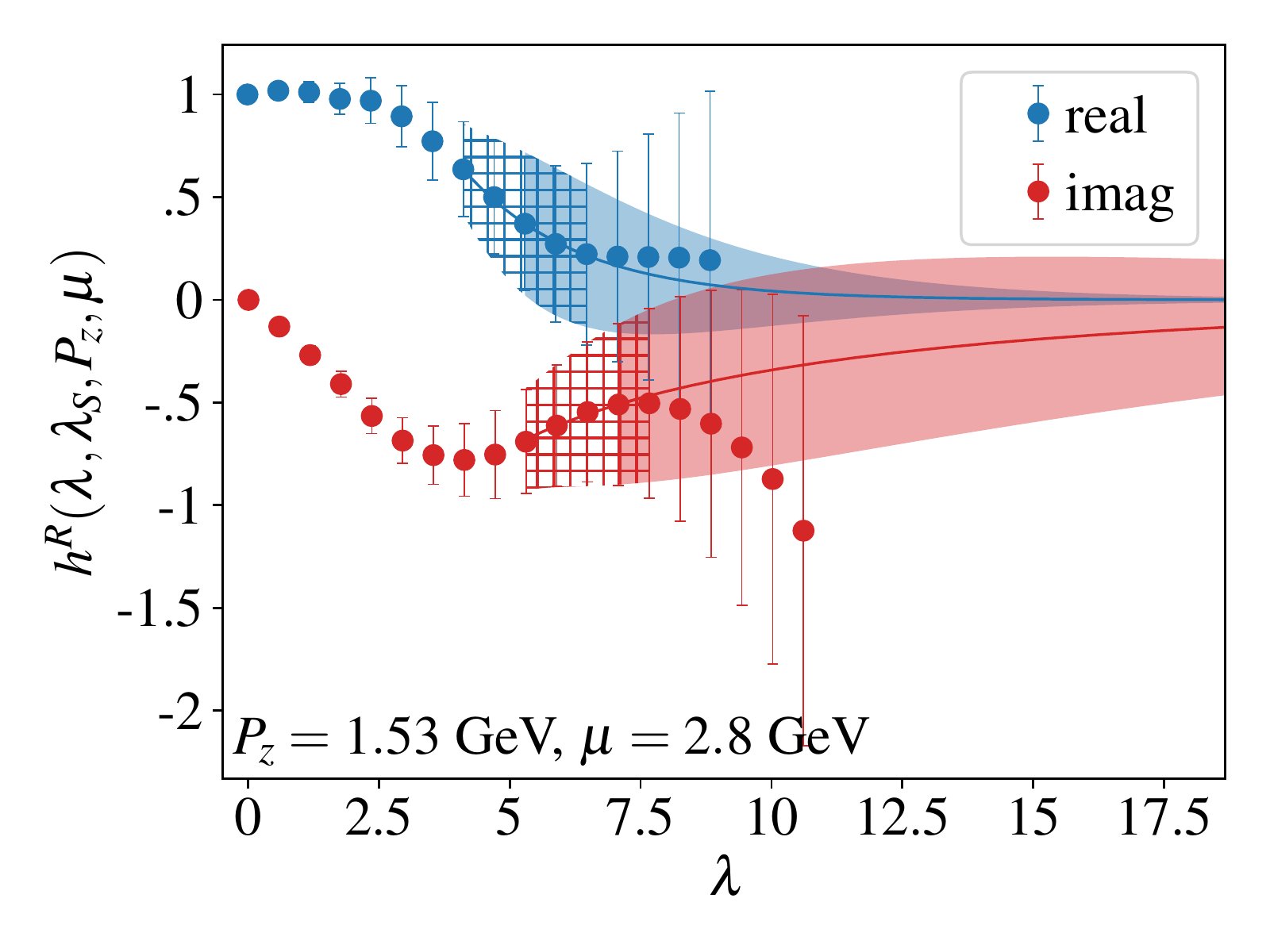}
    \caption{The real (blue) and imaginary (red) parts of the hybrid renormalized matrix elements along with their large-$\lambda$ extrapolations using \Cref{eq:finalmodel}.
    The three rows from top to bottom correspond to momenta $P_z=0.26$, $1.02$, and $1.53$ GeV; the three columns from left
    to right correspond to energy scales $\mu=1.4$, $2.0$, and $2.8$ GeV. The data points themselves come from the lattice
    calculation, while the bands correspond to the extrapolation fit result in the range where we use the extrapolation rather
    than the data. The hatches show the range of $\lambda$ used for the extrapolation fit.}
    \label{fig:largeLambda}
\end{figure*}

Results of these fits for various $P_z$ and $\mu$ are shown in \Cref{fig:largeLambda}, where the hatches indicate the range of $\lambda$ used for the fit while the shaded bands start from $\lambda_L$. The value of $\lambda$ corresponding to a data point that most closely resembles the extrapolation at that value of $\lambda$ becomes the chosen value of $\lambda_L$.
Note that our criteria for choosing $\lambda_L$ can lead to different values chosen for the real and imaginary parts.
The resulting choices for $z_L$ are shown in \Cref{tab:zL}.

\begin{table}
\centering
\begin{tabular}{c|c|c|c}
\hline
\hline
$n_z$ & $\mu$ (GeV) & $z_L^{\rm re}/a$ & $z_L^{\rm im}/a$ \\
\hline
1 & 1.4 & 16 & 16 \\
  & 2.0 & 17 & 16 \\
  & 2.8 & 17 & 16 \\
\hline
4 & 1.4 & 12 & 13 \\
  & 2.0 & 12 & 14 \\
  & 2.8 & 12 & 14 \\
\hline
6 & 1.4 & 8  & 11 \\
  & 2.0 & 9  & 12 \\
  & 2.8 & 9  & 12 \\
\hline
\hline
\end{tabular}
\caption{The values of $z_L$ chosen for the real ($z_L^{\rm re}$) and
         imaginary ($z_L^{\rm im}$) parts of the hybrid renormalized
         matrix elements for each $P_z = \frac{2 \pi}{L}n_z$ and $\mu$
         considered in this work.}
\label{tab:zL}
\end{table}

\subsubsection{Fourier transform}
\label{subsec:FT}

Next, we obtain the quasi-PDF from the Fourier transform of the renormalized matrix elements
\begin{equation}
    \label{eq:qPDF}
    \tilde{q}(y,z_s,P_z,\mu) = \int \frac{dz P_z}{2 \pi} e^{iy P_z z} h^R_{\gamma_t} (z, z_s, P_z, \mu) .
\end{equation}
To perform this integral, we exploit the symmetry property $h^{R}_{\gamma_t}(-z,z_s,P_z,\mu)=h^{R}_{\gamma_t}(z,z_s,P_z,\mu)^*$. Expanding the integrand of \Cref{eq:qPDF}, we find that
\begin{equation}
\begin{split}
    \Re e^{iyP_z z}&h^{R}_{\gamma_t}(z,z_s,P_z,\mu)=\cos(yP_z z)\Re h^{R}_{\gamma_t}(z,z_s,P_z,\mu)\\
    -&\sin{(yP_z z)}\Im h^{R}_{\gamma_t}(z,z_s,P_z,\mu),\\
    \Im e^{iyP_z z}&h^{R}_{\gamma_t}(z,z_s,P_z,\mu)=\sin(yP_z z)\Re h^{R}_{\gamma_t}(z,z_s,P_z,\mu)\\
    +&\cos{(yP_z z)}\Im h^{R}_{\gamma_t}(z,z_s,P_z,\mu).
\end{split}
\end{equation}
Thus, the real and imaginary parts of the integrand in \Cref{eq:qPDF} are symmetric and antisymmetric, respectively. Therefore, the imaginary part vanishes under integration. After simplifying the integral, we divide it into a part that sums over the data up to some maximum $z_L$ and a part that performs the integral analytically from $z_L$ to infinity using the extrapolated fit function
\begin{equation}
\label{eq:qPDF_split}
\begin{split}
    \tilde{q}&(y,z_s,P_z,\mu) = \\
     &\Bigg[\sum_{z=0}^{z_L^{\rm re}/a} \frac{z_L^{\rm re} P_z}{\pi N_{z_L}^{\rm re}} + \int_{z_L^{\rm re}}^{\infty} \frac{dz P_z}{\pi} \Bigg] \Re h^R_{\gamma_t}(z,z_s,P_z,\mu) \cos (z P_z y) \\
    - &\Bigg[\sum_{z=0}^{z_L^{\rm im}/a} \frac{z_L^{\rm im} P_z}{\pi N_{z_L}^{\rm im}} + \int_{z_L^{\rm im}}^{\infty} \frac{dz P_z}{\pi} \Bigg] \Im h^R_{\gamma_t} (z,z_s,P_z,\mu) \sin (z P_z y),   
\end{split}
\end{equation}
where $z_L^{\rm re}$ and $z_L^{\rm im}$ correspond to the values of $z_L$ for the real and imaginary parts, respectively, and $N^{\rm re/im} \equiv z_L^{\rm re/im}/a + 1$.

\subsubsection{Light-cone matching}
\label{subsec:Matching}
The final step of the analysis is to match our momentum space quasi-PDF to the light cone. The light-cone PDF $q(x,\mu)$ is related to the quasi-PDF $\tilde{q}(y,z_s,P_z,\mu)$ via
\begin{equation}
\label{eq:matching}
\begin{split}
    q(x,\mu)&=\int^{\infty}_{-\infty}\frac{d{y}}{|y|}\mathcal{C}^{-1}\left(\frac{x}{y},\frac{\mu}{yP_z},|y|\lambda_s\right)\tilde{q}(y,z_s,P_z,\mu)\\
    &\phantom{==} +\mathcal{O}\left(\frac{\qcdcutoff^2 }{x^2P_z^2},\frac{\qcdcutoff^2}{(1-x)^2P_z^2}\right)\\
    &\equiv\mathcal{C}^{-1}\left(\frac{x}{y},\frac{\mu}{yP_z},|y|\lambda_s\right)\otimes\tilde{q}(y,z_s,P_z,\mu)\\
    &\phantom{==} +\mathcal{O}\left(\frac{\qcdcutoff^2 }{x^2P_z^2},\frac{\qcdcutoff^2}{(1-x)^2P_z^2}\right),
\end{split}
\end{equation}
where $\mathcal{C}^{-1}\left(\frac{x}{y},\frac{\mu}{yP_z},|y|\lambda_s\right)$ is the inverse matching kernel and we use the notation ``$\otimes$" as a short-hand for the integral. We can write the full matching kernel as a series expansion in the strong coupling $\alpha_s$
\begin{equation}\label{eq:full_matching}
    \begin{split}
        \mathcal{C}\left(\frac{x}{y},\frac{\mu}{yP_z},|y|\lambda_s\right)=\delta & \left(\frac{x}{y}-1\right)\\
        &+\sum_{n=1}^{\infty}\alpha_s^n\mathcal{C}^{(n)}\left(\frac{x}{y},\frac{\mu}{yP_z},|y|\lambda_s\right)
    \end{split}
\end{equation}
with its inverse being defined as
\begin{equation}\label{eq:inverse_matching}
\mathcal{C}^{-1}\left(\frac{x}{z},\frac{\mu}{zP_z},|z|\lambda_s\right)\otimes\mathcal{C}\left(\frac{z}{y},\frac{\mu}{yP_z},|y|\lambda_s\right)=\delta\left(\frac{x}{y}-1\right) ,
\end{equation}
where $\delta\left(\frac{x}{y}-1\right)$ is the Dirac delta function. We can obtain a series solution for $\mathcal{C}^{-1}$ by combining \Cref{eq:full_matching,eq:inverse_matching}
\begin{equation}
    \begin{split}
        \mathcal{C}^{-1} & \left(\frac{x}{y},\frac{\mu}{yP_z},|y|\lambda_s\right)=\delta\left(\frac{x}{y}-1\right)\\
        &-\alpha_s\mathcal{C}^{(1)}\left(\frac{x}{y},\frac{\mu}{yP_z},|y|\lambda_s\right)\\
        &+\alpha_s^2\mathcal{C}^{(1)}\left(\frac{x}{z},\frac{\mu}{zP_z},|z|\lambda_s\right)\otimes\mathcal{C}^{(1)}\left(\frac{z}{y},\frac{\mu}{yP_z},|y|\lambda_s\right)\\
        &-\alpha_s^2\mathcal{C}^{(2)}\left(\frac{x}{y},\frac{\mu}{yP_z},|y|\lambda_s\right)+\mathcal{O}(\alpha_s^3).
    \end{split}
\end{equation}

As demonstrated in Appendix C.1 of \cite{Gao:2021dbh}, the replacement of the full integral in \Cref{eq:matching} by matrix multiplication reduces the computational cost of light-cone matching with negligible loss of accuracy. We thus form a matching matrix $C_{xy}$ with the $x$ and $y$ indices corresponding to those in \Cref{eq:full_matching}.

Treating the quasi-PDF as the LO approximation in LaMET, we construct two matrices, $C^{\rm NLO}_{xy}$ and $C^{\rm NNLO}_{xy}$, to achieve NLO and NNLO results, respectively, for the light-cone PDF
\begin{eqnarray}
    q^{\rm LO}_x&=&\qtilde_x,\\
    q^{\rm NLO}_x&=&\qtilde_x-\delta y\sum_{y}C^{\rm NLO}_{xy}\qtilde_y,\\
    q^{\rm NNLO}_x&=&\qtilde_x-\delta y\sum_{y}C^{\rm NLO}_{xy}\qtilde_y-\delta y\sum_{y}C^{\rm NNLO}_{xy}\qtilde_y,
\end{eqnarray}
where $\delta y=0.001$, corresponding to a discretization of the integral in \Cref{eq:matching}.

The corrections to the matching in \Cref{eq:matching} mean that our LaMET calculation breaks down as $x\to 0$ and $x\to 1$. Our range of validity is explained in \Cref{sec:validx}.

\subsection{Results}

\begin{figure}
    \centering
    \includegraphics[width=\columnwidth]{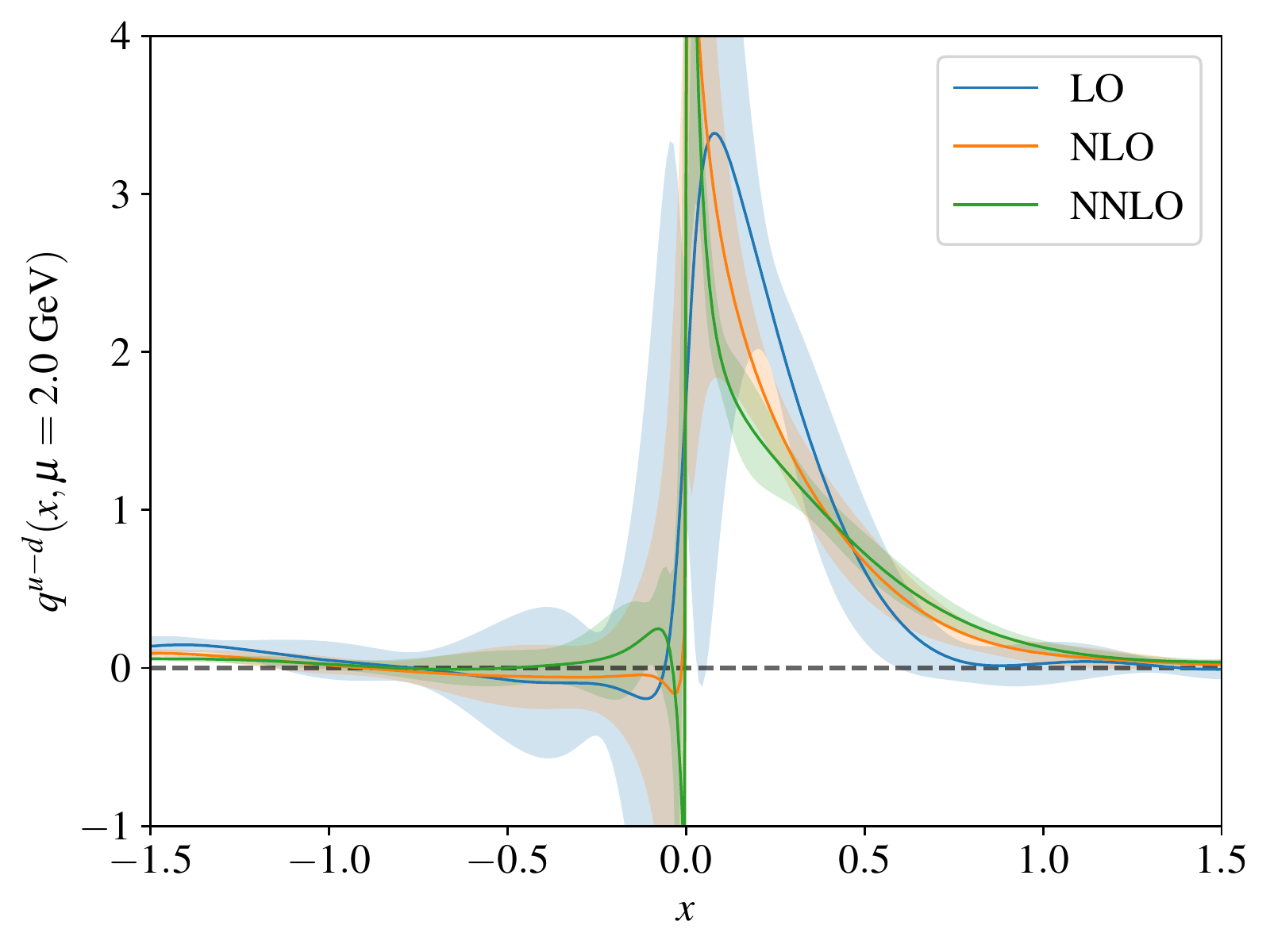}
    \caption{Dependence of the $x$-space matched isovector-quark PDF on the perturbative order used in the matching kernel.
    The results shown use the largest value of momentum computed in this work (i.e. $P_z = 1.53$ GeV).}
    \label{fig:pdf_order_dep}
\end{figure}

\begin{figure}
    \centering
    \includegraphics[width=\columnwidth]{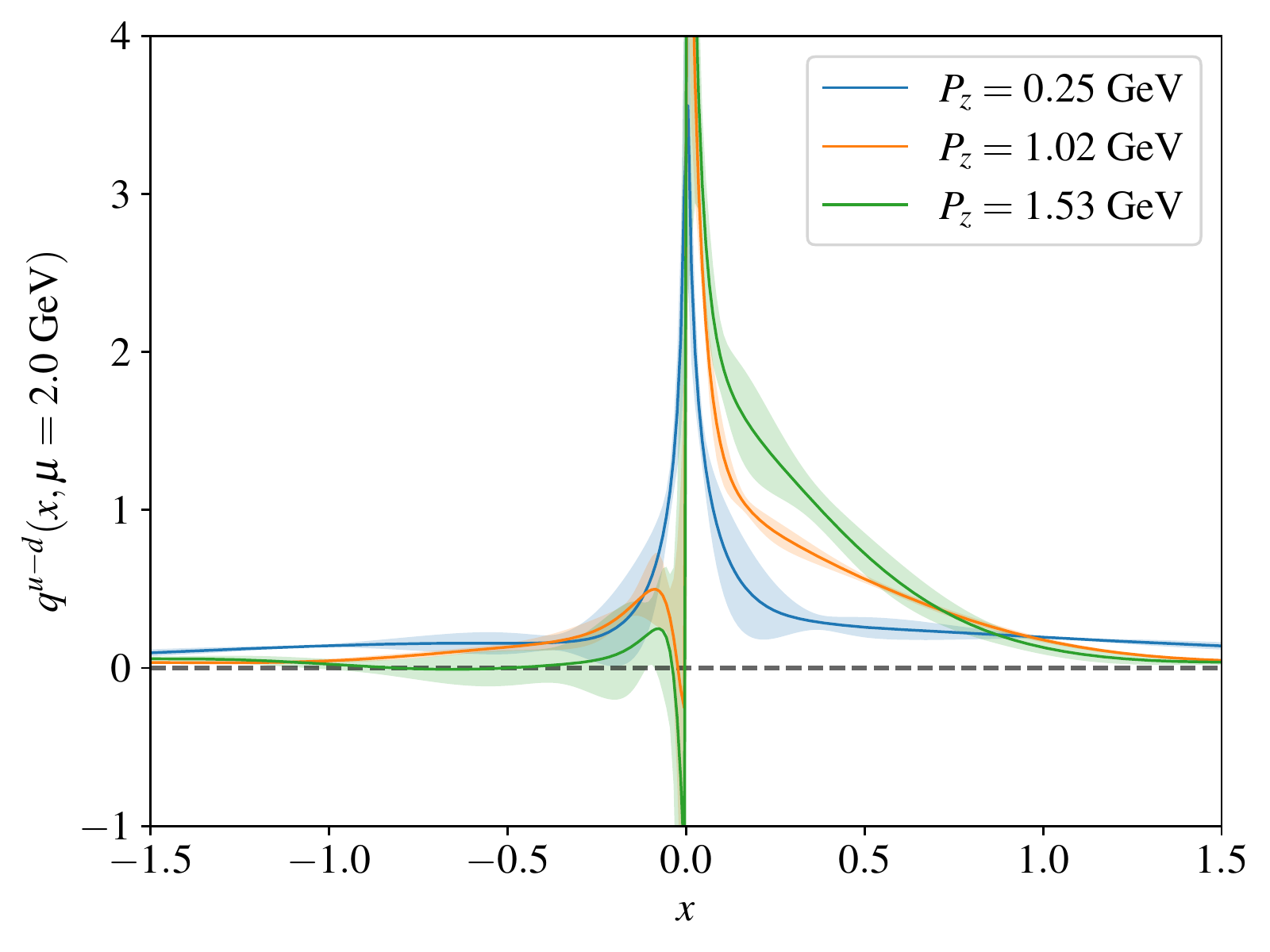}
    \caption{Dependence of the $x$-space matched isovector-quark PDF on the momentum of the nucleon.
    The results shown use the matching kernel computed at NNLO.}
    \label{fig:pdf_pz_dep}
\end{figure}

As a first attempt to understand some of the systematics involved with the LaMET approach,
we start by studying the perturbative matching order dependence with our largest value of $P_z = 1.53$ GeV,
which is shown in \Cref{fig:pdf_order_dep}.
There does appear to be convergence going from NLO to NNLO, at least in the middle-$x$ region where LaMET is expected to hold.
Additionally, we also show the dependence on momentum using the NNLO matching in \Cref{fig:pdf_pz_dep}.
There is a significant dependence on the momentum for most values of $x$, which indicates that the momenta are still not large enough to sufficiently suppress the power corrections. Therefore, we present the PDF obtained at $P_z=1.53$ GeV as our final result.

\subsubsection{Valid range of \texorpdfstring{$x$}{x}}\label{sec:validx}

The LaMET approach is only valid in the middle ranges of $|x| \in [x_{\rm min}, x_{\rm max}]$, which arises from the power corrections in \Cref{eq:matching}. The power counting in \Cref{eq:matching} is based on the argument that the active and spectator partons must carry hard momentum, and that power corrections in the $\overline{\rm MS}$ scheme usually begin at quadratic order. In QCD, they are closely related to the renormalon ambiguities in the leading-twist coefficient functions~\cite{Braun:2018brg,Liu:2020rqi,Ji:2022xxx}. In Ref.~\cite{Braun:2018brg}, the power corrections in the quasi-PDF were estimated by a renormalon analysis, and it was concluded that they behave as ${\cal O}(\Lambda_{\rm QCD}^2/(x^2(1-x)P_z^2))$, whereas an independent analysis led to ${\cal O}(\Lambda_{\rm QCD}^2/(x^2P_z^2))$~\cite{Liu:2020rqi}. On the other hand, the large infared logarithms in the end-point regions, i.e., the DGLAP logarithms $\ln \frac{\mu^2}{4 x^2 P_z^2}$ at small $x$ and threshold logarithms $\ln \frac{\mu^2}{4 (1-x) P_z^2}$ at large $x$~\cite{Gao:2021hxl,Su:2022fiu}, also indicate that nonperturbative effects become important when $2xP^z$ and $2\sqrt{1-x}P^z$ become close to $\Lambda_{\rm QCD}$.

As our statistical precision is not adequate to properly assess all systematics rigorously, we determine the values of $x_{\rm min}$ and $x_{\rm max}$ by requiring that $x_{\rm min}P^z\sim \Lambda_{\rm QCD}$ and $\sqrt{1-x_{\rm max}}P^z \sim \Lambda_{\rm QCD}$.
In our matching coefficient, the strong coupling $\alpha_s$ is defined in the $\overline{\rm MS}$ scheme with $\Lambda_{\rm QCD}^{\overline{\rm MS}}=332$ MeV~\cite{Petreczky:2020tky}.
Therefore, using our largest value of $P_z \approx 1.53$ GeV, we find $x_{\rm min} \approx 0.217$ and $x_{\rm max} \approx 0.953$.
\begin{figure}
    \centering
    \includegraphics[width=\columnwidth]{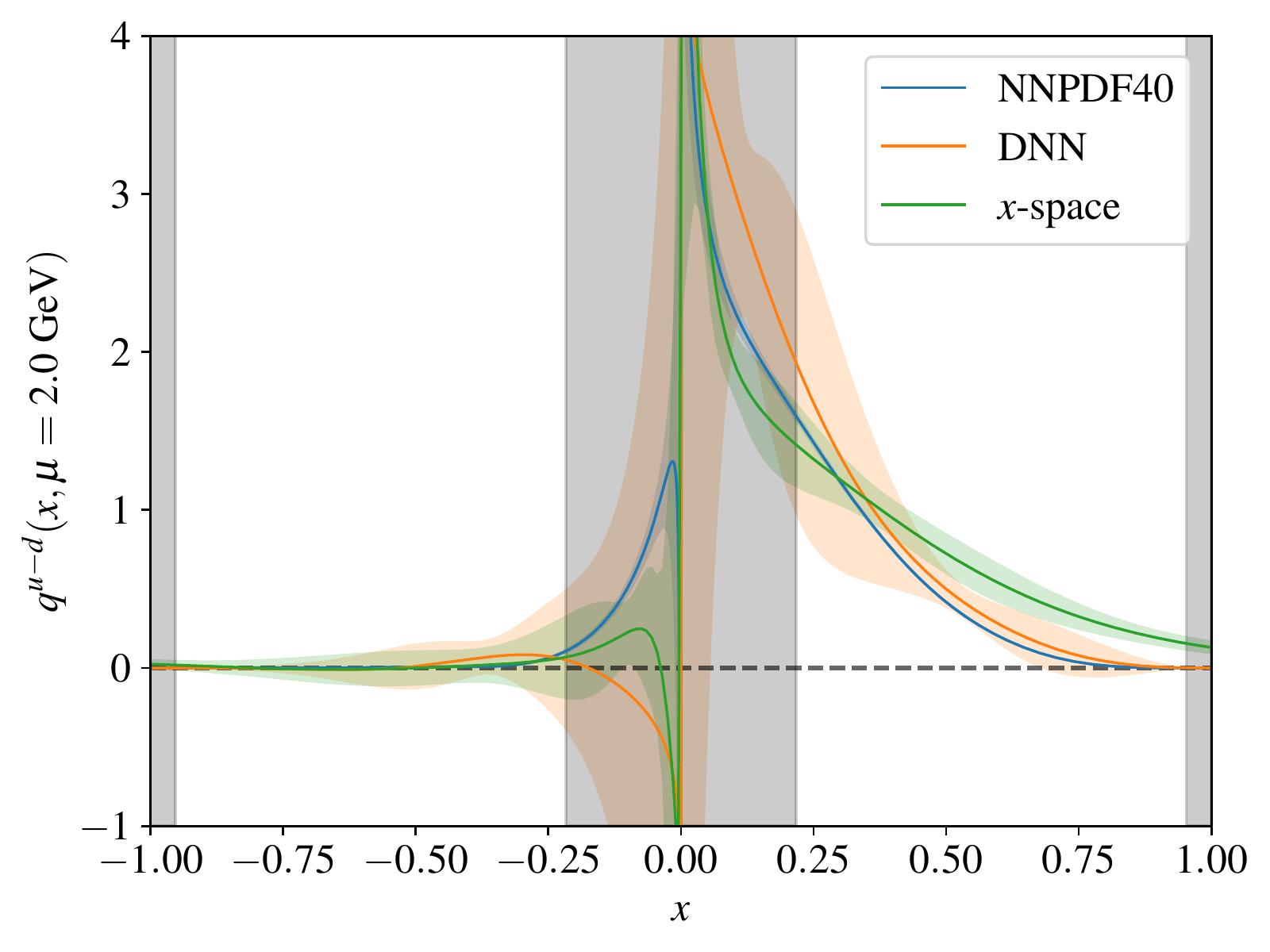}
    \caption{Comparison of the $x$-dependence of the isovector-quark PDF from the global analysis of NNPDF4.0, the DNN, and $x$-space matching.
    The gray bands correspond to the regions of $x$ where we do not rigorously trust the results of LaMET.}
    \label{fig:pdf_comparison}
\end{figure}

In \Cref{fig:pdf_comparison}, we show a comparison of our final estimate at $\mu = 2.0$ GeV from $x$-space matching with $P_z = 1.53$ GeV, the DNN with $z_{\rm max} = 0.76$ fm and $\delta=0.2$, and the global analysis from NNPDF4.0~\cite{NNPDF:2021njg}.
Our $x$-space matching results in the region $0.5 < x < 0.953$ still have noticeable differences from the NNPDF4.0 analysis, 
as we also observe in \Cref{fig:pdf_order_dep} that the NLO and NNLO corrections are significant compared to the LO result.
This indicates that the systematic uncertainties, most likely the unresummed large logarithms at $x\sim1$~\cite{Gao:2021hxl}, the $P^z$-dependent power corrections, and the lattice discretization effects are not well under control. Achieving higher momentum and taking the continuum limit will be necessary to improve the calculation.

As one final check on the systematics, we would like to understand any relationship between the discrepancy observed for $\braket{x}$ and that observed for the PDF.
Therefore, we use the moments determined from the NNPDF4.0 global analysis to reconstruct the reduced pseudo-ITD using the ratio of OPEs in \Cref{eq:fitrpITD}.
Next, when forming the hybrid-renormalized matrix elements in \Cref{eq:hybrid_renorm}, we use this reconstructed reduced pseudo-ITD for $z \leq z_S$, rather than using the lattice data itself.
Finally, we proceed in the same manner to extract the PDF with this newly constructed hybrid-renormalized matrix elements.
If the origin of the discrepancy we see in $\braket{x}$ was the same as that for the PDF itself, this new PDF should agree better with the PDF from the NNPDF4.0 analysis.
However, while we do see a shift in the extracted PDF towards the results from NNPDF4.0 for large $x$, the difference is less than $1\sigma$ and, therefore, cannot fully explain the discrepancy we see in the PDFs.
\section{Conclusions}
\label{sec:conc}

In this paper, we have extracted information on the unpolarized isovector quark PDF of the proton from lattice QCD using various methods.
This is the first such work to utilize an NNLO matching coefficient in a calculation performed with physical quark masses directly. The significance of the NNLO term is demonstrated in Figs. \ref{fig:fixed_z} and \ref{fig:pdf_order_dep} for the Mellin moments and the PDF, respectively. The difference between matching at NLO and NNLO (within the valid $x$-range, in the case of the PDF) is small but non-negligible. This demonstrates good convergence of the matching process. With a physical pion mass, excited state contamination in the three-point functions is significant, and we therefore had to utilize several fitting methods to ensure that effects from these unwanted states are under control.
From the bare matrix elements, we then used the ratio scheme along with the leading-twist OPE approximation to extract the lowest four moments of the PDF.
We found the lowest moment $\braket{x}$ to lie above the result from the global phenomenological fits performed by NNPDF4.0~\cite{NNPDF:2021njg}, which is quite typical in several other lattice calculations.
The effect is likely due to unaccounted-for systematics (e.g. discretization effects, higher-twist contributions, etc.).

Next, we determined the $x$-dependent PDFs utilizing a DNN to solve the inverse problem arising in the pseudo-PDF approach.
Although we found agreement with the results from NNPDF4.0, the errors were significantly larger, which suggests that our statistical errors dominated any potential systematic error.

Our final method involved performing the matching directly in momentum space using the hybrid renormalization scheme with LaMET.
The results show some tension with both our DNN results and the global results from NNPDF4.0, even in the region of $x$ in which we expect power corrections to be under control.
This is not entirely unexpected, as we did not observe a convergence in the PDF results as $P_z$ was increased, suggesting a need for larger $P_z$.
Further, since these calculations were performed with a single ensemble, the size of discretization effects has not been determined. Despite these limitations, the tension we see in our results is not substantial enough to question the validity of the methods and is a promising first step towards a more thorough investigation.

Our future work will include more statistics, more ensembles, and larger momenta, which will allow for control over the remaining unaccounted-for systematics in our current calculations.

\section*{Acknowledgments}

ADH acknowledges helpful discussions with Konstantin Ottnad and Andr\'e Walker-Loud. JH acknowledges helpful discussions with Rui Zhang.

This material is based upon work supported by The U.S. Department of Energy, Office of Science, Office of Nuclear Physics through \textit{Contract No.~DE-SC0012704}, \textit{Contract No.~DE-AC02-06CH11357}, and within the frameworks of Scientific Discovery through Advanced Computing (SciDAC) award \textit{Fundamental Nuclear Physics at the Exascale and Beyond} and the Topical Collaboration in Nuclear Theory \textit{3D quark-gluon structure of hadrons: mass, spin, and tomography}.
SS is supported by the National Science Foundation under CAREER Award PHY-1847893 and by the RHIC Physics Fellow Program of the RIKEN BNL Research Center.
YZ is partially supported by an LDRD initiative at Argonne National Laboratory under Project~No.~2020-0020.

This research used awards of computer time provided by the INCITE program at Argonne Leadership Computing Facility, a DOE Office of Science User Facility operated under Contract No.~DE-AC02-06CH11357. 
This work used the Delta system at the National Center for
Supercomputing Applications through allocation PHY210071 from the
Advanced Cyberinfrastructure Coordination Ecosystem: Services \& Support
(ACCESS) program, which is supported by National Science Foundation
grants \#2138259, \#2138286, \#2138307, \#2137603, and \#2138296.
Computations for this work were carried out in part on facilities of the USQCD
Collaboration, which are funded by the Office of Science of the
U.S. Department of Energy.
Part of the data analysis are carried out on Swing, a high-performance computing cluster operated by the Laboratory Computing Resource Center at Argonne National Laboratory.

The computation of the correlators was carried out with the \texttt{Qlua} software suite~\cite{qlua}, which utilized the multigrid solver in \texttt{QUDA}~\cite{Clark:2009wm,Babich:2011np}.
The analysis of the correlation functions was done with the use of \texttt{lsqfit}~\cite{lsqfit:11.5.1} and \texttt{gvar}~\cite{gvar:11.2}.
Several of the plots were created with \texttt{Matplotlib}~\cite{Hunter:2007}.

\appendix

\section{\texorpdfstring{$\chi^2$}{chi2} and the details of the DNN training}\label{app:DNNchisq}

In this work, our ratio scheme renormalized matrix elements $\mathcal{M}_{\gamma_t} (\lambda, z^2; P_z^0)$ used $P_z^0=0$ which is the standard reduced pseudo-ITD~\cite{Orginos:2017kos,Joo:2019jct,Joo:2020spy,Bhat:2020ktg,Karpie:2021pap,Egerer:2021ymv,Bhat:2022zrw}, denoted here as $\widetilde{Q}(\lambda, z^2) =\mathcal{M}_{\gamma_t} (\lambda, z^2; P_z^0=0)$, which can be related to the PDF by,
\begin{align}
\begin{split}
    \widetilde{Q}(\lambda, z^2)
= &\int_{-1}^1d\alpha\int_{-1}^1 dy\frac{\mathcal{C}(\alpha, \mu^2z^2)}{C_0(\mu^2z^2)} e^{-iy \alpha\lambda} q(y,\mu)\\
= &\int_{-1}^1 dy q(y)
    \frac{\int_{-1}^1 d\alpha e^{-iy \alpha\lambda} \mathcal{C}(\alpha,\mu^2z^2)}{C_0(\mu^2z^2)} \\
=&\int_{-1}^1dyq(y) 
    \overline{\mathcal{C}}(y \lambda,\mu^2z^2)\,,
\end{split}
\end{align}
where $C_0(\mu^2z^2)$ is the zeroth-order Wilson coefficient, and
\begin{align}
\overline{\mathcal{C}}(\lambda,\mu^2z^2) 
\equiv\;&
    \frac{\int_{-1}^1 d\alpha e^{-i\lambda\alpha}\ \mathcal{C}(\alpha,\mu^2z^2)}{C_0(\mu^2z^2)} .
\end{align}
One then can study the real and imaginary part of $\widetilde{Q}(\lambda, z^2)$ separately. Here we only discuss the real part while the imaginary part can be derived by changing $\cos(\lambda\alpha)$ to $\sin(\lambda\alpha)$. Up to NNLO, we express the convolution kernel as
\begin{align}
\begin{split}
&C_\text{NNLO}(\lambda, \mu^2 z^2)
\equiv   \int_{-1}^1\!\! d\alpha \cos(\lambda\alpha)\, \mathcal{C}(\alpha,\mu^2z^2) \\
&=\cos(\lambda)\, d_\text{NNLO}(\mu^2 z^2)\\
&+   \int_{0}^1\!\! d\alpha \big[\cos(\lambda\alpha )-\cos(\lambda)\big]\, n_\text{NNLO}(\alpha, \mu^2 z^2) ,
\end{split}
\end{align}
with
\begin{align}
\begin{split}
C_\text{NNLO,0}(\mu^2 z^2) 
\equiv\;&
    C_\text{NNLO}(\lambda=0, \mu^2 z^2) 
=   d_\text{NNLO}(\mu^2 z^2)\,,
\end{split}
\end{align}
where
\begin{align}
\begin{split}
d_\text{NNLO}(\mu^2 z^2) 
=\;&
    1 + \sum_{i=1}^{2}\sum_{j=0}^{i} 
    \Big(\frac{\alpha_s}{2\pi}\Big)^i\,
    d_{i,j} \,
    \Big[L(\mu^2 z^2)\Big]^j \,,
\end{split}\\
\begin{split}
n_\text{NNLO}(\alpha, \mu^2 z^2) 
=\;&
    \sum_{i=1}^{2}\sum_{j=0}^{i} 
    \Big(\frac{\alpha_s}{2\pi}\Big)^i\,
    n_{i,j}(\alpha) \,
    \Big[L(\mu^2 z^2)\Big]^j \,, 
\end{split}
\end{align}
where
\begin{align}
&L(\mu^2 z^2) \equiv 2\gamma_E + \ln \frac{\mu^2 z^2}{4},
\end{align}
\begin{align}
&d_{1,0}=\frac{5C_F}{2},
\end{align}
\begin{align}
&d_{1,1}=\frac{3C_F}{2},
\end{align}
\begin{align}
\begin{split}
&d_{2,0}=C_F\Bigg[\frac{469\beta_0}{48}+\frac{223C_F-94C_A}{96}\\
&+\frac{\pi^2 (8C_F-15C_A)}{36}+2(C_A-4C_F)\zeta(3)\Bigg],
\end{split}
\end{align}
\begin{align}
\begin{split}
&d_{2,1}=\Bigg[C_F^2\left(-\frac{5}{8}+\frac{2\pi^2}{3}\right)+C_FC_A\left(\frac{49}{24}-\frac{\pi^2}{6}\right) \\
&-\frac{5}{6}C_Fn_fT_F \Bigg]+ \frac{5 C_F}{2}\Big(\frac{3}{2}C_F + \beta_0 \Big),\\
\end{split}
\end{align}
\begin{align}
&d_{2,2}=\frac{3C_F}{4}\Big(\frac{3}{2}C_F + \beta_0 \Big),\\
&n_{1,0}(\alpha)=C_F\Big[2(1-\alpha)-\frac{1+\alpha^2}{1-\alpha}-\frac{4\ln(1-\alpha)}{1-\alpha}\Big],\\
&n_{1,1}(\alpha)=-C_F\frac{1+\alpha^2}{1-\alpha} ,
\end{align}
as well as $n_{2,0}(\alpha)$, $n_{2,1}(\alpha)$ and $n_{2,2}(\alpha)$ in more complicated forms. The constants in the formulas are $C_F=4/3$, $T_F=1/2$, $C_A=3$, $n_f=3$ (3 flavor in this work) and $\beta_0=(11C_A-4n_fT_F)/6$. Then we denote, and numerically compute, that
\begin{align}
    c_{i,j}(\lambda) \equiv \int_{0}^1\!\! d\alpha \big[\cos(\lambda\alpha)-\cos(\lambda)\big] n_{i,j}(\alpha)\,,
\end{align}
which gives,
\begin{align}
\begin{split}
&\overline{C}_\text{NNLO}(\lambda,\mu^2 z^2) \\
=\;&
    \cos(\lambda) + 
    \frac{\int_{0}^1\!\! d\alpha \big[\cos(\lambda\alpha)-\cos(\lambda)\big]\, n_\text{NNLO}(\alpha, \mu^2 z^2)}{d_\text{NNLO}(\mu^2 z^2)} \\
=\;&
    \cos(\lambda) + 
    \frac{\sum_{i=1}^{2}\sum_{j=0}^{i}  (\alpha_s/2\pi)^i\,c_{i,j}(\lambda) \, L^j(\mu^2 z^2)}{1 + \sum_{i=1}^{2}\sum_{j=0}^{i}  (\alpha_s/2\pi)^i\,d_{i,j} \, L^j(\mu^2 z^2)}
\,.
\end{split}
\end{align}

In this work, we express the PDF $q^-(x)$ by DNN together with extra factors,
\begin{align}
    q^-_\text{DNN}(y) \equiv N\,y^{\alpha^-}(1-y)^{\beta^-}\big[1+\delta\sin(f^-_\text{DNN}(y,\boldsymbol{ \theta}^-))\big]\,, %N\,y^{-e^\alpha}\,(1-y)^\beta\,\exp\big[f_\text{DNN}(\boldsymbol{\theta};y)\big]\,,
\end{align}
with $\delta$ being some number that limits the contribution of subleading terms. The normalization condition yields
\begin{align}
\begin{split}
    &N(\alpha^-,\beta^-,\boldsymbol{\theta}^-) 
    \\=& \bigg\{2\int_0^1 dy\,y^{\alpha^-}\,(1-y)^{\beta^-}\,\big[1+\delta\sin(f^-_\text{DNN}(y,\boldsymbol{\theta}^-))\big]\bigg\}^{-1}\,.
\end{split}
\end{align}
The loss function is defined as
\begin{align}
J(\alpha^-,\beta^-,\boldsymbol{\theta}^-) 
\equiv \;&
    \frac{\eta}{2} \boldsymbol{\theta}^-\cdot\boldsymbol{\theta}^-
    +\frac{1}{2} \chi^2(\alpha^-,\beta^-,\boldsymbol{\theta}^-),
\end{align}
where
\begin{align}
\begin{split}
\chi^2(\alpha^-,\beta^-,\boldsymbol{\theta}^-)
&\equiv
    \sum_{i,j}  \big[\text{Cov}^{-1}\big]_{ij} \Bigg[\widetilde{Q}_\text{DNN}(P_i,z_i) - \widetilde{Q}(P_i,z_i)\Bigg]\\
    &\times
    \Bigg[\widetilde{Q}_\text{DNN}(P_j,z_j) - \widetilde{Q}(P_j,z_j)\Bigg].
\end{split}
\end{align}

\newpage

\bibliography{refs}
\end{document}